\documentclass{jfm}

\usepackage{graphicx}
\usepackage{newtxtext}
\usepackage{caption}
\usepackage{natbib}
\usepackage{newtxmath}
\usepackage{subfigure}
\usepackage[final]{changes}

\usepackage{xcolor}
\usepackage[normalem]{ulem} 
\newcommand\hl{\bgroup\markoverwith
	{\textcolor{yellow}{\rule[-.5ex]{2pt}{2.5ex}}}\ULon}

\DeclareMathAlphabet{\mathpzc}{OT1}{pzc}{m}{it}
\usepackage{bm}      
\usepackage{bbm}     
\usepackage{dsfont}  
\usepackage{yfonts}  

\newcommand{\RomanNumeralCaps}[1]

\RequirePackage{natbib}
\RequirePackage{etoolbox}
\usepackage{lineno}
\RequirePackage{hyperref}
\hypersetup{
	colorlinks = true,
	urlcolor   = blue,
	citecolor  = blue,
	linkcolor=blue
}

\makeatletter

\patchcmd{\NAT@citex}
{\@citea\NAT@hyper@{%
		\NAT@nmfmt{\NAT@nm}%
		\hyper@natlinkbreak{\NAT@aysep\NAT@spacechar}{\@citeb\@extra@b@citeb}%
		\NAT@date}}
{\@citea\NAT@nmfmt{\NAT@nm}%
	\NAT@aysep\NAT@spacechar\NAT@hyper@{\NAT@date}}{}{}

\patchcmd{\NAT@citex}
{\@citea\NAT@hyper@{%
		\NAT@nmfmt{\NAT@nm}%
		\hyper@natlinkbreak{\NAT@spacechar\NAT@@open\if*#1*\else#1\NAT@spacechar\fi}%
		{\@citeb\@extra@b@citeb}%
		\NAT@date}}
{\@citea\NAT@nmfmt{\NAT@nm}%
	\NAT@spacechar\NAT@@open\if*#1*\else#1\NAT@spacechar\fi\NAT@hyper@{\NAT@date}}
{}{}

\makeatother

\title{Transition reversal over a blunt plate at Mach 5}

\author{Peixu Guo\aff{1}, Jiaao Hao\aff{1} \and Chih-Yung Wen\aff{1}}

\affiliation{\aff{1}Department of Aeronautical and Aviation Engineering, The Hong Kong Polytechnic University, Kowloon, Hong Kong SAR}

\begin{document}
	\maketitle

\begin{abstract} 
In this work, the stability and transition to turbulence over \replaced{a}{an experimental} blunt flat plate with different leading-edge radii are investigated\added{ computationally. The benchmark experimental work for comparative studies is conducted by Borovoy \etal\ (\textit{AIAA J.}, vol. 60, 2022, pp. 497--507)}. The freestream Mach number is 5, the unit Reynolds number is $6\times10^7$  m$^{-1}$, and the maximum nose-tip radius 3 mm exceeds the experimental reversal value. High-resolution numerical simulation and stability analysis are performed. Three-dimensional broadband perturbation is added on the farfield boundary to initiate the transition. The highlight of this work is that the complete physical process is considered, including the three-dimensional receptivity, linear and nonlinear instabilities, and transition. The experimental reversal phenomenon is favourably reproduced in the numerical simulation for the first time. Linear stability analysis shows that unstable first and second modes are absent in the blunt-plate flows owing to the presence of the entropy layer, although these modes are evident in the sharp-leading-edge case. Therefore, the transition on the blunt plate is due to nonmodal instabilities. Numerical results for all the blunt-plate cases reveal the formation of streamwise streaky structures downstream of the nose (stage I) and then the presence of intermittent turbulent spots in the transitional region (stage II). In stage I, a preferential spanwise wavelength of around 0.9 mm is selected for all the nose-tip radii, and low-frequency components are dominant. In stage II, high-frequency secondary instabilities appear to grow, which participate in the eventual breakdown. By contrast, leading-edge streaks are not remarkable in the sharp-leading-edge case, where transition is induced by oblique first and Mack second modes. The transition reversal beyond the critical nose-tip radius arises from an increasing magnitude of the streaky response in the early stage, while the transition mechanism keeps similar qualitatively.
\end{abstract}


\section{Introduction}
\label{sec:intro}
Hypersonic boundary layer stability and transition have drawn extensive attention due to fundamental and engineering importance. The drag force and the surface heat flux of hypersonic vehicles can be increased by several times after the transition to turbulence. Hence, it is of great interest to predict or control the hypersonic boundary layer transition. Possible transition processes include receptivity to external disturbances, transient growth, eigenmodal growth, parametric resonance and mode--mode interactions, breakdown to turbulence, and bypass mechanisms \citep{morkovin1994transition}. The detailed transition path depends on the level of environmental disturbances. 

During the design of hypersonic vehicles, the leading edge is usually blunted to mitigate local heating. In the meantime, approximate nose bluntness also delays the boundary layer transition, which favourably reduces aerothermal load. In terms of the physical mechanism, theoretical and computational studies have demonstrated that the appearance of the entropy layer near the bow shock has a stabilisation effect on the main instability modes of boundary layers  \citep*{reshotko1980stability,malik1990effect,zhong2002receptivity,zhong2006boundary}. According to linear stability analysis, first and second modes are usually the dominant instability modes in supersonic and hypersonic boundary layers, respectively \citep{mack1984boundary}. The two modes are of vortical and acoustic nature, respectively. With increased nose bluntness, these two modes tend to be stabilised near the nose region due to increasingly favourable pressure gradient, decreased local Reynolds number and lower Mach number. Furthermore, another inviscid-type instability can be supported inside the entropy layer, which is called the entropy-layer mode \citep{Dietz1999Entropy}. Some literatures have found that the entropy-layer normal mode is not remarkable enough to resist the overall stabilisation effect of increasing nose bluntness \citep{paredes2019nose,wan2018response,wan2020receptivity,chen2021receptivity}. Unstable entropy-layer
mode exists in the nose region with low frequencies and small \replaced{growth}{grow} rates\added{, which is a less significant discrete mode in the blunt-body flow} \citep{wan2020receptivity,chen2021receptivity}. It should be \replaced{cautioned}{cautious} that the entropy-layer normal mode is not identical to the entropy-layer instability, where the latter may be of nonmodal nature.

With regard to experimental efforts, \cite{Stetson1983} performed a systematic study of the nose bluntness effect on the transition, which is a continuation of an earlier publication \citep{Stetson1967}. Three facilities, including two wind tunnels and one shock tunnel, produced the same data trend of the bluntness effect. The slender cone models were tested at various Mach numbers and unit Reynolds numbers with different nose-tip radii. The transition locations were obtained from measurements of surface heat transfer. The informative data from Stetson display two distinct regimes. One is the small-bluntness regime, where the transition location moves
downstream with increasing bluntness. The other is the large-bluntness regime, where
the transition location moves upstream rapidly. This phenomenon is called transition reversal. The reversal performance is clear in a $\Rey_\textit{R}$--$\Rey_t$ plot, where $\Rey_\textit{R}$ and $\Rey_t$ are the horizontal and the vertical axis, respectively. Here, $\Rey_t$ is the Reynolds number based on the freestream condition and the transition onset location $x^{\ast}_t$, $\Rey_\textit{R}$ is based upon the nose-tip radius $R_\textit{n}^{\ast}$, and the asterisk denotes dimensional quantities. The critical reversal value for the horizontal axis is referred to as $\Rey_{\textit{R},\textit{c}}$. Owing to the stabilisation effect of the entropy layer, the small-bluntness regime is conceivable. In the large-bluntness regime ($\Rey_\textit{R}>\Rey_{\textit{R},\textit{c}}$), the transition onset becomes more sensitive to the roughness effect. Stetson speculated that early frustum
transition for the large-bluntness regime was dominated by disturbances originating near the nose tip, and that these disturbances were closely related to the roughness effect. Stetson also reminded that model nose-tips were polished before each
run. However, this operation only removed surface material \replaced{protrusion}{protrudion}, while cavities remained after repetitive runs. Therefore, the roughness effect was not eliminated in a rich dataset.

Although the large-bluntness regime may be sensitive to roughness, there is no solid evidence that roughness is a necessary condition for transition reversal. In addition to Stetson's work, the transition measurements on blunt cones \citep{Softley1969Experimental,Ericsson1988Effect,Zanchetta1996kinetic,Aleksandrova2014,Marineau2014Mach,paredes2019nose}, ogive-cylinders \citep{Hill2022Experimental} and blunt flat plates \citep{Lysenko1990Influence,borovoy2022laminar} have reported transition reversal in different facilities. Part of them explicitly claimed that model surfaces had been polished in certain runs \citep{Zanchetta1996kinetic,paredes2019nose,borovoy2022laminar}. The root-mean-square (r.m.s.) roughness height was within/around a micrometer after polishing. Among these studies, \cite{Zanchetta1996kinetic} detected less frequent events of turbulent bursts after surface polishing. \cite{paredes2019nose} found that polishing led to a laminar flow at large bluntness ($R_\textit{n}^{\ast}=15.24\ \rm{mm}$), whereas the existence of distributed roughness gave rise to early transition. However, their results did not exclude the possibility that $\Rey_{\textit{R},\textit{c}}$ might be shifted to a higher value at runs with polishing. With polished models, \cite{borovoy2022laminar} reported continuous occurrence of transition reversal with different leading-edge shapes. In other words, transition reversal was not eliminated over a relatively smooth wall. Experimental scientists also pointed out that the transition reversal behaviour can be affected by multiple factors, such as freestream turbulent intensity, surface roughness, pressure gradient,
wall temperature, flow separation, etc. Nevertheless, the dominant factor is difficult to determine.

In terms of mechanism-related measurements, \cite{Marineau2014Mach} examined the
boundary-layer instabilities over sharp and blunt cones by PCB sensors at Mach 10. Beyond the critical nose-tip radius, the transition occurred ahead of the appearance of second-mode instabilities, and the pressure signature was weak. At large bluntness, both \cite{Stetson1983} and \cite{Marineau2014Mach} observed the onset of transition that was away from the entropy swallowing point and close to the nose region.  Schlieren images of \cite{Jagde2019Visualizations} and \cite{Kennedy2019Visualizations} revealed that the second-mode rope-like structures for sharp cones were replaced by wisp-like structures above the boundary layer edge for largely blunted cones.  Laser-induced-fluorescence-based
schlieren measurements by \cite{Grossir2014Hypersonic} over blunt cones also reported flow structures that were dissimilar to second-mode ones over sharp cones. More recent work of \cite{Kennedy2022Characterization} ascribed the appearance of elongated structures above the boundary layer at large bluntness to nonmodal instabilities.

Regardless of the confirmed reversal phenomenon, it is difficult to uncover the physical mechanism by the wind-tunnel experiment only. Challenges include insufficient resolution of flow field, decreased reproducibility of large-bluntness early transition, and reliable theoretical tools that correlate the experimental data satisfactorily. Parallel and nonparallel stability analyses have shown that modal amplification of first-mode, second-mode,
and entropy-layer instabilities is not strong enough to account for transition reversal including Stetson's experiment \citep{malik1990effect,Marineau2017prediction,Jewell2017Boundary,paredes2019nose,Paredes2020mechanism}. Due to the non-orthogonality of the linearised Navier--Stokes (N--S) equation, nonmodal instabilities may exist even if the flow is modally stable. \cite{reshotko2000blunt,Reshotko_2004} used the optimal transient growth theory to explain wind-tunnel observations such as nose-tip transition. The transient growth analysis provides the upper bound of the propagating disturbance \replaced{kinetic energy}{amplitude}, which can be escalated by two to four orders of magnitude \citep{reshotko2001transient}. Following that, \cite{paredes2019nose} attempted to explain transition reversal in this framework. Stationary disturbances, originating from the nose tip and propagating in the entropy layer, were reported to experience significant nonmodal amplification over largely blunted cones at Mach 6. Over various blunt bodies, transient growth analysis has also shown pronounced nonmodal instabilities that may be connected with experimental early transition \citep{paredes2017blunt,paredes2018blunt,paredes2019nonmodal,quintanilha2022transient}. 

Another popular framework is the resolvent analysis (input--output analysis), which seeks the optimal response of the linear system to external input forcings \citep{monokrousos2010global,bugeat20193d,bae2020resolvent,lugrin2021transition,hao2023response,Guo_2023,caillaud2024separation}. By computational fluid dynamics (CFD) and resolvent analysis, \cite{Melander2022Nose} identified 3-D low-frequency streamwise streaks near the nose region of the blunt cone. Furthermore, the streaky response becomes stronger with a larger nose-tip radius. In spite of the \deleted{shed} new insights, the transient growth analysis or resolvent analysis does not provide direct evidence for transition reversal. In real-life situations, the input forcing or the inflow optimal disturbance that leads to maximal energy amplification may be `not physically realizable' \citep*{kamal2023global}. Recently, \cite{cook2022free,Cook2024} have included the shock/disturbance interaction in the resolvent analysis. At Mach 5.8, the improved framework demonstrated that the receptivity of the blunt-cone flow to freestream disturbances is of highly three-dimensional (3-D) nature. In those cases, energetic first-mode and entropy-layer instabilities were identified with tens of kilohertz. 

In recent decades, CFD has been an effective tool to throw light on flow mechanisms. For the concerned high-speed blunt-body flows, numerous direct numerical simulations (DNS) have been conducted to reveal the receptivity process \citep{zhong2002receptivity,zhong2006boundary,kara2011effects,balakumar2015receptivity,balakumar2018transition,he2021hypersonic,ba2023hypersonic} and the breakdown scenario \citep{Paredes2020mechanism,Hartman2021Nonlinear,Goparaju2022Role,zhu2023direct}. The nonlinear interaction induced by oblique waves, propagating inside the entropy layer, was found to be significant to the final transition.   \cite*{Hartman2021Nonlinear} compared the numerical inclined structure in the entropy layer with the experimental schlieren image, and reached a qualitative agreement. The lift-up and Orr-like mechanisms were deduced to be responsible for the amplification of entropy-layer instabilities \citep{Goparaju2022Role}. The mechanisms causing the initiation of entropy-layer instabilities were not studied. Among the DNS studies on nonlinear stages, the blunt-cone flow by \cite{zhu2023direct}  and the ogive-cylinder forebody flow by \cite{aswathy2024effects} included the effect of a varying nose-tip radius. However, with perturbations added downstream of the shock, they did not capture the transition reversal. The authors suspect that the failure to reproduce transition reversal is attributed to the absence of upstream receptivity. Applying two-dimensional (2-D) DNS, \cite*{goparaju2021effects} seeded random pressure noise in front of the detached shock of an experimental blunt flat plate at Mach 6. Beyond the critical nose-tip radius, the frequency of the most amplified disturbance deviates from the second-mode semi-empirical value. The resulting growth rate is increased compared to small-bluntness cases, and the temperature fluctuation turns to peak outside the boundary layer. By employing tunnel-like noise on the freestream boundary of 3-D DNS, \cite{liu2022interaction} reported inclined structures that resembled experimental schlieren observations over a blunt cone at Mach 8.
Inspired by recent progress, the authors attach importance to the inclusion of upstream receptivity when studying the transition reversal.

In summary, current studies only provide indirect supports for the reason of transition reversal. A more direct connection between the informative CFD/theoretical results and the experimental transition measurement \replaced{needs}{is expected} to be established. To the best knowledge of the authors, no high-fidelity numerical simulation has successfully reproduced and explained the transition reversal in wind-tunnel experiments, which is the objective of this paper. To this end, the complete physical process from the shock/diturbance interaction to the transition to turbulence is explored. The detailed flow physics is envisaged to be revealed. The article is organised as follows. The investigated physical problem and flow conditions are described in \S\space\ref{sec:Problem description}. The numerical and theoretical tools are introduced in \S\space\ref{sec:Methodology}. The simulation strategy and numerical details are shown in \S\space\ref{sec:strategy}. A comparison with experimental transition data is first depicted in \S\space\ref{sec:Phenomenon}. The flow mechanisms and the related discussions are displayed in \S\space\ref{sec:mechanism}. Concluding remarks are given in \S\space\ref{sec:Conclusions}. A brief study of mesh convergence is provided in Appendix \ref{appA}, and additional information is shown in Appendixes \replaced{\ref{appB}--\ref{appE}}{\ref{appB} and \ref{appC}}.

\section{Problem description}\label{sec:Problem description}

An experimental flat plate model with a cylindrically blunted or sharp leading edge is investigated. The tests were conducted in a Ludwig-type wind tunnel UT-1M in Russia by \cite{borovoy2022laminar}. Figure \ref{fig1} gives a schematic drawing of the simulated problem. The difference from the blunt-cone flow is that, entropy swallowing does not seem to occur within a practical distance in the considered blunt-flat-plate flow. Therefore, the impact of the entropy layer is persistent, and the effect of entropy swallowing is excluded. Furthermore, no surface roughness is placed in the present numerical simulation. Thus, the nose bluntness effect tends to be isolated in the flat-plate configuration, which simplifies the problem. The angles of attack and sideslip are considered to be zero. A Cartesian coordinate system ($x$, $y$, $z$) is constructed with the origin at the centre of the cylindrical nose, corresponding to the streamwise, wall-normal and spanwise velocities ($u$, $v$, $w$). An orthogonal body-fitted coordinate system ($\xi$, $\eta$, $z$) is also defined, which is along the wall-tangent, wall-normal and spanwise directions, respectively. The freestream conditions of the tunnel are given as follows: Mach number $M_{\infty} =5$, total temperature $T^{\ast}_{0} = 468$ K, static temperature $T^{\ast}_{\infty} = 78$ K and unit Reynolds number $\Rey^{\ast}_{\infty} =6\times10^7$  m$^{-1}$. The subscript `$\infty$' refers to the freestream quantity. The suggested surface temperature is the room temperature $T^{\ast}_\textit{w} =293$ K, which corresponds to $T^{\ast}_\textit{w}/T^{\ast}_{0} \approx0.626$. The subscript `$\textit{w}$' represents the quantity at the wall.

In this paper, the primitive variables are nondimensionalised by the corresponding freestream quantities except that the pressure $p$ is by the freestream $\rho^{\ast}_{\infty } {u}^{\ast 2}_{\infty }$, where $\rho$ represents density. The reference length scale for nondimensionalisation is $L^{\ast}_\textit{ref}=1$ mm, which is in the same order of magnitude as the downstream boundary layer thickness. Under the current setting of $\Rey^{\ast}_{\infty} =6\times10^7$  m$^{-1}$, the critical nose-tip radius for transition reversal in the experiment is estimated to be about \replaced{$R_{\textit{n},\textit{critical}}^{\ast}=1.19$}{$R_{\textit{n},\textit{critical}}^{\ast}=2.37$}  mm. Different nose-tip radii $R_\textit{n}^{\ast}$ are considered, namely 0 (sharp leading edge), \added{1.8 mm, }2 mm, 2.7 mm and 3 mm. Thus, nose-tip radius that is either smaller or larger than the critical value has been considered.

\section{Methodology overview}\label{sec:Methodology}

The 3-D compressible N--S equations in the Cartesian coordinate system can be written in a dimensionless conservation form:
\begin{equation}\label{NSeq}
\frac{{\partial {\bm{Q}}}}{{\partial t}} + \frac{{\partial {\bm{F}}}}{{\partial x}} + \frac{{\partial {\bm{G}}}}{{\partial y}} + \frac{{\partial {\bm{H}}}}{{\partial z}} = \frac{1}{\Rey}\left(\frac{{\partial {\bm{F}_v}}}{{\partial x}} + \frac{{\partial {\bm{G}_v}}}{{\partial y}} + \frac{{\partial {\bm{H}_v}}}{{\partial z}}\right),
\end{equation}
where $t$ denotes time, ${\bm{Q}} = (\rho, \rho u, \rho v, \rho w, \rho E)^\text{T}$ is the vector of conservative variables, ${\bm{F}}$, ${\bm{G}}$ and ${\bm{H}}$ represent the vectors of inviscid fluxes, and ${\bm{F}_v}$, ${\bm{G}_v}$ and ${\bm{H}_v}$ refer to the vectors of viscous fluxes. Detailed expressions of the fluxes can be found in \cite{anderson1995computational}. The symbol $E$ represents the total energy per unit mass, and the superscript `T' indicates matrix transpose. A calorically perfect gas (air) model is assumed with a constant specific heat ratio of $\gamma = 1.4$. Sutherland's law is adopted to calculate the dynamic viscosity $\mu$, and then the thermal conductivity $\kappa$ is computed with a constant Prandtl number $\textit{Pr}=0.72$. Simulations of the 2-D laminar base flow, the 2-D instability and the full 3-D transitional flow are performed using an in-house finite-volume-based solver called PHAROS \citep{Hao_2016,Hao_Wen_2020}. This solver has been well applied and validated in relevant physical problems including the 3-D instability of double-cone flows \citep{Hao_2022} and transitional flat-plate boundary layers \citep*{Guo_2023}. 

\subsection{Numerical simulation}\label{sec:DNS}
\begin{figure}
	\centering{ \includegraphics[height=4.8cm]{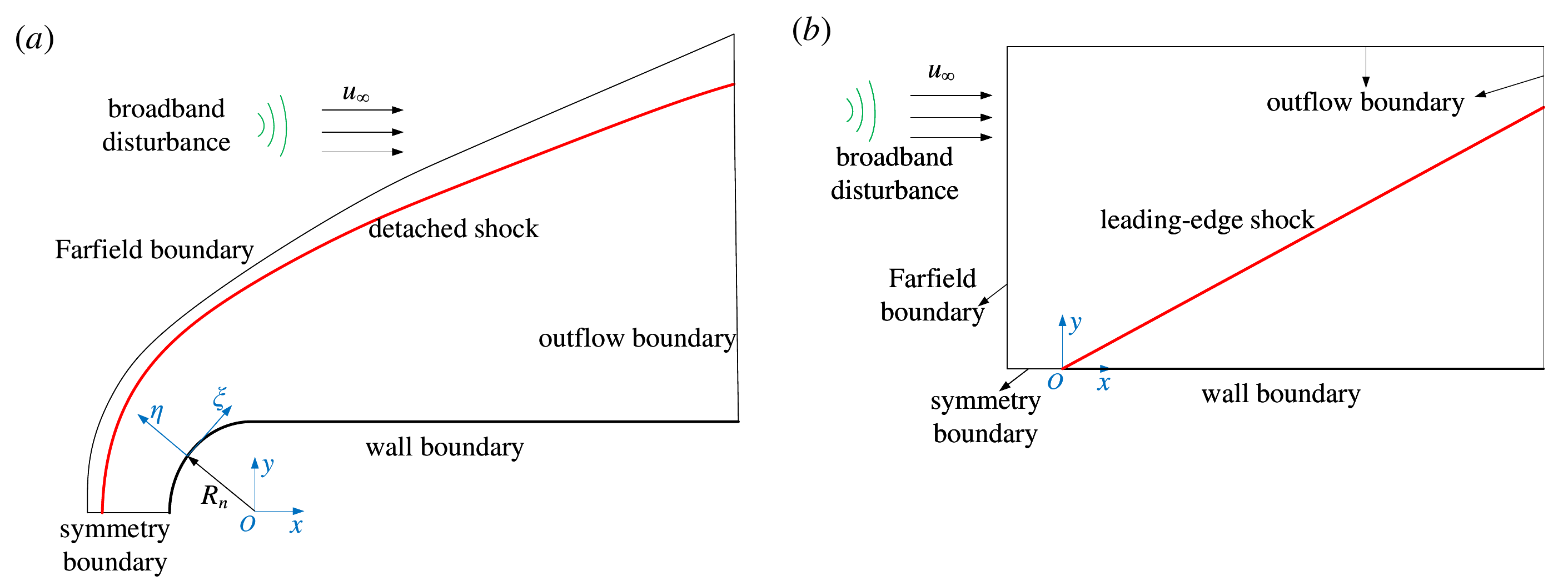}}
	\caption{Schematic drawing of the simulated flow over a (\textit{a}) blunt plate and (\textit{b}) sharp-leading-edge plate (not to scale).}
	\label{fig1}
\end{figure}

In wind-tunnel experiments such as \cite{Stetson1983} and \cite{borovoy2022laminar}, the transition onset Reynolds number at the critical nose-tip radius is generally in the order of $10^7$. As a result, it is computationally expensive to perform DNS with several nose-tip radii. In this work, we conduct affordable direct numerical simulations, and follow the DNS setup of the finest resolution of \cite*{Pirozzoli2004Direct} in a supersonic turbulent flat-plate boundary layer. In detail, a combination of dimensionless grid spacings $\Delta x^+=10\sim20$, $\Delta y_\textit{w}^+<1$ and $\Delta z^+<10$ and a seventh-order upwind-biased numerical scheme for construction of inviscid fluxes will be applied. Here, during the evaluation of $\Delta x^+$, $\Delta y_\textit{w}^+$ and $\Delta z^+$, the friction velocity $u_{\tau}$ is calculated based on the turbulent skin friction coefficient $C_f$ via van Driest II correlation\added{, which is taken at a reference downstream location $x=400$}. The correlation of $C_f$ is a function of the momentum thickness Reynolds number $\Rey_{\theta}$. The detailed evaluation procedures can been found in \cite{Franko_Lele_2013} and \cite{Guo_Heat_2022}. 

Note that the pre-transitional and transitional regions rather than the fully developed turbulent region are the main research concerns of this paper. A mesh convergence study is conducted, and the results are displayed in Appendix \ref{appA}. In general, the adopted mesh resolution is sufficient to achieve the research objective. It should be remarked that the frequency spectra and the amplitude of incoming disturbances were not given in most experimental studies, including the considered one by \cite{borovoy2022laminar}. The frequency-dependent spectra are usually difficult to measure accurately in the wind tunnel. Thus, with insufficient physical information, it is demanding and fortuitous to reproduce the accurate transition locations via numerical simulation. As mentioned by \cite{Stetson1983}, what is useful should be the relative change and trend of data rather than the transition Reynolds number itself. As a consequence, the present numerical simulation is expected to reproduce the reversal phenomenon (tendency) rather than the exact transition locations.

\subsection{Linear stability analysis}
Compressible linear stability theory (LST) is utilised to identify the normal-mode instability. The instantaneous flow field can be represented as the sum of mean and fluctuating quantities. The variable vector $\bm{\phi}=(u,v,p,w,T)^\text{T}$, can be decomposed into
\begin{equation}
\bm{\phi}(\xi,\eta,z,t)=\bar{\bm{\phi}}(\xi,\eta)+\bm{\phi}'(\xi,\eta,z,t),
\end{equation}
where the overbar represents the time-averaged flow. In the present laminar flow analysis, the base flow $\bar{\bm{\phi}}$ is two-dimensional, i.e., $\bar{w}=0$. With the normal-mode ansatz, the small-amplitude disturbance can be expressed by 
\begin{equation}
\bm{\phi}' = \hat{\bm{\phi}}(\eta)\exp{\left[ {{\rm{i}}\left( {\alpha\xi +\beta z - \omega t} \right)} \right]}+ {\rm{c}}{\rm{.c}}{\rm{.}}
\end{equation}
Here, $\hat{\bm{\phi}}$ represents the eigenfunction, $\alpha$ is the complex streamwise wavenumber, $\beta$ is the spanwise wavenumber, $\omega$ is the angular frequency, and c.c. denotes complex conjugate. The linearised N--S equation can be derived from (\ref{NSeq}). Under the quasi-parallel flow assumption, the linear stability equation can be then derived and written as  \citep{malik1990numerical}
\begin{equation}\label{LSTequation}
{\mathsfbi H}_{\eta\eta}\frac{{\partial^2\hat{\bm{\phi}} }}{{\partial\eta^2}}+{\mathsfbi H}_{\eta}\frac{{\partial\hat{\bm{\phi}} }}{{\partial\eta}}+{\mathsfbi H}_{0}\hat{\bm{\phi}}  = \bm{0},
\end{equation}
where ${\mathsfbi H}_{\eta\eta}, {\mathsfbi H}_{\eta}, {\mathsfbi H}_{0}$ are $5\times5$ matrices. The boundary condition is given by
\begin{equation}
\left\{ \begin{array}{l}
\hat u = \hat v = \hat w = \hat T = 0,\eta = 0\\[2pt]
\hat u = \hat v = \hat w = \hat T = 0,\eta \to \infty, 
\end{array} \right.
\end{equation}
while $\hat p$ is solvable from the wall-normal momentum equation on the boundary. The stability equation is finally transformed to a complex eigenvalue problem with respect to $\alpha$. The operators of equation (\ref{LSTequation}) are related to $\alpha$, $\beta$ and the local base flow. The local growth rate $\sigma=-\alpha_i$ is positive if the mode is unstable. The linear stability analysis is performed by our in-house code, which has been well validated by benchmark cases \citep{guo2020linear,guo2021sensitivity,Guo_Heat_2022,Guo_2023,Cao2023stability}. A global spectral collocation method is utilised to obtain the global spectrum, and a local algorithm is employed to improve the eigenmodes \citep{malik1990numerical}.

\section{Simulation strategy}\label{sec:strategy}
\subsection{Numerical method}

Figure \ref{fig1} displays the simulation strategy. A shock-capturing method is adopted, following a recent practice in the cross-flow-dominated hypersonic transition subject to free-stream acoustic noise \citep{cerminara2020transition}. For cases with blunted leading edges, the structured mesh is iteratively designed to be entirely aligned with the shock shape near the detached shock. The wall-normal distribution of grid points is clustered near the shock and the wall using a hyperbolic tangent function. In the wall-normal direction, at least 140 points are located in all the fully developed turbulent boundary layers. The grid spacing is uniform in the spanwise direction and mostly in the streamwise direction. In the vicinity of the leading edge, the grid is clustered in the streamwise direction. 

As a  first step, the 2-D laminar base flow is obtained by the same numerical scheme as 3-D simulations, \replaced{except}{expect} that the data parallel-line relaxation method is used for efficient convergence to a steady state \citep{wright1998data,Hao_2016,Hao_2022}. For 3-D cases, mixed numerical schemes are employed. The inviscid flux is reconstructed by the seventh-order upwind scheme in the smooth region away from the shock and away from the nose-tip region in the range $x>0$. The inviscid flux is then calculated using the Harten-Lax-van Leer-Contact (HLLC) Riemann solver \citep*{toro1994restoration}. In the remaining spatial domain, i.e., near the discontinuity detected by a Ducros sensor \citep{Ducros1999} or near the nose tip ($x\le0$), the inviscid flux is reconstructed by the second-order monotonic upstream-centered scheme for conservation laws (MUSCL) scheme with limiters \citep{VANLEER1979101}, and then calculated by the modified Steger--Warming scheme \citep{maccormack2014numerical}.\added{ The modified Steger--Warming scheme reduces to the original one near a strong shock based on a pressure-gradient sensor \citep{scalabrin2005development}.} The above setup ensures that the complete shock region is calculated by the second-order MUSCL scheme. The purpose is to obtain a high resolution in the downstream smooth region in conjunction with a robust solution near the strong discontinuity and the large-gradient flow near the nose tip. The viscous fluxes are discretised by the second-order central scheme, which has been applied in previous hypersonic transition simulations \citep{Guo_Heat_2022,Guo_2023}. To perform time-accurate simulations, the three-stage third-order total variation diminishing Runge--Kutta
method is employed for time marching. The boundary conditions are given as follows:
free-stream conditions are imposed on the far-field boundary, extrapolation is used on the
outflow boundary, and isothermal, no-slip and no-penetration conditions are enforced on
the wall boundary.\added{ The symmetry boundary condition (slip wall condition) is given on the $x<0, y=0$ plane.} The periodic condition is utilised on the spanwise boundary. Next to the outflow boundary, sponge zones are placed to minimise the reflection
of disturbances \citep{mani2012analysis}.

\subsection{Case description}

\begin{table}
	\begin{center}
		\def~{\hphantom{0}}
		\begin{tabular}{lccccccccc}
			Case  &$R_\textit{n}$ &$L_x$ &$L_z$ &$n_x$ &$n_y$ &$n_z$ &$\Delta x^+$ &$\Delta y^+_\textit{w}$ &$\Delta z^+$\\[5pt]
			R0 &0	&170	&12 	&1769	&221	&289	&19.2	&0.58	&8.0	\\
			\added{R1.8} &\added{1.8}	&\added{400}		&\added{12}		&\added{4121}		&\added{241}		&\added{289} 		&\added{19.2}		&\added{0.58}		&\added{8.0}\\
			R2 &2	&260		&12		&2737		&251		&289 		&19.2		&0.58		&8.0\\
			R2.7 &2.7   &260		&12		&2745		&261  	&289  	&19.2		&0.58		&8.0 	\\
			R3 &3   &340		&12		&3549		&271  	&289  	&19.2		&0.58		&8.0 	\\
			R3F &3   &140		&12		&2913		&271  	&289  	&9.6		&0.58		&8.0 	\\
			R2C &2	&260		&8		&2273		&291		&110 		&24.0		&0.58		&14.1\\
		\end{tabular}
		\caption{Case details for numerical simulations.}
		\label{table1}
	\end{center}
\end{table}

Case information about the computational domain and mesh resolution is given in table \ref{table1}. The symbols $L_x$ and $L_z$ represent the length and the width of the computational domain, and $n_x$, $n_y$ and $n_z$ refer to the mesh node numbers in the three directions. Cases R0, \added{R1.8, }R2, R2.7 and R3 are simulated to reveal the bluntness effect on the flow transition. In the early pre-transitional stage of blunt-plate cases, about 22 points are used in the spanwise direction for each spacing of the most amplified streamwise streak. The spanwise width of the computational domain is able to contain around 13 most amplified streaks. We have also conducted a prior test for case R2 by either reducing the spanwise width to two-thirds of the baseline value or increasing $n_y$ from 251 to 351. No visible change was found in the characterisation of steady and unsteady flow fields. The mesh convergence with respect to $n_x$ and $n_y$ is also confirmed in the 2-D receptivity study in Appendix \ref{appA}. In the streamwise direction, around 15 points are employed for the pronounced high-frequency short-wavelength structure contained in the wave packet of blunt-plate cases. The streamwise length of the computational domain for case R2 was found to be insufficient to establish the fully developed turbulence. As a consequence, the domain of case R3 is further extended. 

To examine the mesh resolution effect, a new case R3F is simulated with an evidently shorter streamwise length and a finer resolution. The grid spacings $(\Delta x^+,\Delta y^+,\Delta z^+)$ of case R3F approach the DNS setup for hypersonic wall-bounded flows by \cite{Huang2017} and \cite{duan2019characterization}. The results in Appendix \ref{appA} indicate that the current mesh resolution is sufficient for 3-D transitional studies. Another case R2C is simulated with a smaller total number of grid points and thus less computational cost. Case R2C presents the same flow phenomena as the baseline case R2. Data collected from a long-time simulation of case R2C will be used for spectral proper orthogonal decomposition (SPOD) analysis. 

\subsection{Broadband disturbance model}
To trigger the transition and mimic a real-life environment, broadband disturbances are added on the farfield boundary in front of the shock. As a consequence, the receptivity to freestream disturbances is included in the simulation. The receptivity process has been extensively investigated for hypersonic flows over flat plates, wedges, cones, etc. at different flow conditions, e.g., by \cite{balakumar2015receptivity}. The instability waves in boundary layers are found to be about three to five times more receptive to slow
acoustic waves than fast acoustic, vorticity and entropy waves. Furthermore, the
intensity of acoustic disturbances increases rapidly with Mach
number. The more pronounced receptivity to slow acoustic waves has also been verified in response to broadband disturbances \citep{He2022The}. Acoustic disturbances, radiated mostly from the nozzle-wall turbulent boundary layer, tend to dominate the overall
disturbance environment of wind tunnels at Mach 2.5 or above \citep{Laufer1961,Laufer1964,schneider2001effects,schneider2008development}. A more recent combined experimental and numerical study confirmed that the slow acoustic wave is the dominant acoustic mode in noisy hypersonic wind tunnels \citep{Wagner2018Combined}. Therefore, the slow acoustic wave is adopted as a representative disturbance model for the freestream boundary condition. 

In this paper, the three-dimensional broadband acoustic-wave model of \cite{cerminara2020transition} is employed. The merit of the broadband model is that the flow is able to select the preferential frequency and wavenumber naturally rather than adopt artificially imposed ones. The dimensionless pressure perturbation for a Fourier mode ($m$, $n$) with respect to the frequency and the spanwise wavenumber is given by 
\begin{equation}
{p'_{m,n}} = {A_m}\left[ {\cos ({\beta _n}z + {\psi _{m,n}}) + \cos ( - {\beta _n}z + {\psi _{m,n}})} \right]\cos ({\alpha _m}x - {\omega _m}t + {\varphi _m}),
\label{eq4.1}
\end{equation}
where
$m = 1,2,...,{M_f }$ and $n = 0,1,...,{N_\beta}$. Here, ${M_f }$ and ${N_\beta }$ are the total numbers of frequencies and nonzero spanwise wavenumbers, respectively. The symbols ${\psi _{m,n}}$ ($n\ne0$) and ${\varphi _m}$ represent random constant phase angles. In other words, once all the phase angles are randomly generated at $t=0$, the angle values will not change with time. The phase angles are also unchanged for cases with different nose-tip radii, which excludes the effect of the initial phase difference. For $n = 0$, it is enforced that ${\beta _0} = {\psi _{m,0}} = 0$. The symbols ${\alpha _m}$ and ${\omega _m}$ represent the streamwise wavenumber and the angular frequency for the \textit{m}-th frequency component, and ${\beta _n}$ is the spanwise wavenumber for the \textit{n}-th wavenumber component. The Fourier modes are actually higher-order harmonics, which is indicated by
\refstepcounter{equation}
$$
\begin{array}{l}
{\omega _m}=m{\omega _1},\quad{\beta _n}=n{\beta_1}.
\end{array}
\eqno{(\theequation{\mathit{a},\mathit{b}})}\label{eq4.2}
$$
Moreover, ${\omega _m} = 2\pi {f_m}$, where ${f_m}$ is the frequency. 
In accord with \cite{cerminara2020transition}, the streamwise wavenumber and the angular frequency is linked through ${\alpha _m} = {\omega _m}/(1 \pm 1/{M_\infty })$. Here, $1 \pm 1/{M_\infty }$ is the dimensionless phase speed of the fast/slow acoustic wave, and the slow-acoustic-wave one is finally chosen. In this paper, the incidence angle of the acoustic wave on the $x$--$y$ plane is assumed to be zero, which was also adopted by \cite{cerminara2020transition}. The oblique wave angle on the $x$--$z$ plane is given by ${\theta _{m,n}} = {\arctan}({\beta _n}/{\alpha _m})$.

In equation (\ref{eq4.1}), the dimensionless amplitude is ${A_m} = A_m^ \ast /\left( {\rho _\infty ^ \ast u_\infty ^{ \ast 2}} \right)$ is equally set for each spanwise-wavenumber component. In other words, no preferential spanwise wavenumber is imposed. The frequency-dependent dimensional amplitude $A_m^ \ast$ is determined by the following relation
\begin{equation}
A_m^ \ast/{p_\infty ^ \ast }  = \left\{ \begin{array}{ll}
\sqrt {{C_L}f_m^{ *  - 1}\Delta {f^ * }/2}, &f_m^ *  \le 40\ {\rm{kHz}}\\[5pt]
\sqrt {{C_U}f_m^{ *  - 3.5}\Delta {f^ * }/2}, &\text{otherwise}, 
\end{array} \right.\label{eq_amp}
\end{equation}
which is fitted from the measured frequency spectra of noise in the Arnold Engineering Development Complex (AEDC) Hypervelocity Wind Tunnel 9 \citep{marineau2015investigation,balakumar2018transition}. The law of $f_m^{ *  - 3.5}$ at high frequencies has been reported by the measured noise data in various tunnel conditions with $M_\infty$ ranging from 6 to 14, including the Hypersonic Ludwieg Tube Braunschweig, the Purdue Boeing/AFOSR Mach-6 Quiet Tunnel, the NASA 20-Inch Mach 6, the Sandia Hypersonic Wind Tunnel at Mach 8 and the AEDC Tunnel 9 \citep{duan2019characterization}. The law of $f_m^{ *  - 1}$ at low frequencies was also verified by the DNS data of the tunnel noise at different Mach numbers \citep{duan2019characterization}. The amplitude constants are ${C_L} = 3.953 \times {10^{ - 4}}$ and ${C_U} = 126.5 \times {10^6}$ in SI units. In the present model, the dimensional frequency ranges from 10 kHz to 1000 kHz with interval of $\Delta {f^ {\ast}} = 5$ kHz.\added{ Harmonics with frequencies less than 10 kHz are not included, because the tunnel measurement result was not displayed in such range for a reliable amplitude fitting in (\ref{eq_amp}) \citep{marineau2015investigation, balakumar2018transition}. Lower-frequency components may benefit the generation of relevant responses such as stationary streaks.} Meanwhile, the spanwise wavenumber ${\beta^{\ast} _n}$  ranges from 600 m$^{-1}$ and 20600 m$^{-1}$ with interval of $\Delta {\beta ^ * }=$ 800 m$^{-1}$, which is expected to achieve a broadband state. The mode numbers are finally $M_f = 199$ and ${N_\beta } = 26$. The corresponding spanwise wavelength varies from 0.3 mm to 10.5 mm, and the oblique wave angle ${\theta _{m,n}}$ on the $x$--$z$ plane spans from 4$^{\circ}$ to 89$^{\circ}$.

Following the dispersion relations of slow acoustic waves by \cite*{egorov2006numerical} and \cite{cerminara2020transition}, we obtain the perturbations of other primitive variables by
\refstepcounter{equation}
$$
\begin{array}{l}
{u'_{m,n}} =  - {p'_{m,n}}{M_\infty }\cos {\theta _{m,n}},\\[2pt]
{v'_{m,n}} = {w'_{m,n}} = 0,\\[2pt]
{T'_{m,n}} = (\gamma  - 1)M_\infty ^2{p'_{m,n}}.
\end{array}
\eqno{(\theequation{\mathit{a},\mathit{b},\mathit{c}})}\label{eq_dispersion}
$$
\added{Without an angle of incidence, the wall-normal velocity fluctuation is naturally zero, which is also consistent with the symmetry boundary condition along $y=0$ in figure \ref{fig1}(\textit{a}). In terms of the freestream $u',T'$ and $p'$, the symmetrical slip boundary has no noticeable impact on them. }Finally, the total perturbation of the variable $\bm{\phi}=(u,v,p,w,T)^\text{T}$ is given by
\begin{equation}\label{ABpara}
{\bm{\phi}} '(x,z,t) = B_\textit{rescaled} A_\textit{rescaled}\sum\limits_{n = 0}^{{N_\beta }} {\sum\limits_{m = 1}^{{M_f}} {{{\bm{\phi}} '_{m,n}}} },
\end{equation}
while the instantaneous density $\rho$ is directly calculated from the equation of state for ideal gas. For 2-D receptivity studies ($N_{\beta}=0$), the parameters are set to $A_\textit{rescaled}=B_\textit{rescaled}=1$ in equation (\ref{ABpara}). For 3-D transitional simulations, the amplitude rescaling parameter is set to $A_\textit{rescaled}=0.366$, such that the spanwise averaged $p^{\ast\prime}_{\infty,\textit{rms}}$ of the 3-D wave is numerically equal to the 2-D counterpart determined by (\ref{eq_amp}). The detailed intensity of the pressure fluctuation is $p^{\ast\prime}_{\infty,\textit{rms}}/\bar{p}^{\ast}_{\infty}=2.85$ \%. Note that the amplitude and spectrum of incoming disturbances were not given in the experimental study of \cite{borovoy2022laminar}. \deleted{A precursor simulation of case R2 reports that the designated amplitude leads to a significant advance of the transition onset compared to the experimental data. This observation might indicate that the intensity of real tunnel noise in Borovoy \etal (2022) should be further lower. To approach the experimental data, another} \added{Another} rescaling parameter is\added{ artificially} multiplied and given by $B_\textit{rescaled}=0.6$.\added{ This parameter can be adjusted to match the experimental transition onset.} Finally, the resulting amplitude of the pressure fluctuation is $p^{\ast\prime}_{\infty,\textit{rms}}/\bar{p}^{\ast}_{\infty}=1.71$ \%. At each time step of the simulation, the instantaneous quantity on the farfield boundary is forced to be the superimposition of the base-flow quantity and the perturbation.

\section{Phenomenon of transition reversal}\label{sec:Phenomenon}
\begin{figure*}
	\centering{ \includegraphics[height=5cm]{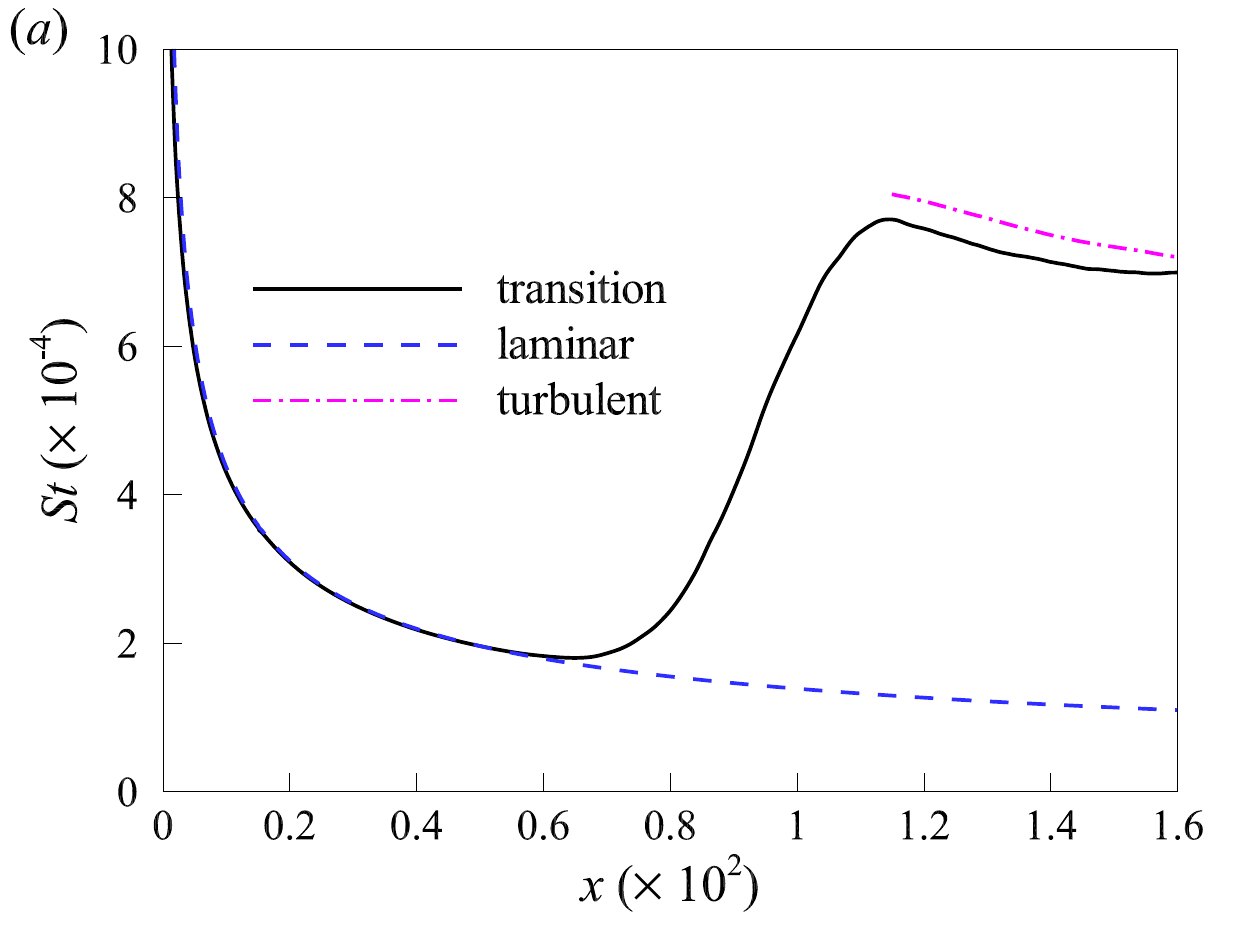}}
	\centering{ \includegraphics[height=5cm]{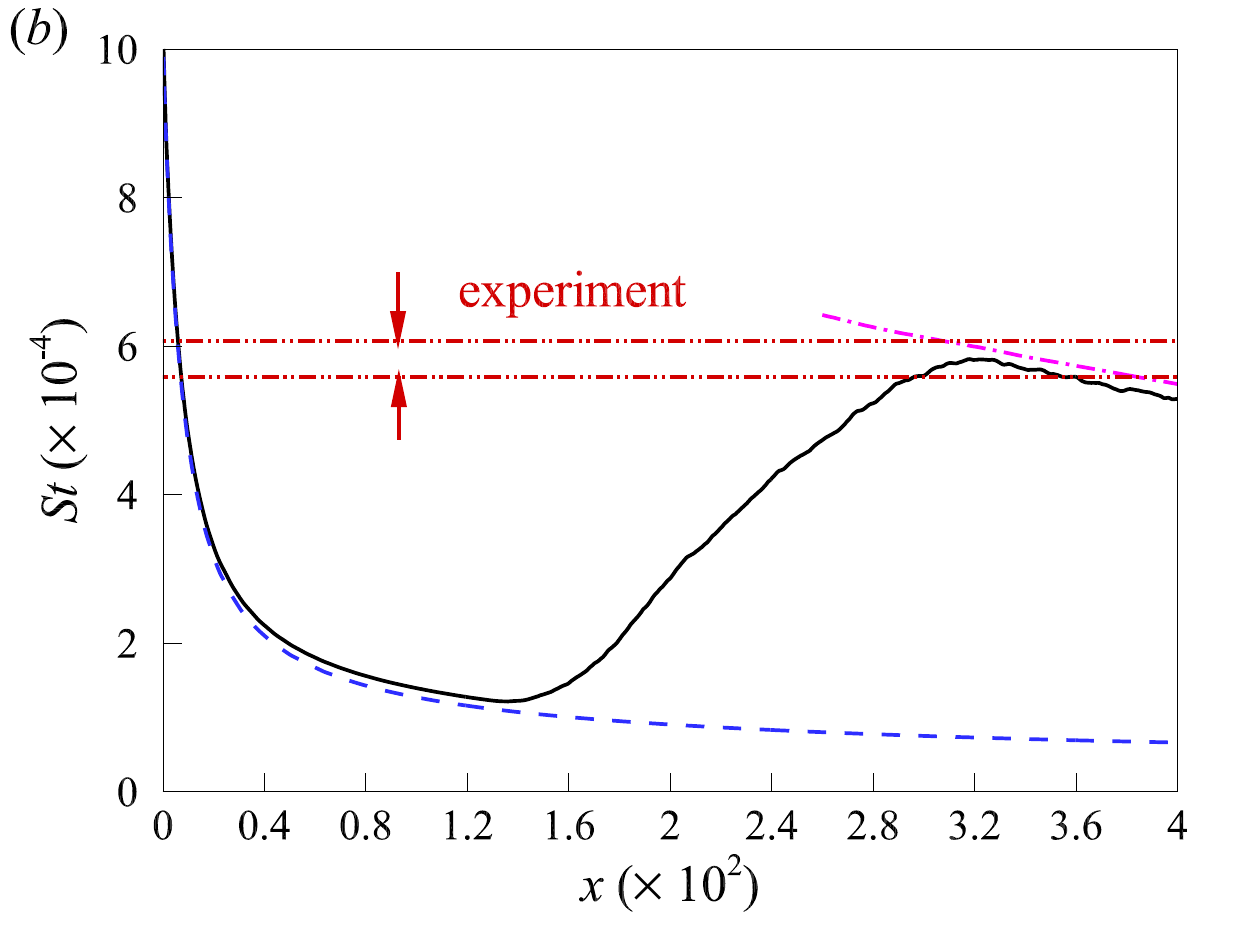}}
	\centering{ \includegraphics[height=5cm]{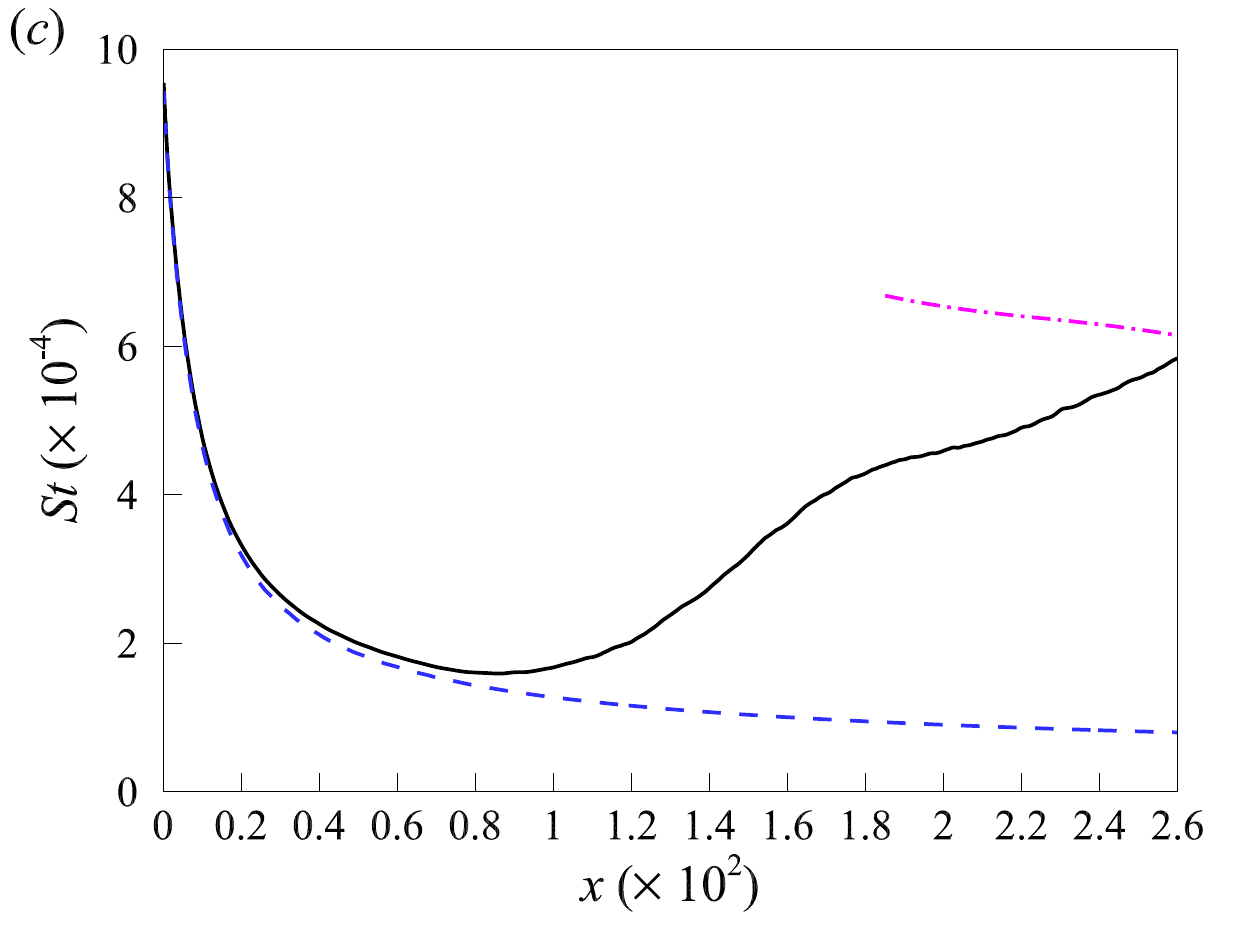}}
	\centering{ \includegraphics[height=5cm]{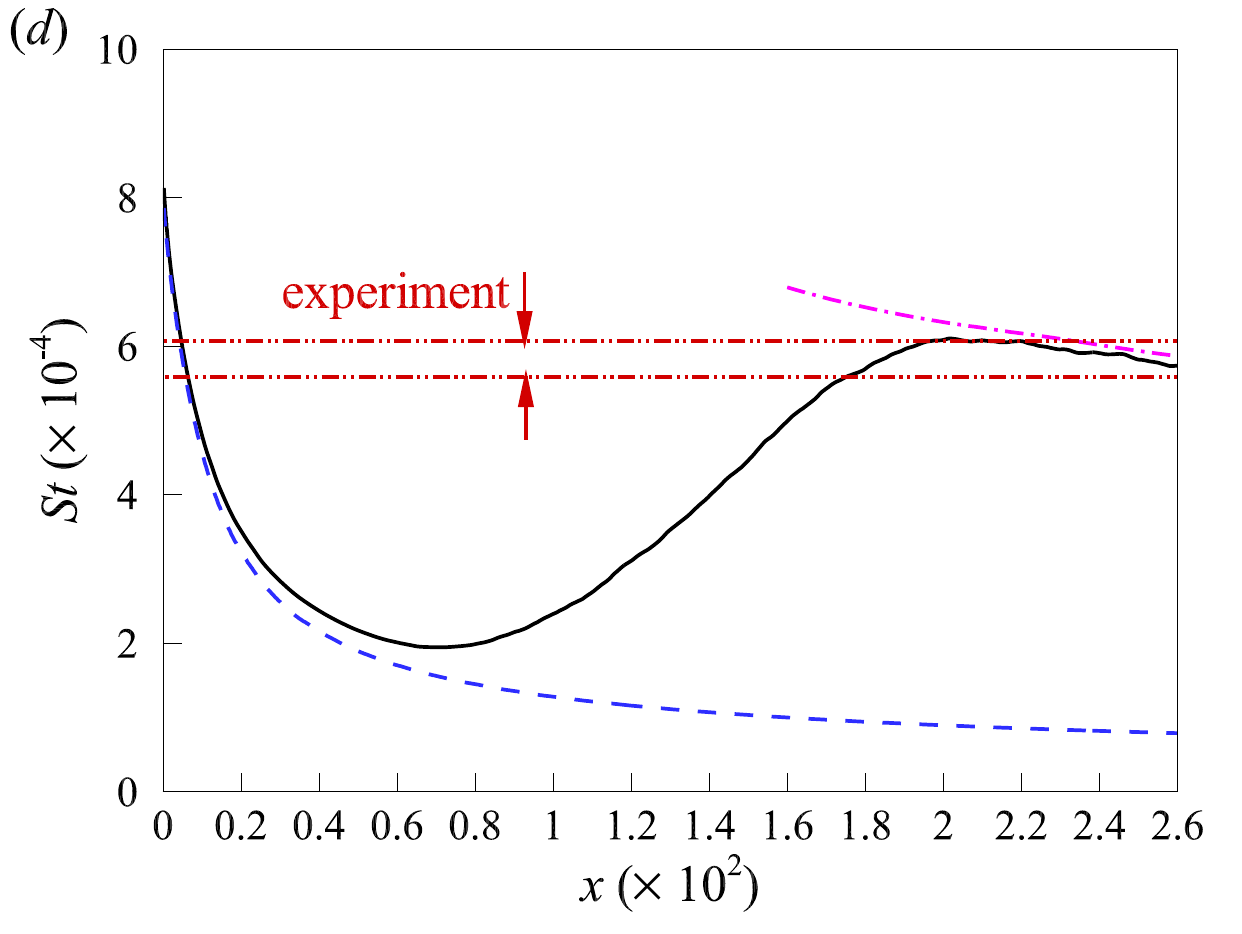}}
	\centering{ \includegraphics[height=5cm]{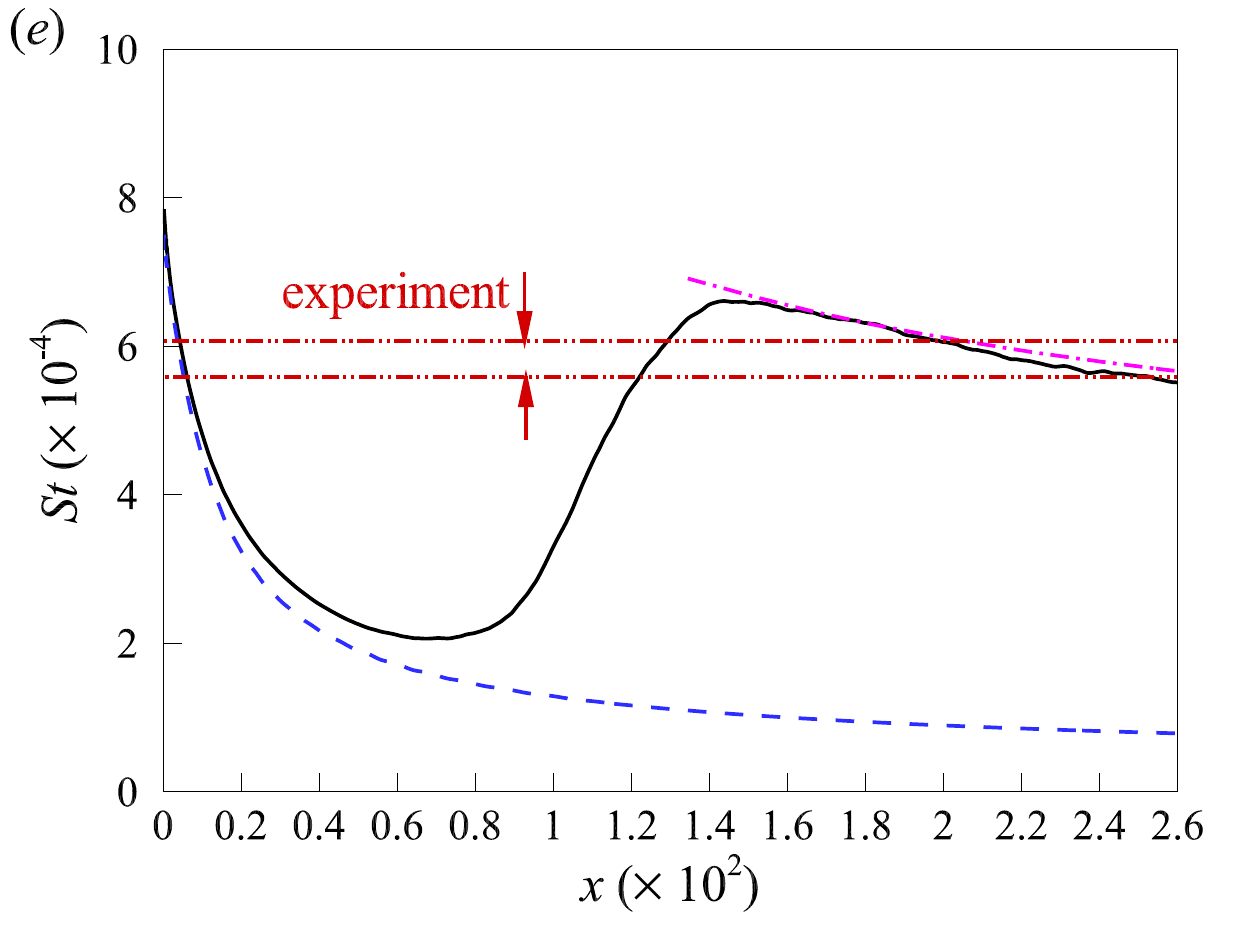}}
	\caption{Spanwise- and time-averaged Stanton number of (\textit{a}) R0, \replaced{(\textit{b}) R1.8, (\textit{c}) R2, (\textit{d}) R2.7 and (\textit{e}) R3}{(\textit{b}) R2, (\textit{c}) R2.7 and (\textit{d}) R3}. The horizontal dashed-double-dotted lines represent the upper and lower bounds of the experimental turbulent value with  \replaced{$\Rey_\textit{R}=1.975\times10^4$}{$\Rey_\textit{R}=3.95\times10^4$} \citep{borovoy2022laminar}. Note that the total temperature rather than the recovery wall temperature was used in the definition of $St$ in the experiment, and a data transformation has been performed. The evaluation procedure of the turbulent Stanton number can been found in \cite{Franko_Lele_2013} and \cite{Guo_Heat_2022}.}
	\label{fig_St_curves}
\end{figure*}

\begin{table}
	\begin{center}
		\def~{\hphantom{0}}
		\begin{tabular}{lccccc}
			cases &R0 &\added{R1.8} &R2 &R2.7 &R3	\\[5pt]
			$\Rey_\textit{R}(\times10^5)$ &0	&\added{1.08} &1.2	&1.62 	&1.8	\\
			$\Rey_\textit{t}(\times10^6)$ &4.34	&\added{8.93} &5.80	&4.98 	&4.85	\\
			$x_\textit{t}$ &72.3	&\added{148.9} &96.6	&83.0	&80.8	\\
		\end{tabular}
		\caption{Transition onset Reynolds numbers and locations determined from numerical simulations.}
		\label{table2}
	\end{center}
\end{table}
\begin{figure*}
	\centering{ \includegraphics[height=6.5cm]{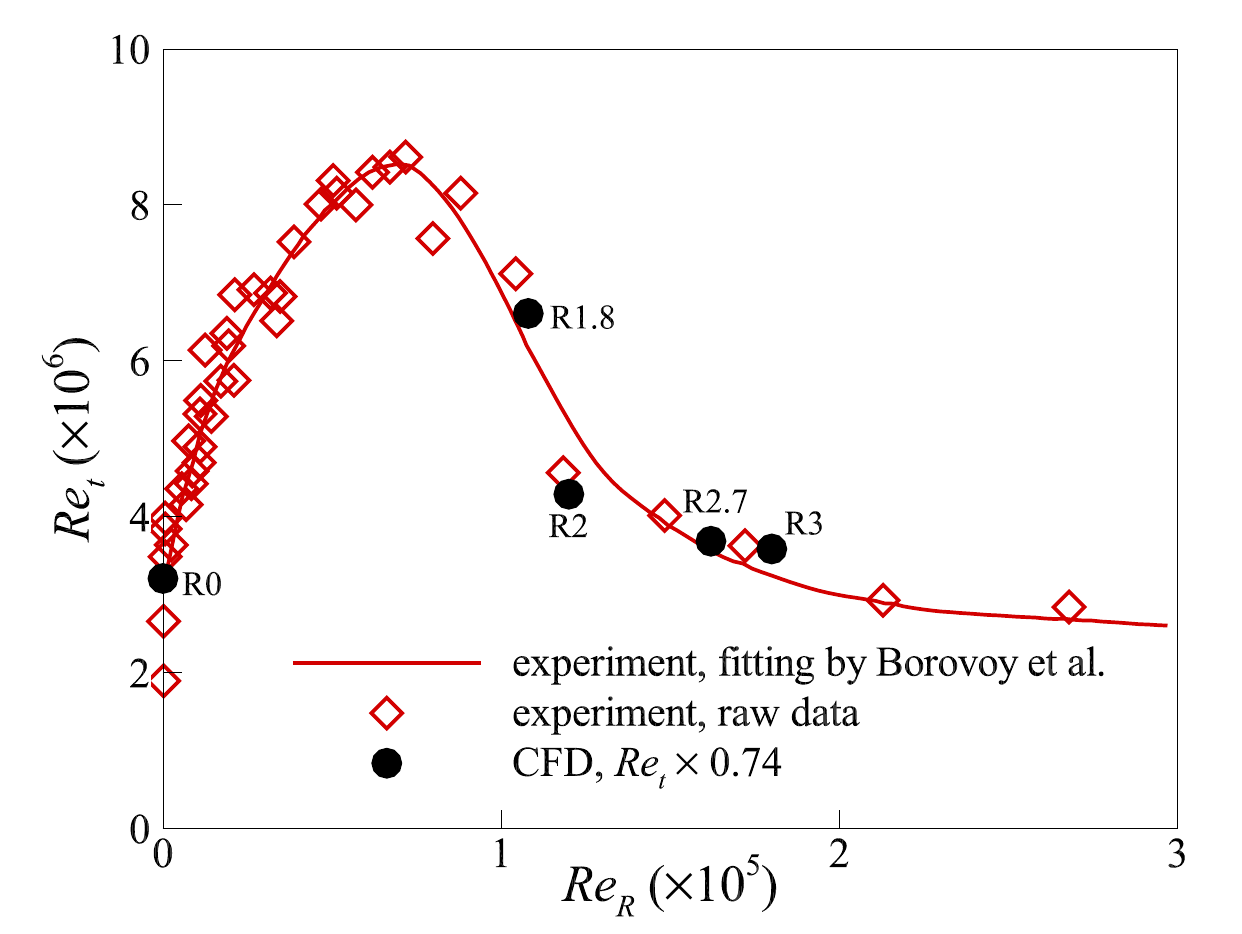}}
	\caption{Plot of transition onset Reynolds number versus nose\added{-tip} radius Reynolds number. \added{The original experimental Reynolds number based on the leading-edge thickness $b=2R_\textit{n}$ is transformed into the nose-tip radius-based Reynolds number.}}
	\label{fig_comparison_with_exp}
\end{figure*}

Since the research subject is the transition reversal, we first present the reversal phenomenon manifested by experimental and computational data. Figure \ref{fig_St_curves} shows the spanwise- and time-averaged Stanton number for the present cases R0,\added{ R1.8,} R2, R2.7 and R3. The Stanton number is defined by
\begin{equation}
St={q_\textit{w}^\ast } / \left[ {\rho _\infty ^ \ast u_\infty ^ \ast c_{\textit{p}\infty }^ \ast \left( {T_\textit{aw}^ \ast  - T_\textit{w}^ \ast } \right)} \right],
\end{equation}
where $q_\textit{w}^\ast $ is the wall heat flux, and $T_\textit{aw}^ \ast$ is the adiabatic (recovery) wall temperature for turbulent flows. The legend `laminar' refers to the curve of the 2-D laminar flow, while the `turbulent' Stanton number is obtained from Reynolds analogy. To figure out the `turbulent' Stanton number, the skin friction coefficient $C_f$ is calculated based on van Driest II correlation, as described in \S\ \ref{sec:DNS}. The Reynolds analogy factor $2St/C_f$ was incipiently recommended to be $\textit{Pr}^\text{-2/3}$ by \cite{colburn1964method}, whereas a following high-speed experiment by \cite{hopkins1971evaluation} indicated a lower value. Consistent with \cite{Franko_Lele_2013} and \cite{Guo_Heat_2022},  the Reynolds analogy factor is set to 1.0 in this paper. As shown by figure \ref{fig_St_curves}, the $St$-curves after transition collapse well onto the turbulent empirical formula as well as the experimental range of \cite{borovoy2022laminar}\added{ in the turbulent region}. An exception is that the length of the computational domain for case R2 is insufficient to see the transition end. A further investigation on the mean velocity profile and the associated law of the wall is given in Appendix \ref{appB}. A fully developed turbulent state is indicated for case R3 in an approximate range $x>180$. 

To directly compare with the experimental result, the determination approach of the transition onset location is kept unchanged to \cite{borovoy2022laminar}. This approach has been commonly applied in hypersonic transition measurements. In detail, the mean Stanton number result is mapped onto a $\log_{10}(\Rey_x)$--$\log_{10}(St)$ plot, where $\Rey_x$ is the $x$-based Reynolds number. The transition onset is then defined by the intersection point of approximate laminar and transitional lines. A graphical example is provided in Appendix \ref{appC}. The transition onset Reynolds number is then given in table \ref{table2}. Obviously, the transition onset reverses when the nose-tip radius exceeds \replaced{1.8}{2} mm. 

\replaced{Given}{Despite} that a rescaling parameter $B_\textit{rescaled}$ is employed in equation (\ref{ABpara}), the transition onset is \replaced{now slightly later}{still earlier} than the experimental one. As aforementioned, the data trend instead of the exact transition Reynolds number is of interest. We keep the current computational setup unchanged instead of adjusting $B_\textit{rescaled}$ repeatedly to match the experimental result. For a straightforward comparison, the transition onset Reynolds numbers $\Rey_t$ for all the CFD cases are multiplied by a factor \replaced{0.74}{1.45}.  This operation \added{calibrates the $Re_t$ of case R0 to match the experimental data, whereas it }does not alter the tendency arising from the bluntness effect. Figure \ref{fig_comparison_with_exp} compares the CFD and experimental data. A fitting curve was given in conjunction with the raw data by \cite{borovoy2022laminar}. \replaced{The}{Regardless of the more deviated case R0, the} overall tendency of the nose bluntness effect is reproduced by the present numerical simulation. \deleted{Borovoy \etal\ (2022) mentioned that the data scattering was increased evidently with small nose bluntness, which might partly account for the deviation of the case-R0 state.} In the following section, the detailed flow mechanism will be analysed by numerical and theoretical tools.

\added{Please note that for small-bluntness regime cases, the dimensional distance between the shock and the wall is remarkably reduced. This fact gives rise to a very small maximum time step size for Runge--Kutta time marching. As a consequence, it is computationally expensive to conduct a series of sensitivity case studies and report the transition for both small-bluntness and large-bluntness cases, which are left for future studies. The final displayed states are R0, R1.8, R2, R2.7 and R3, which show a good agreement with the trend of the experimental transition onset.}

\section{Flow mechanism}\label{sec:mechanism}
\subsection{Two-dimensional response}
\begin{figure*}
	\centering{ \includegraphics[height=2cm]{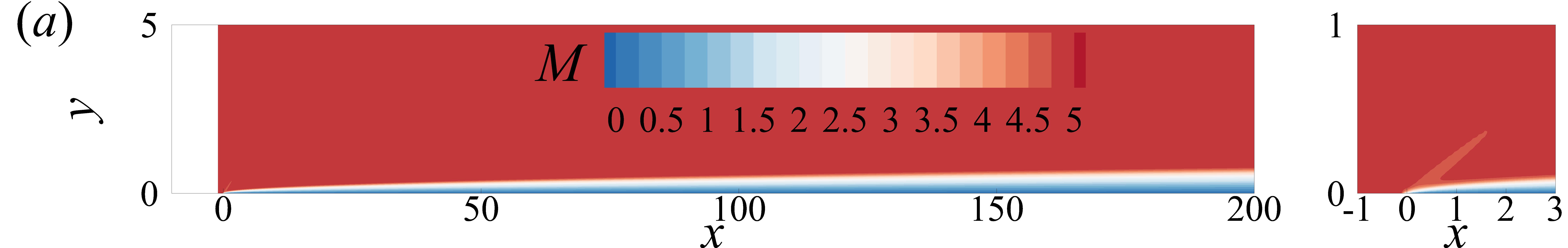}}
	\centering{ \includegraphics[height=2cm]{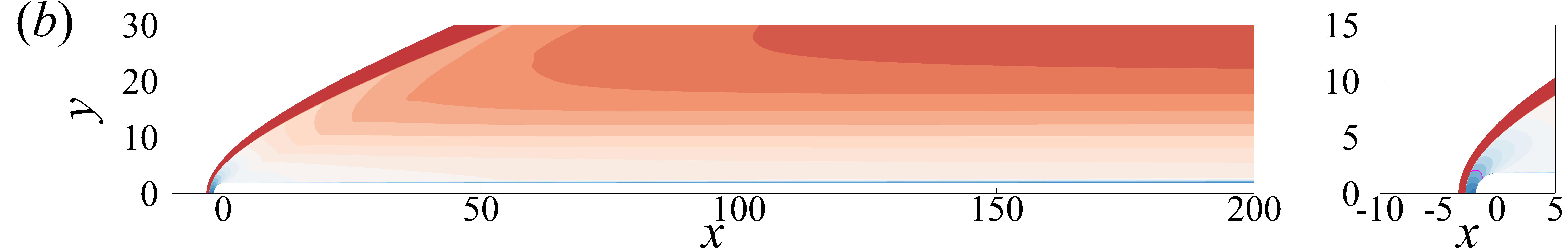}}
	\centering{ \includegraphics[height=2cm]{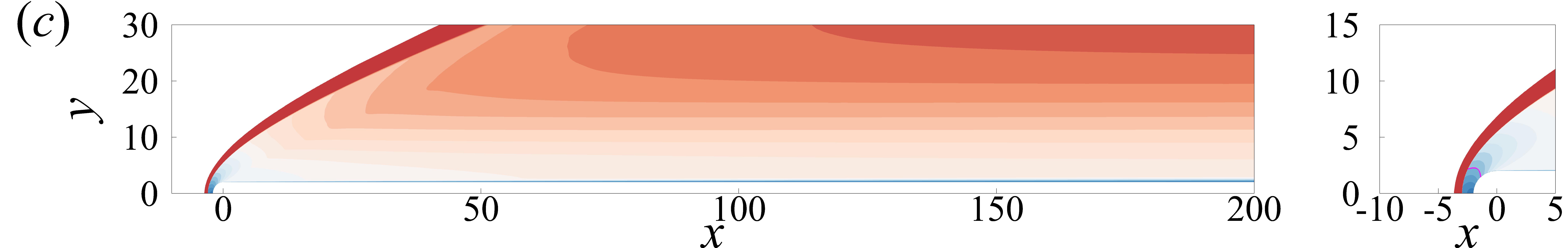}}
	\centering{ \includegraphics[height=2cm]{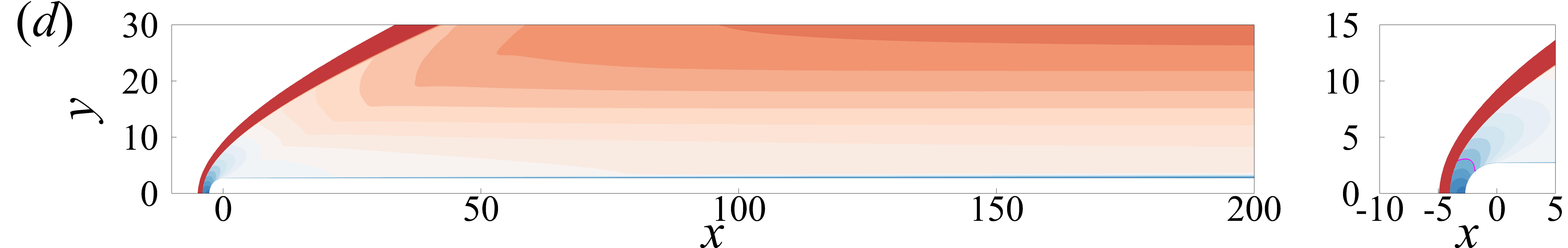}}
	\centering{ \includegraphics[height=2cm]{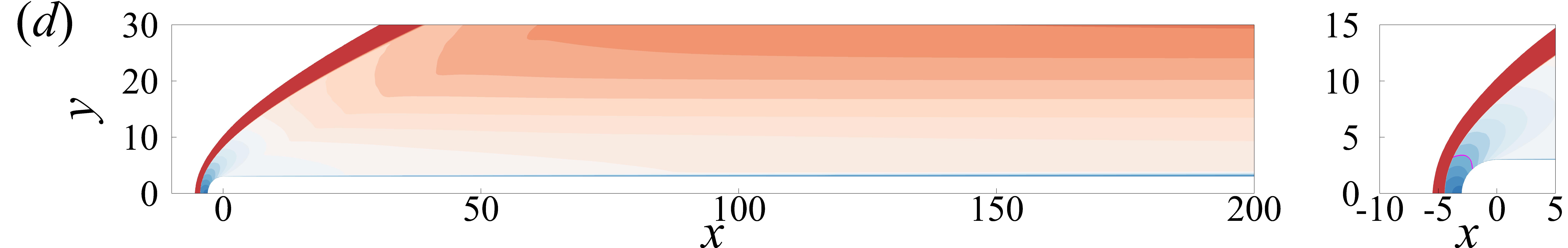}}
	\caption{Mach number contour of the steady laminar flow for cases (\textit{a}) R0, \replaced{(\textit{b}) R1.8, (\textit{c}) R2, (\textit{d}) R2.7 and (\textit{e}) R3. Pink solid lines in the right column represent the sonic lines near the nose.}{(\textit{b}) R2, (\textit{c}) R2.7 and (\textit{d}) R3.}}
	\label{fig_baseflow}
\end{figure*}

\begin{figure*}
	\centering{ \includegraphics[height=5cm]{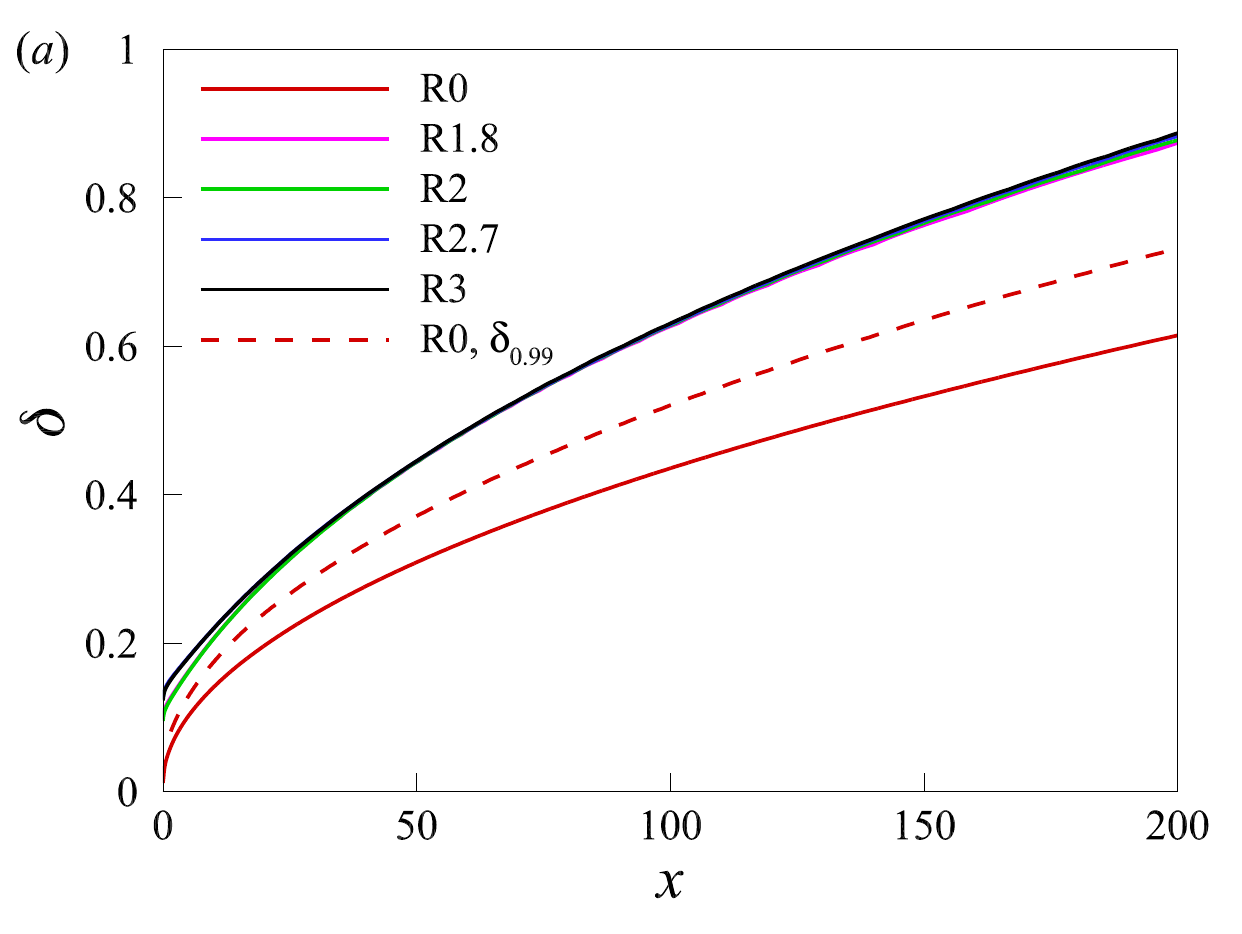}}
	\centering{ \includegraphics[height=5cm]{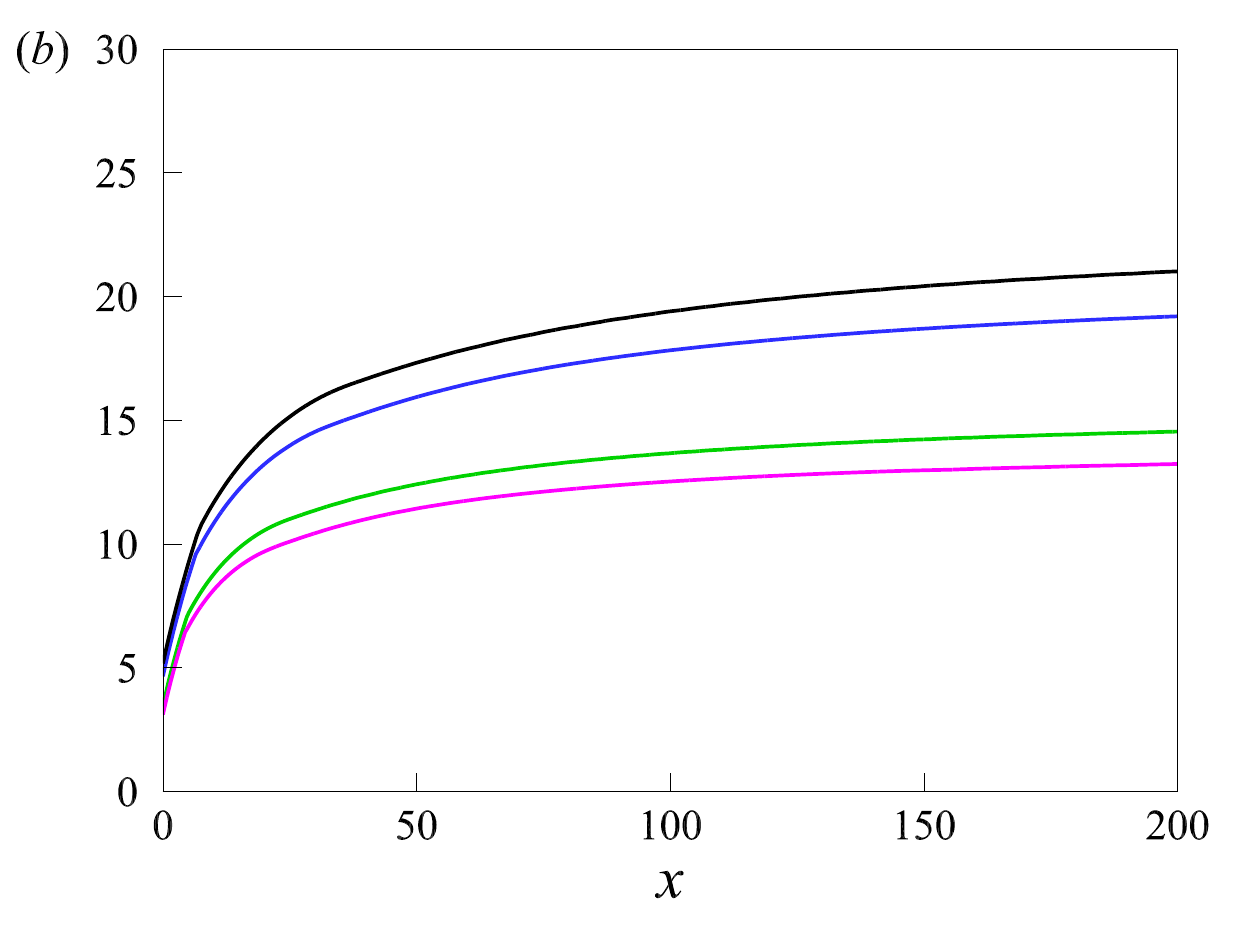}}
	\caption{\added{Thickness of (\textit{a}) boundary layer and (\textit{b}) entropy layer of the steady laminar flow.}}
	\label{fig_thickness}
\end{figure*}

\begin{figure*}
	\centering{ \includegraphics[width=6.5cm]{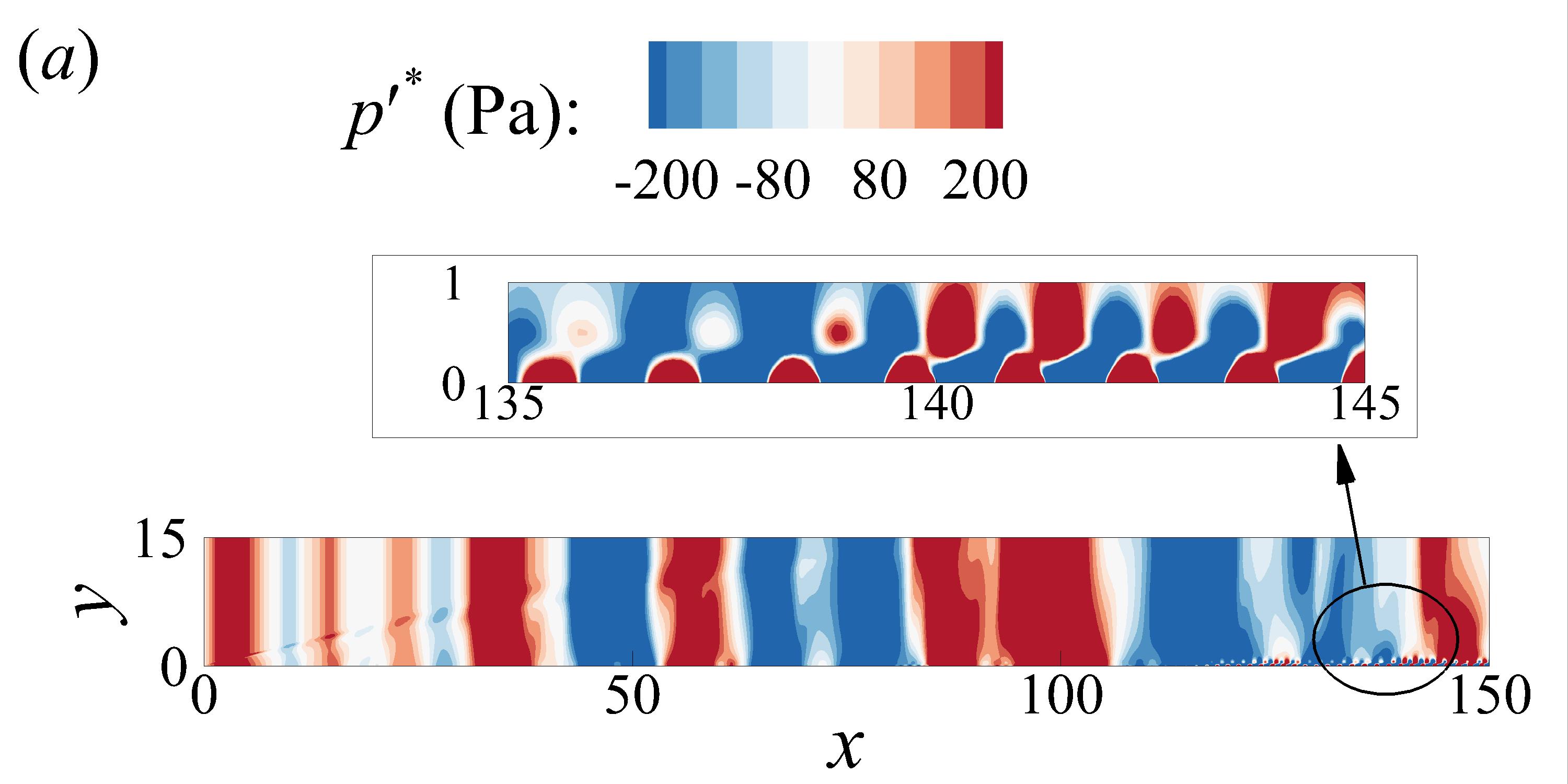}}
	\centering{ \includegraphics[width=6.5cm]{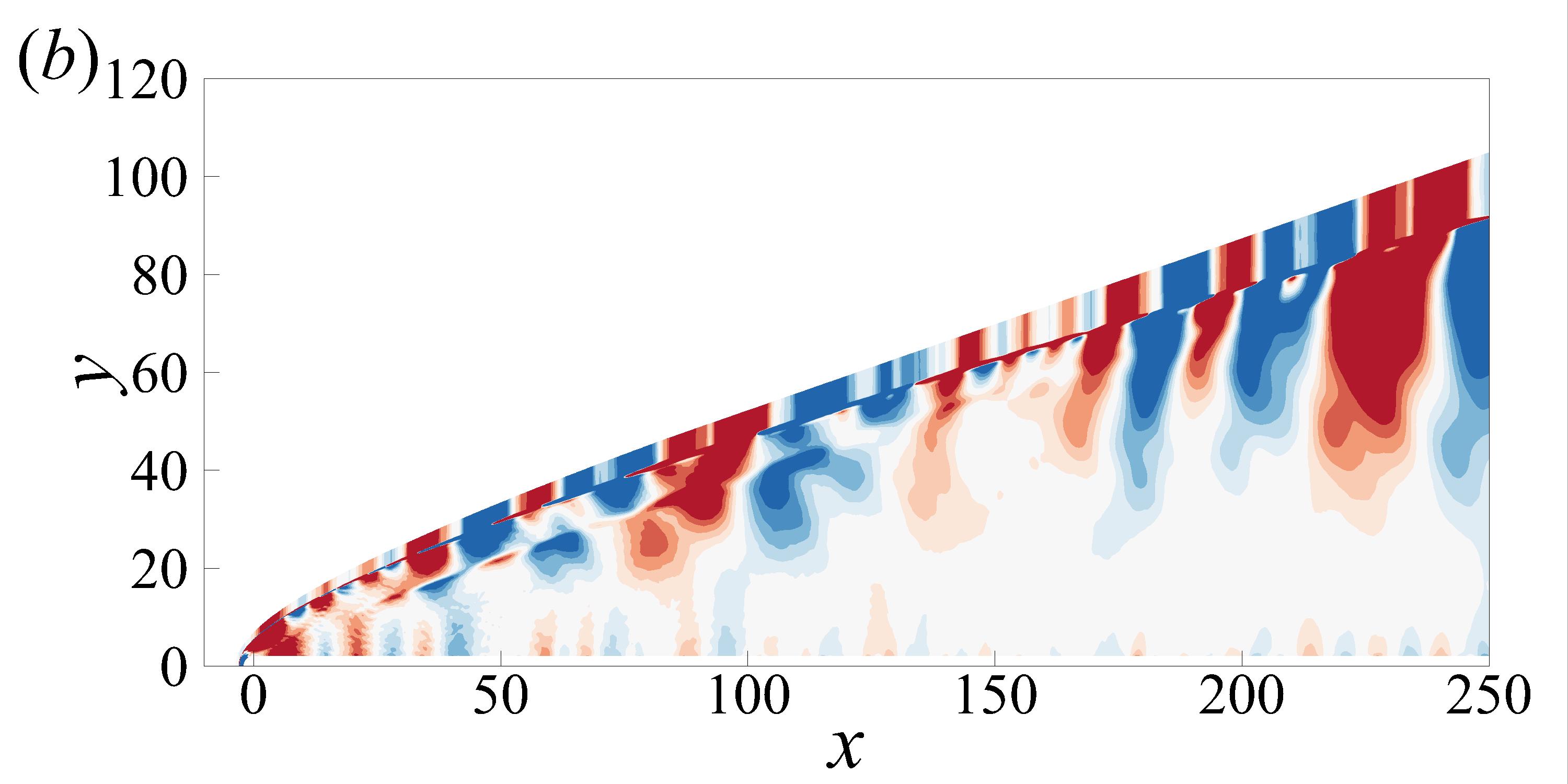}}
	\centering{ \includegraphics[width=6.5cm]{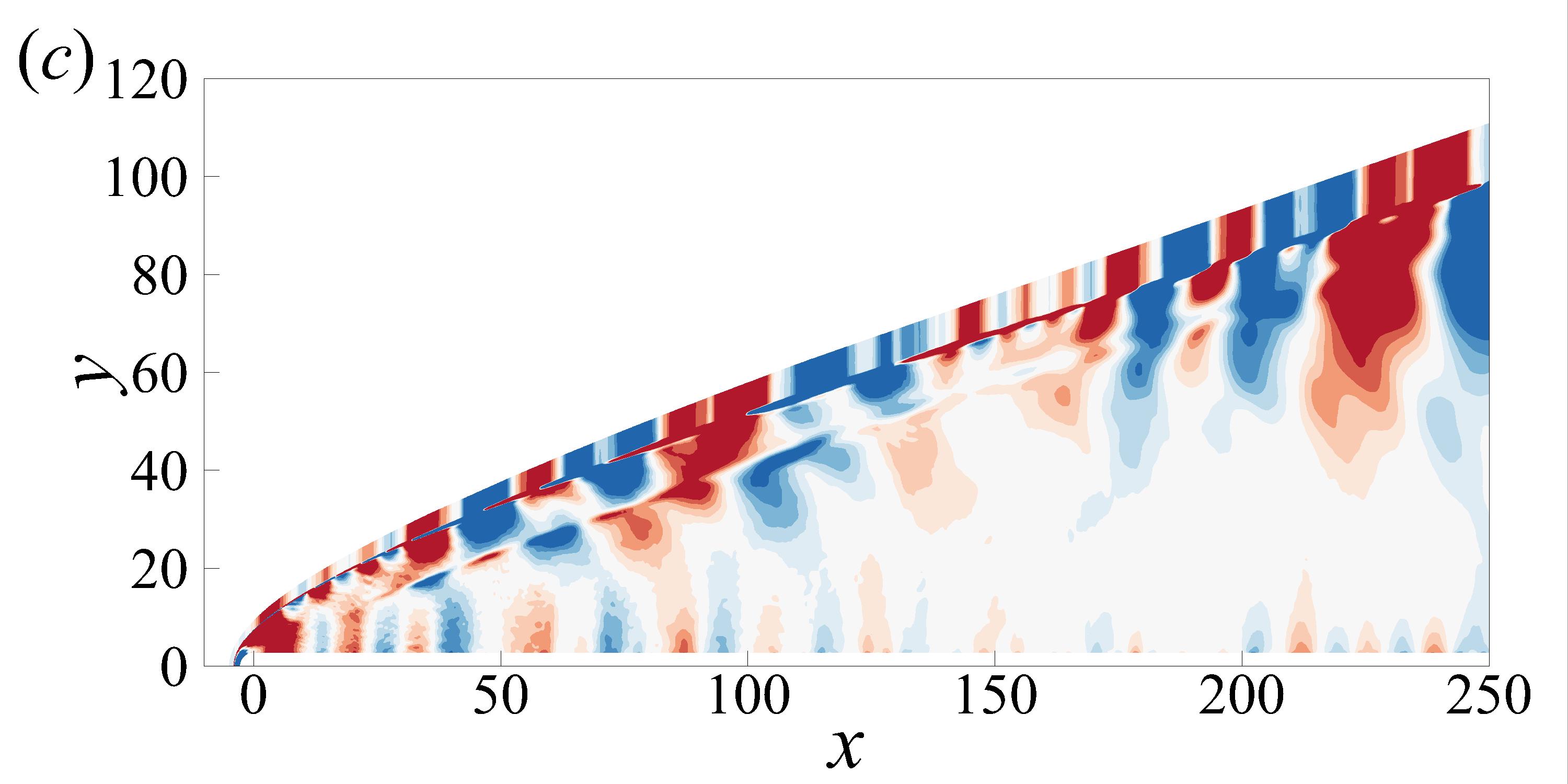}}
	\centering{ \includegraphics[width=6.5cm]{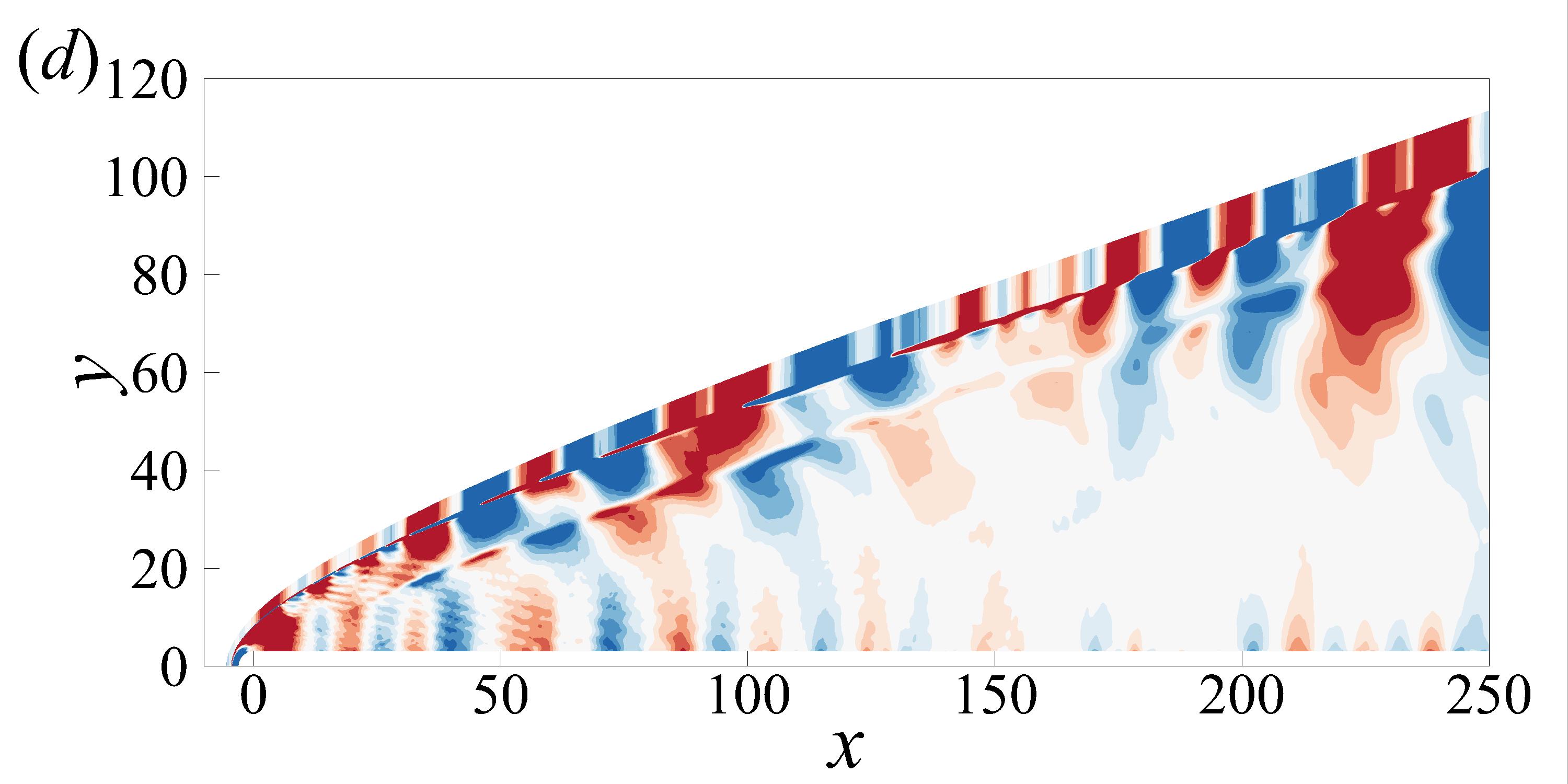}}
	\caption{Pressure fluctuation contour at $t^{\ast}=$ 0.5 ms of 2-D cases (\textit{a}) R0, (\textit{b}) R2, (\textit{b}) R2.7 and (\textit{d}) R3.}
	\label{fig_2d_disturb}
\end{figure*}

\begin{figure*}
	\centering{ \includegraphics[width=6.5cm]{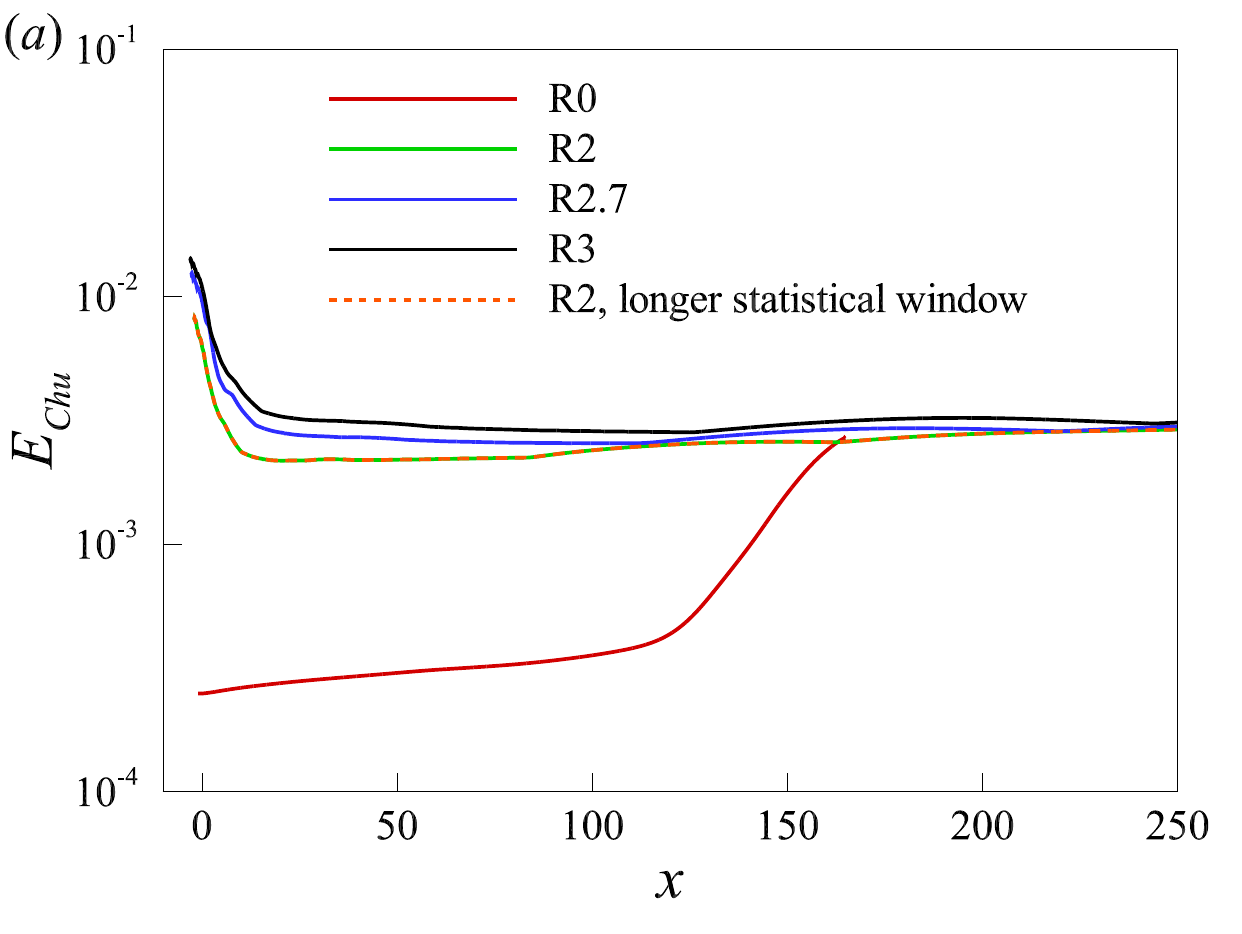}}
	\centering{ \includegraphics[width=6.5cm]{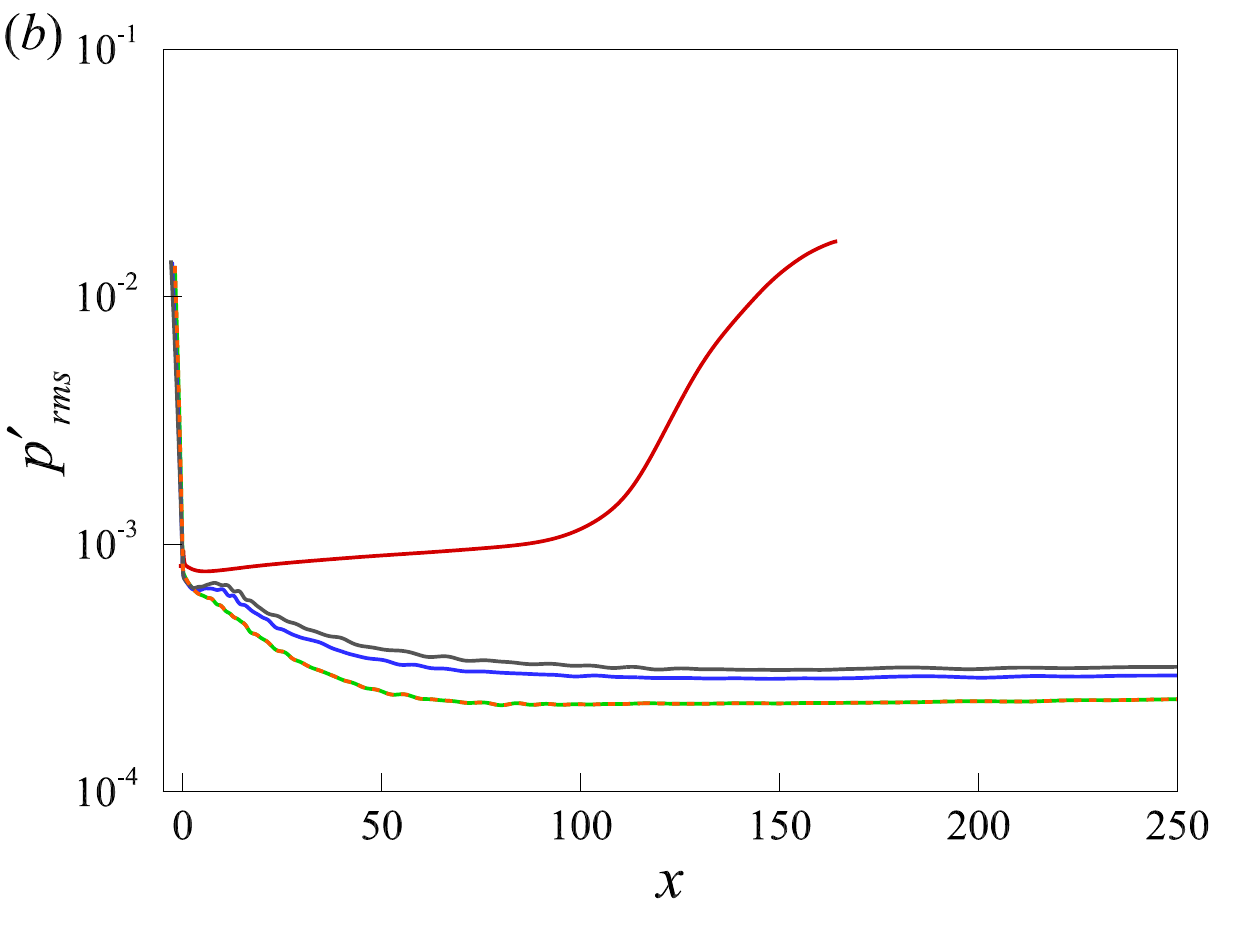}}
	\caption{(\textit{a}) Chu's energy and (\textit{b}) r.m.s. of wall pressure fluctuations for 2-D cases. Statistics is performed in an interval of $t^{\ast}=$ 1 ms (at least 3.5 flow-through time units). The `longer statistical window' corresponds to another statistical interval of 2 ms.}
	\label{fig_2d_stati}
\end{figure*}

The Mach number contour of the two-dimensional (2-D) steady base flow is shown in figure \ref{fig_baseflow} for the four cases with different nose-tip radii.  Following the criterion of \cite{paredes2019nose}, the edge of the boundary layer is defined as the location where
the local total enthalpy is 0.995 times the freestream value. Furthermore, the edge of the entropy layer is defined as the location where the local entropy increment is 0.25 times the value at the wall. Here, the total enthalpy and the entropy increment are defined by $H^{\ast}=c_p^{\ast} T^{\ast}+(u^{\ast 2}+v^{\ast 2})/2$ and $\Delta S^{\ast}=c_p^{\ast}\ln(T^{\ast}/T^{\ast}_{\infty})-{\cal{R}}^{\ast}\ln(p^{\ast}/p^{\ast}_{\infty})$, respectively. The symbols $c_p^{\ast}$ and ${\cal{R}}^{\ast}$ represent the specific heat at constant pressure and the gas constant for air, respectively. The thicknesses of the boundary layer and the entropy layer are plotted against $x$ in figure \ref{fig_thickness}. In addition to the total enthalpy criterion, the nominal thickness $\delta_{0.99}$ of the boundary layer is also shown in figure \ref{fig_thickness}(\textit{a}) for case R0. Clearly, the presence of a blunted leading edge thickens the boundary layer. Away from the nose tip, the value of the nose-tip radius has no visible impact on the boundary layer thickness. Moreover, an increasing nose-tip radius makes the entropy layer thicker. Different from the cone geometry, entropy swallowing is not seen in a distance of interest. In other words, the boundary layer is entirely covered by the entropy layer. This feature may enable a long-distance stabilisation effect of the entropy layer on the normal-mode instability. Furthermore, the Mach number on the boundary layer edge evolves to be around $M=2.7$ downstream for cases \added{R1.8, }R2, R2.7 and R3. Therefore, unstable Mack second mode, usually appears above Mach 4, is probably not supported in the considered blunt-plate flow.

Figure \ref{fig_2d_disturb} shows the instantaneous pressure fluctuation at $t^{\ast}=$ 0.5 ms when the 2-D unsteady flow is fully established. \added{The pressure fluctuation is obtained by subtracting the laminar state from the instantaneous pressure field. The displayed solution is linear, since the isoline of the 2-D time-averaged pressure field visibly accords with the steady-flow counterpart (not shown).} For case R0, the broadband planar acoustic waves interact with the weak leading-edge shock first and the boundary layer subsequently. Clear double-cell signature of the pressure fluctuation for Mack second mode is captured in the downstream boundary layer. For blunt-plate cases, the planar acoustic waves interact with first the detached shock and then the weak expansion wave forming in the vicinity of the junction point $(x,y)=(0,R_n)$.\added{ As shown in figure \ref{fig_baseflow}, the junction is located downstream of the sonic line, and the expansion wave forms downstream of the sonic line on the nose.} No Mack-second-mode like structure is captured throughout the computational domain\added{. This is because no typical double-cell structure is observed in the contour of pressure fluctuations, and no pronounced high-frequency signals are amplified during the simulation}. With varying nose-tip radius, the structure of $p'$ is not substantially altered. This might suggest some similarities in the instability evolution for the considered blunt-plate cases, which will be further examined in 3-D studies. 

Figure \ref{fig_2d_stati} shows the streamwise evolution of Chu's energy and the r.m.s. of the wall pressure fluctuation $p'_\textit{rms}$. The Chu's energy $E_\textit{Chu}$ is defined as the wall-normal integral of Chu's energy density, which is given by \citep{chu1965energy}
\begin{equation}
	E_\textit{Chu} = \frac{1}{2}\int_0^\infty  {\left[ {\bar \rho \left( {\overline{{{u'}^2}} + \overline{{v'}^2} + \overline{{w'}^2}} \right) + \frac{{\bar T}}{{\gamma M_\infty ^2\bar \rho }}\overline{{\rho '}^2} + \frac{{\bar \rho }}{{\gamma \left( {\gamma  - 1} \right)M_\infty ^2\bar T}}\overline{{T'}^2}} \right]} d\eta.
\end{equation}
Different length of the statistical window is used, and no visible difference in the concerned quantity is found in figure \ref{fig_2d_stati}. Thus, statistical convergence is reached. The sharp-leading-edge case supports the amplification of 2-D Mack second mode, while an approximately neutral state is reported on the downstream flat plate for all the blunt-plate cases. The downstream magnitude of $p'_\textit{rms}$ satisfies $p'_{\textit{rms},\text{R2}}<p'_{\textit{rms},\text{R2.7}}<p'_{\textit{rms},\text{R3}}<p'_{\textit{rms},\text{R0}}$. Meanwhile, the final Reynolds number based on the transition onset satisfies  $Re_{\text{R2}}>Re_{\text{R2.7}}>Re_{\text{R3}}>Re_{\text{R0}}$. Note that a higher-amplitude fluctuation usually corresponds to an earlier transition onset if the other factors are unchanged. Thus, the tendency of the response magnitude of the disturbance, dependent on the nose-tip radius, is coincidently consistent with the performance of the transition onset. With regard to Chu's energy, the base flow of case R0 does not possess a thick entropy layer compared to the others. As a result, disturbance energy outside the boundary layer is relatively small for case R0, and the integrated $E_\textit{Chu}$ is considerably lower than the other cases. However, for blunt-plate cases, $E_{\textit{Chu},\text{R2}}<E_{\textit{Chu},\text{R2.7}}<E_{\textit{Chu},\text{R3}}$ still holds true. The above observation may suggest that the bluntness effect has already been reflected, though incompletely, in the 2-D physical problem. 

\subsection{Linear stability analysis}\label{sec:LST}
\begin{figure*}
	\centering{ \includegraphics[width=6.5cm]{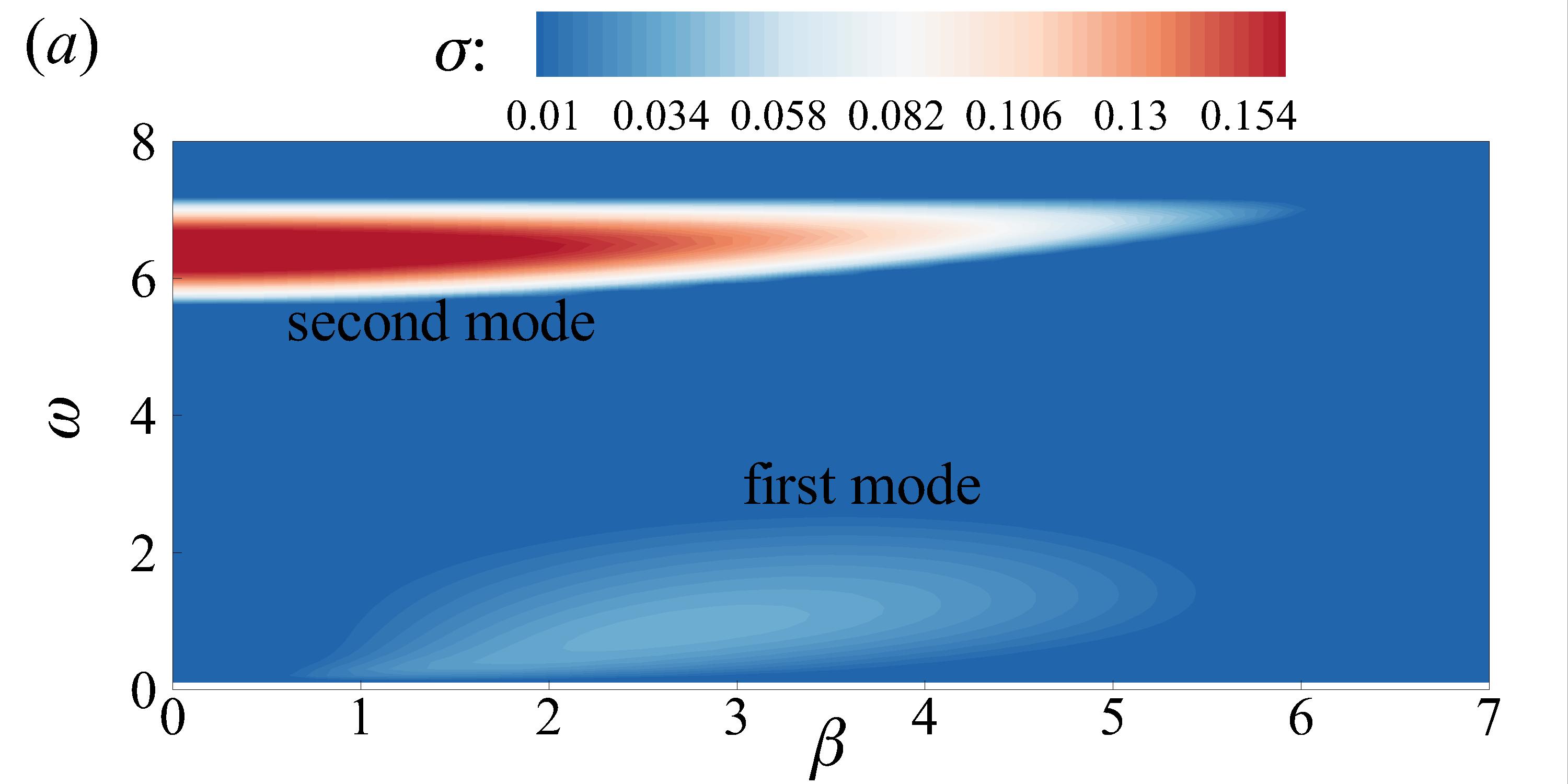}}
	\centering{ \includegraphics[width=6.5cm]{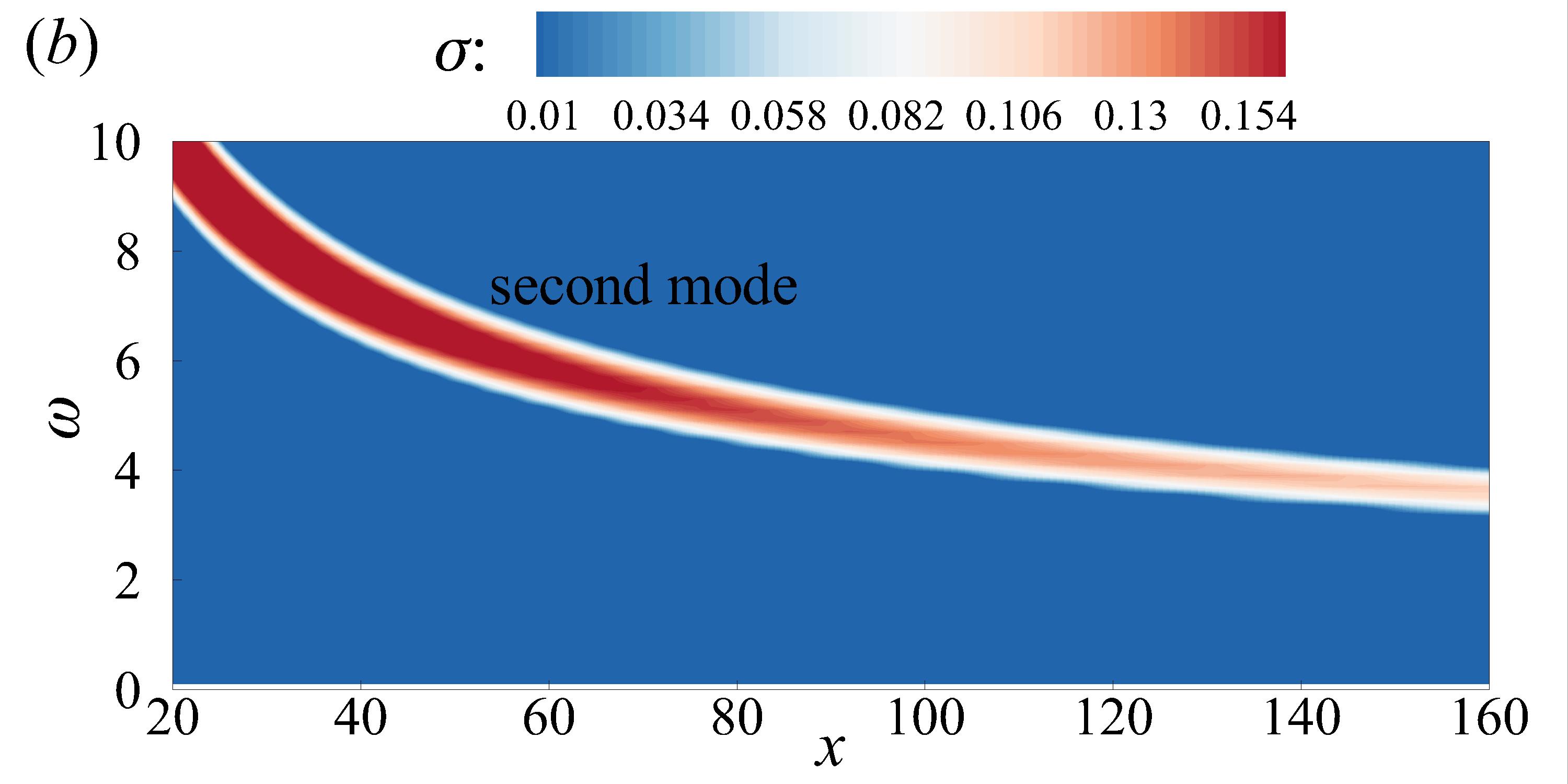}}
	\caption{\added{Contours of (\textit{a}) local growth rate $\sigma$ versus $\beta-\omega$ at $x=50$ and (\textit{b}) $\sigma$ versus $x-\omega$ with $\beta=0$ for the sharp-leading-edge case R0}. No unstable local mode is identified downstream for blunt plate cases and thus not shown, e.g., case R2 at $x=50$ and $x=150$.}
	\label{fig_growthrate}
\end{figure*}

\begin{figure*}
	\centering{ \includegraphics[width=4.2cm]{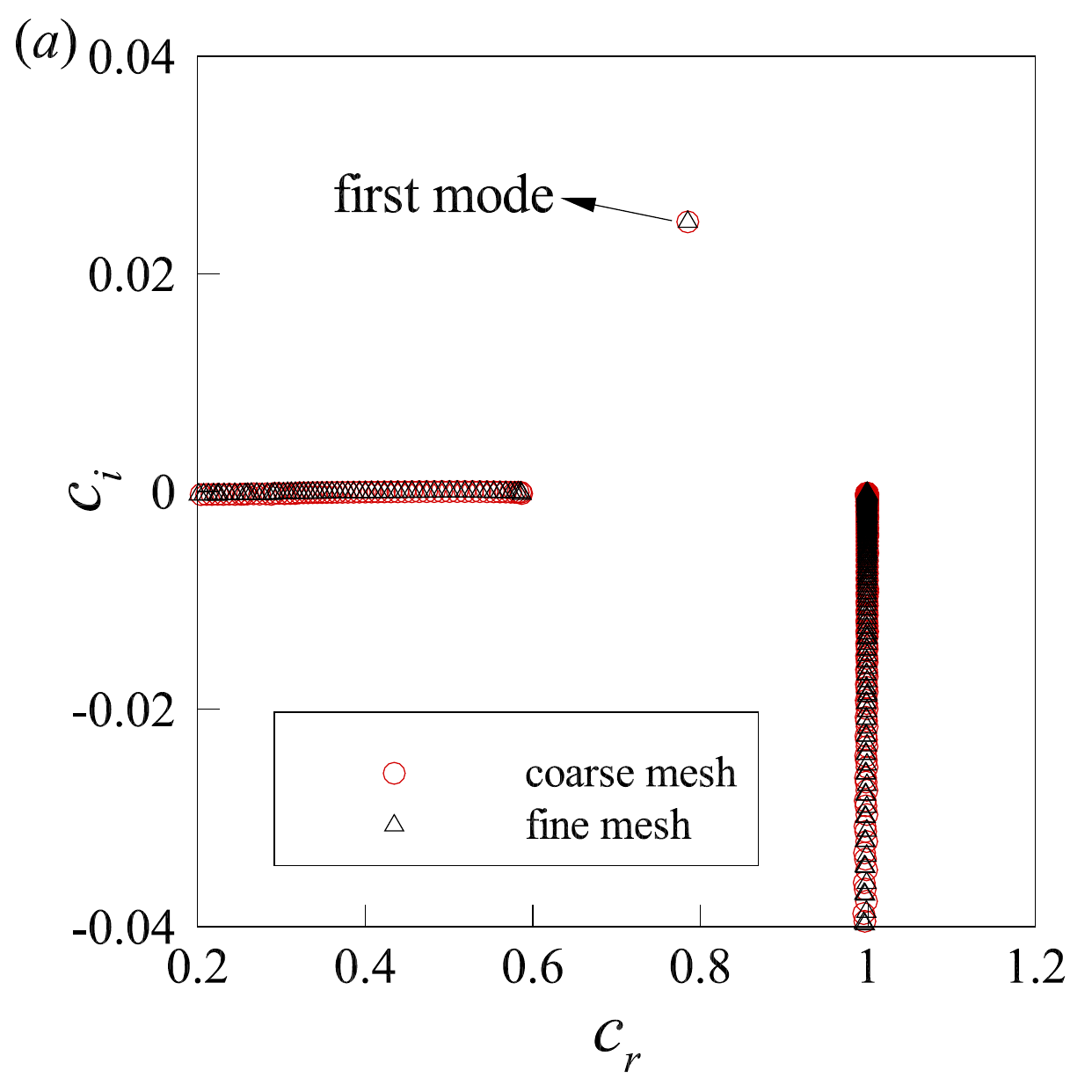}}
	\centering{ \includegraphics[width=4.2cm]{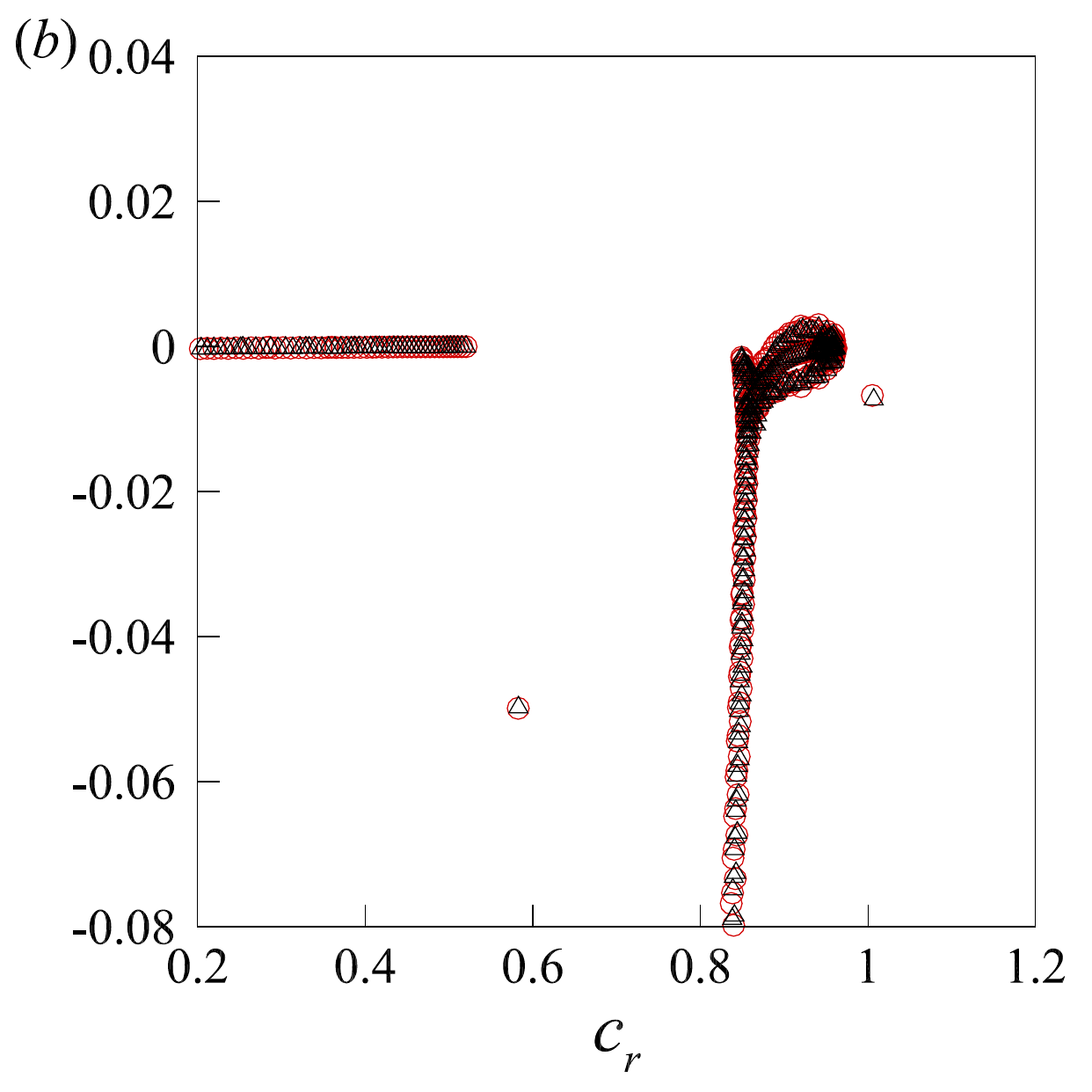}}
	\centering{ \includegraphics[width=4.2cm]{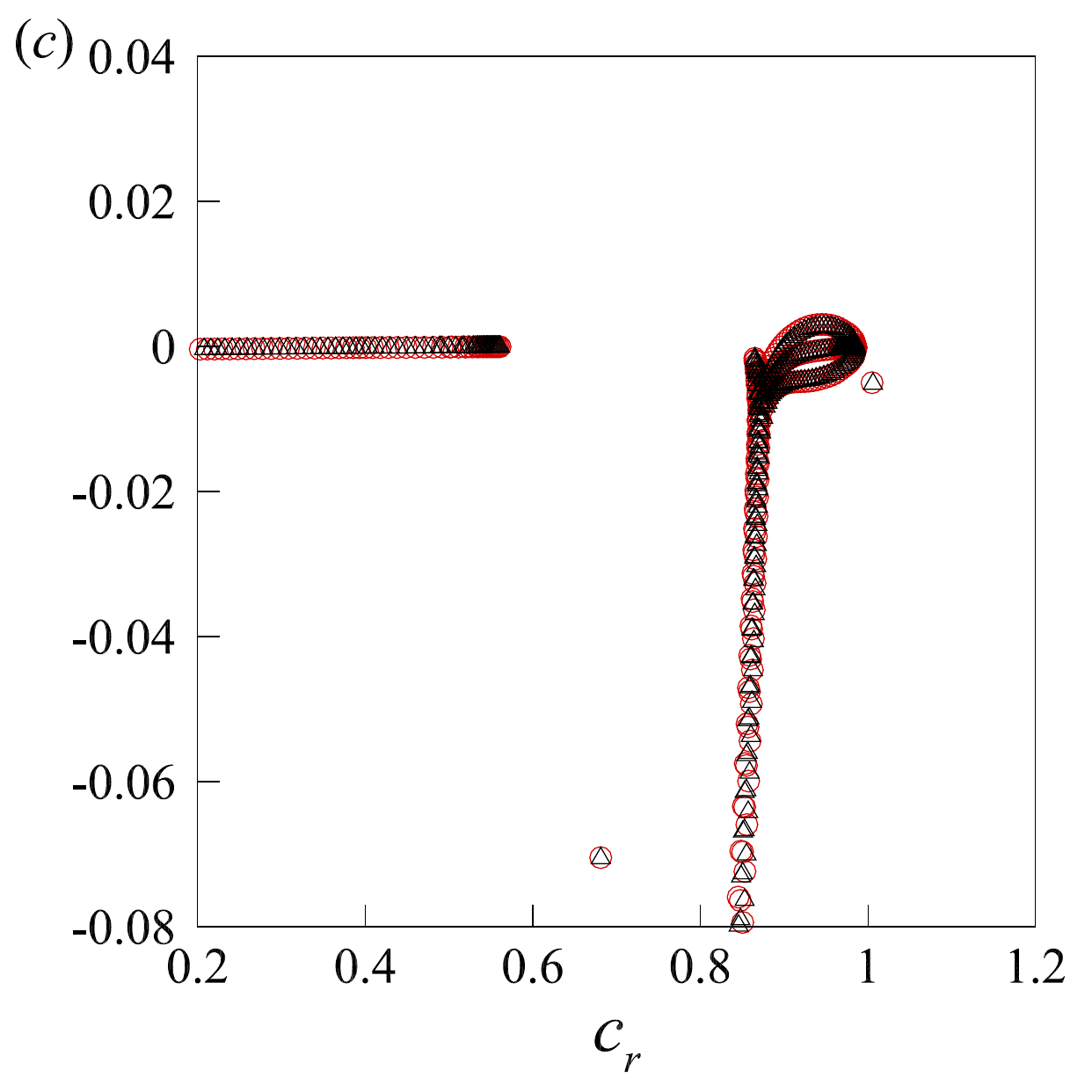}}
	\caption{Eigenspectrum of the complex phase velocity $c=\omega/\alpha$ with $\omega=0.9$ and $\beta=2.75$ for (\textit{a}) case R0 at $x=50$, (\textit{b}) case R2 at $x=50$ and (\textit{c}) case R2 at $x=150$.}
	\label{fig_eigenvalue}
\end{figure*}
To clarify the mechanism of the linear instability, it is necessary to perform a normal-mode stability analysis. For the sharp-leading-edge case, the laminar base flow is extracted at a pre-transitional location $x=50$ for stability analysis. Subsequently, the one-dimensional (1-D) base flow along the wall-normal direction is transformed into another 1-D mesh, which is implemented by cubic spline interpolation via Intel MKL. Algebraic stretching is employed for the new 1-D mesh points, which are more robust for spectral difference in the stability analysis. The algebraic stretching function has been incorporated into the spectral collocation method by \cite{malik1990numerical}. For the new 1-D mesh, $n_\textit{y,new}\ge161$ is found to be sufficient to achieve convergence of eigenvalues. The following `coarse mesh' and `fine mesh' in figure \ref{fig_eigenvalue} represent the mesh point number of $n_\textit{y,new}=201$ and $n_\textit{y,new}=221$, respectively.

The local growth rate $\sigma=-\alpha_i$ is calculated in the space of $(\omega,\beta)$, where $\sigma>0$ suggests an unstable mode. Figure \ref{fig_growthrate}(\textit{a}) clearly shows the unstable range of the second mode and the oblique first mode \added{at $x=50$ for case R0}.  The growth rates of the first and second modes peak at around $(\omega,\beta)=(0.9,2.75)$ and $(\omega,\beta)=(6.4,0)$, respectively. These two states correspond to dimensional frequencies of 127 kHz and 902 kHz, respectively. The wavelength of the most unstable first mode is $\lambda_z^{\ast}=2.28$ mm. However, for all the blunt-plate cases, no discrete mode with $\sigma>0$ is identified in the range $x\in[5, 200]$, $\omega\in[0, 8]$ and  $\beta\in[0, 7]$. It is not surprising that the unstable second mode is absent, since the boundary-layer-edge Mach number is only 2.7 downstream. \added{For case R0, figure \ref{fig_growthrate}(\textit{b}) displays the streamwise dependent unstable region with $\beta=0$, which indicates the presence of the unstable second mode and the absence of the unstable 2-D first mode.} 

With regard to the most unstable first mode, the eigenspectra of case R0 at $x=50$ as well as case R2 at $x=50$ and $x=150$ are shown in figure \ref{fig_eigenvalue}. The complex phase velocity is defined by $c=\omega/\alpha$, where $c_i>0$ indicates an unstable mode. Clearly, the discrete mode becomes stable even when the nose-tip radius is only $R_n=2$. These findings are consistent with existing knowledge of the stabilisation effect of the entropy layer. \added{Note that figures \ref{fig_eigenvalue}(\textit{b}, \textit{c}) appear to show marginally unstable eigenvalues in the continuous branch. However, these eigenvalues cannot achieve convergence with an increasing $n_\textit{y,new}$, and they diverge in the iterative local method of \cite{malik1990numerical}. Therefore, these eigenvalues should correspond to spurious modes. By contrast, the discrete eigenvalue can easily converge, e.g., $c= (0.68, -0.07)$ in figure \ref{fig_eigenvalue}(\textit{c}). The nose bluntness effect on the presence of the most unstable first mode is also examined at $x=50$. The critical nose-tip radius for the disappearance of an unstable first mode is approximately $R_n=0.5$, where the most unstable state is $(\omega,\beta)=(0.5,1)$. In the simulated cases from R1.8 to R3, unstable first mode and second mode are not identified. As a result, the normal-mode instability is not significant even with a small nose-tip radius, which differs from a slender-cone case with early entropy swallowing. The stabilisation effect of the persist blunt-flat-plate entropy layer on the normal modes extends far downstream.} The non-swept plane stagnation flow near the nose tip is also known to be linearly stable to three-dimensional normal-mode instabilities \citep{Wilson1978The,Lyell1985Linear}. The takeaway information in this case is that the transition to turbulence subject to freestream broadband disturbances, if possible, should be probably due to nonmodal instability mechanisms. The considered blunt-plate cases exclude the possibility of normal-mode instabilities.

\subsection{Three-dimensional instability and transition}

\begin{figure*}
	\centering{ \includegraphics[width=8cm]{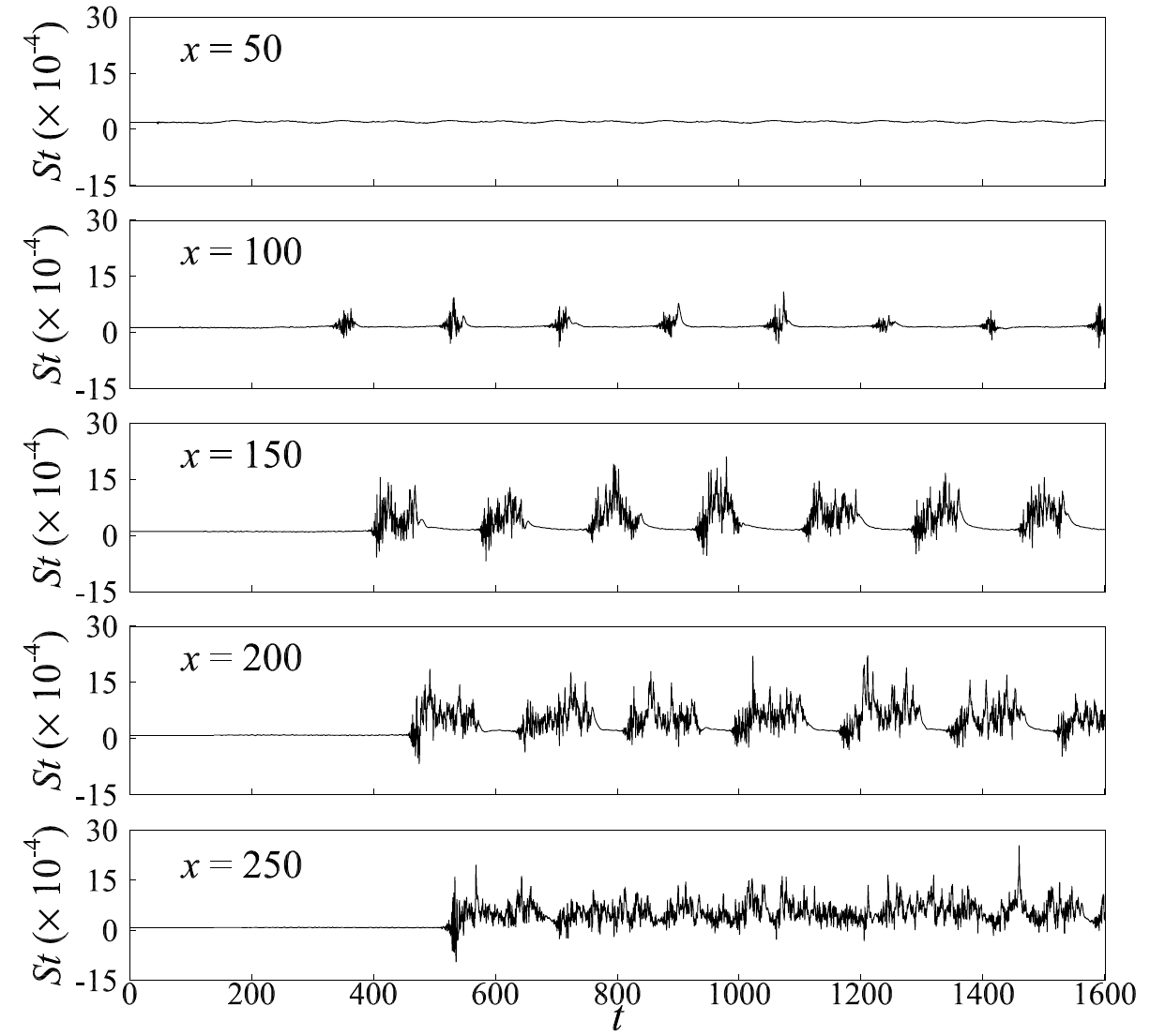}}
	\caption{Time history of the Stanton number at different streamwise locations on the symmetry plane for case R2.}
	\label{fig_st_history}
\end{figure*}
\begin{figure*}
	\centering{ \includegraphics[width=11cm]{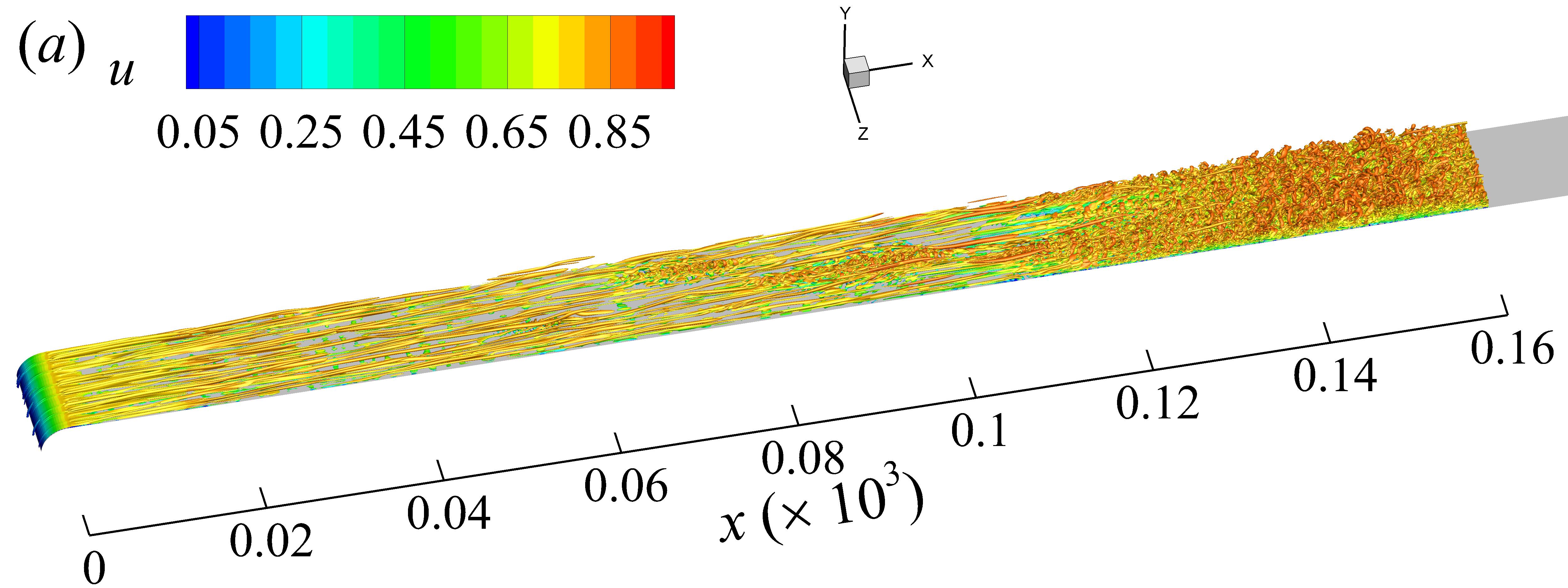}}
	\centering{ \includegraphics[width=11cm]{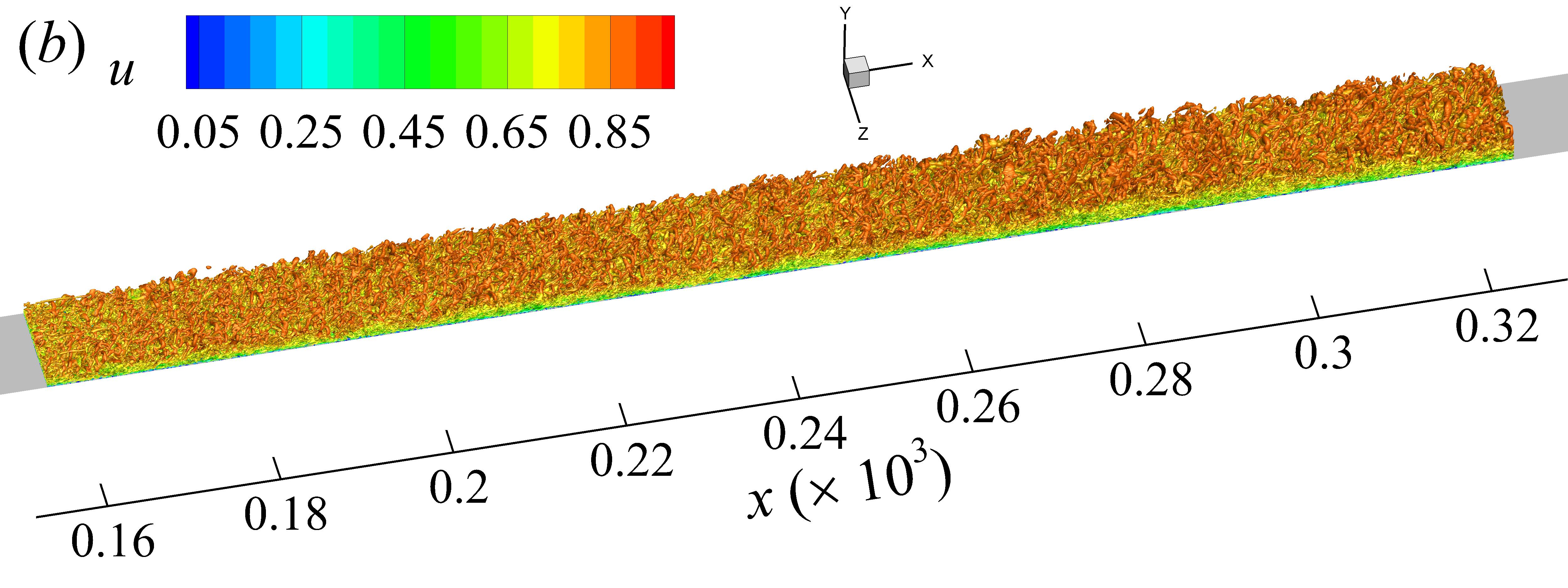}}
	\caption{Isosurface of $Q$-criterion $Q=1\times10^{-3}$ at $t=806$ in the range (\textit{a}) $x<160$ and (\textit{b}) $x>160$ coloured by the streamwise velocity for case R3.}
	\label{fig_Q}
\end{figure*}

\begin{figure*}
	\centering{ \includegraphics[width=6.5cm]{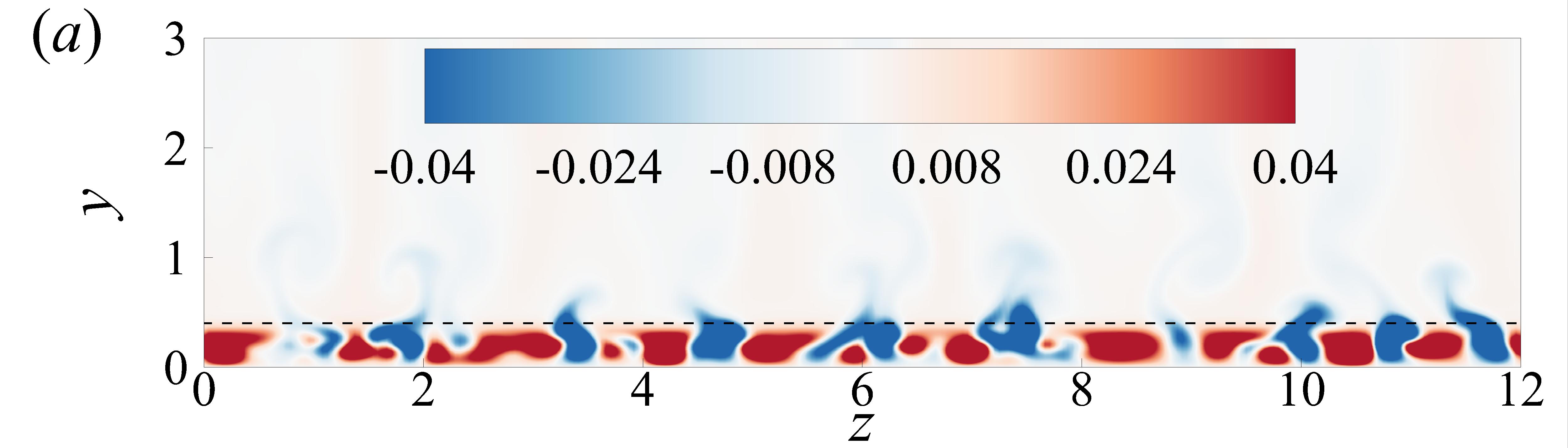}}
	\centering{ \includegraphics[width=6.5cm]{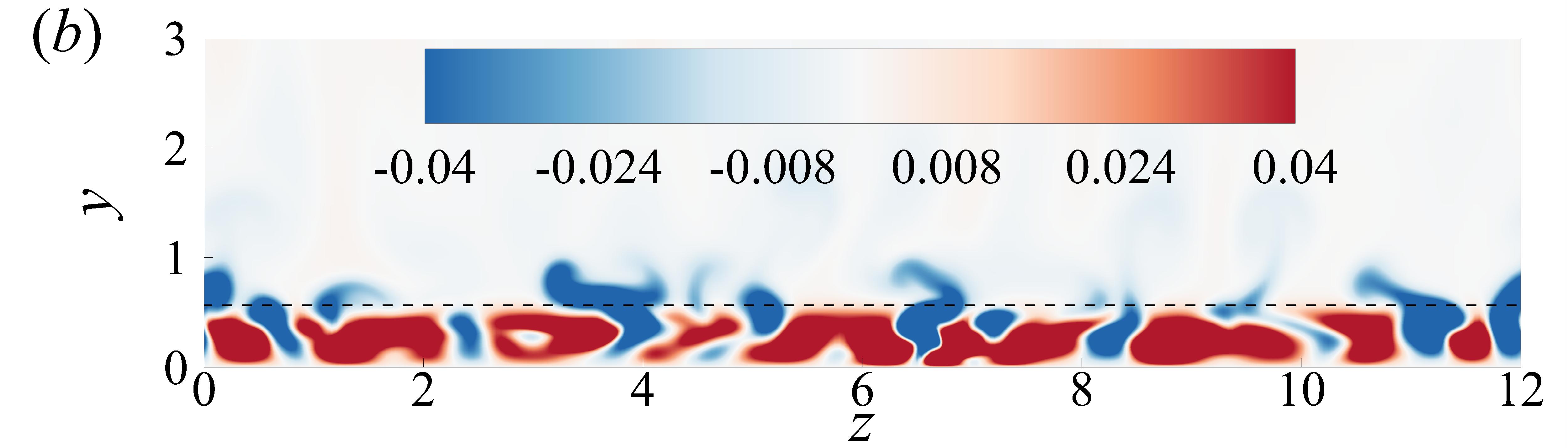}}
	\caption{\added{Contour of the streamwise velocity perturbation $u_{\textit{perturb}}$ compared to the laminar flow of case R3 at $t=806$ for (\textit{a}) $x=40$ and (\textit{b}) $x=80$. Dashed lines mark the position of the 2-D laminar boundary layer edge.}}
	\label{fig_ZY_uperturb}
\end{figure*}

\added{Prior to the simulation with broadband acoustic-wave forcings, a pre-run without the forcing is conducted for case R1.8. The numerical noise is found to have an ignorable impact on the laminar flow, and further details are given in Appendix \ref{appE}. Similar to the 2-D simulation, the detached bow shock first interacts with the incoming perturbation and amplifies the amplitude. For instance, along the symmetry plane $y = 0$ where the bow shock approaches a normal shock, the spanwise-averaged pressure fluctuation r.m.s. is increased by about eight times for cases R1.8 to R3 across the shock. This degree of increase is not sensitive to the nose-tip radius probably due to the dominance of inviscid properties at large nose-radius Reynolds number. The rise in the sound pressure level across the shock is about 19 dB, which approaches the inviscid analysis (20 dB at Mach 5) by \cite{mahesh1995interaction}. Farther downstream, disturbances are excited in the boundary layer and entropy layer. }

The time history of the Stanton number at different streamwise stations of case R2 is recorded and shown in figure \ref{fig_st_history}.\added{ Note that the probe on the symmetry plane $z=L_{z}/2$ is taken for an example, and the freestream forcing is not symmetrical with respect to $z$ due to the random phase angles.} For the probed laminar-flow location $x=50$, the perturbations exert no substantial influence on the mean heat flux compared to the undisturbed laminar value at $t=0$. In the transitional region ($x=100$, $x=150$, $x=200$ and $x=250$), clear signature of intermittency is observed. Windows with high-amplitude perturbations appear 
intermittently when turbulent spot (shown later) is present. During the intermittent turbulent window, the mean heat flux is escalated to around 4$\sim$6 times the laminar value. Furthermore, the degree of intermittency, i.e. the ratio of the turbulent time to the total time, is visibly increased from the early stage $x=100$ to the late stage $x=250$. The presence of intermittency is observed for all the 3-D cases in this paper. Note that a flow-through time unit is $t=260$ for case R2. After a stationary state is reached, statistical calculations of mean and r.m.s. quantities are performed during an interval of $t_2-t_1=1770$ or $t_2-t_1=1010$, and no visible difference is found between the two choices.

\begin{figure*}
	\centering{ \includegraphics[width=12cm]{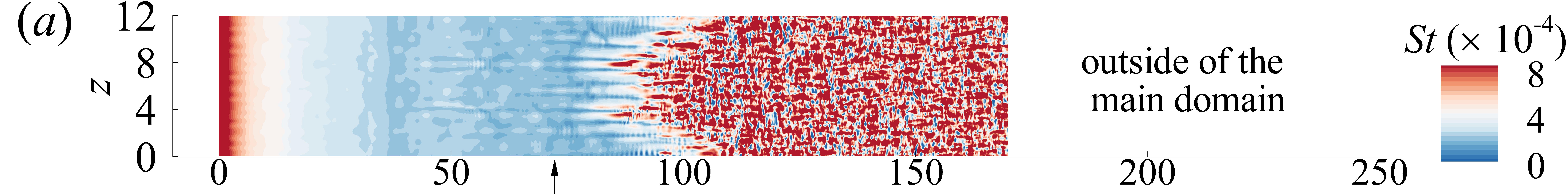}}
	\centering{ \includegraphics[width=12cm]{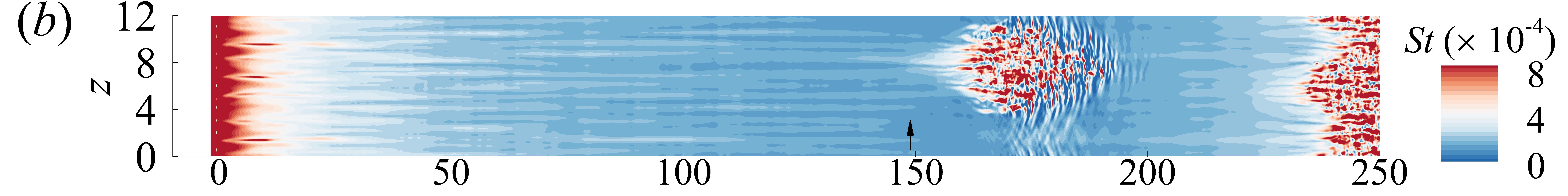}}
	\centering{ \includegraphics[width=12cm]{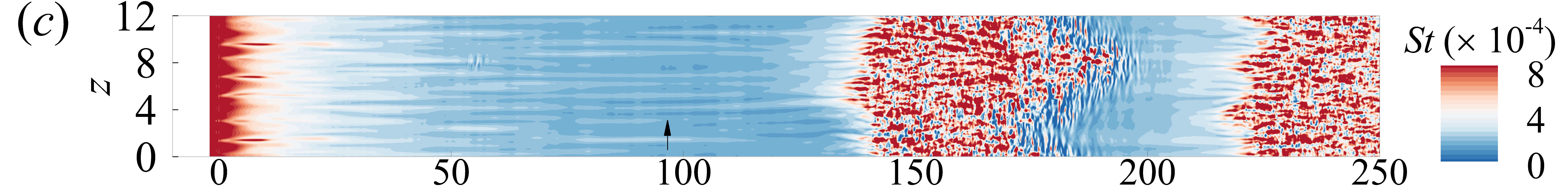}}
	\centering{ \includegraphics[width=12cm]{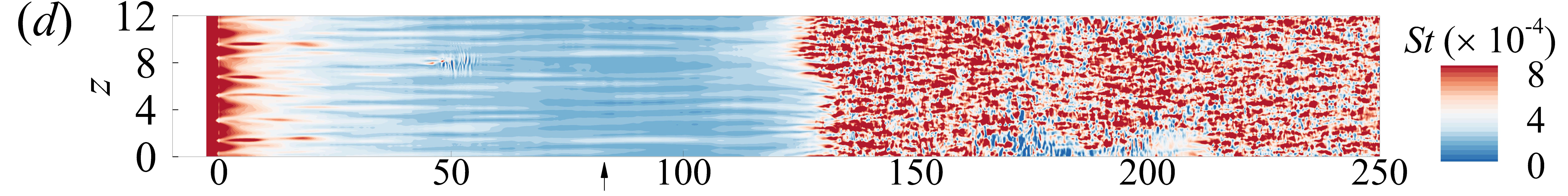}}
	\centering{ \includegraphics[width=12cm]{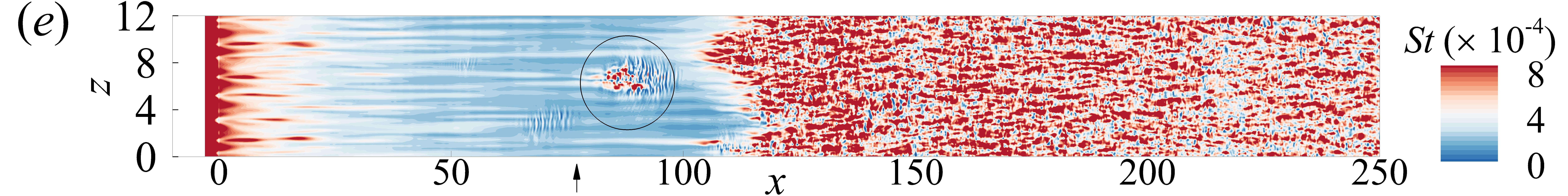}}
	\caption{Contour of instantaneous Stanton number at $t=806$ for cases (\textit{a}) R0, \replaced{(\textit{b}) R1.8, (\textit{c}) R2, (\textit{d}) R2.7 and (\textit{e}) R3}{(\textit{b}) R2, (\textit{c}) R2.7 and (\textit{d}) R3}. Circle in (\textit{d}) mark\added{s} the early turbulent spots. Arrows represent the transition onset locations.}
	\label{fig_st_contour}
\end{figure*}

Figure \ref{fig_Q} gives a $Q$-criterion visualisation of the flow structure for case R3. Figure \ref{fig_ZY_uperturb} further shows the streamwise velocity perturbation $u_{\textit{perturb}}$. Here, the perturbation is the difference between the instantaneous field and the laminar flow, i.e., $u_{\textit{perturb}}(x,y,z,t)=u(x,y,z,t)-u_{\textit{laminar}}(x,y)$. As shown by figure \ref{fig_Q}, streamwise-vortices like structures are formed immediately downstream of the blunt nose. Figure \ref{fig_ZY_uperturb} depicts that the development of streamwise vortices or
streamwise streaks is mostly beneath the boundary layer edge (dashed line)\deleted{ and far away from the entropy layer edge (dashed-dotted line)}. The outer entropy layer is only marginally disturbed\added{ (see figure \ref{fig_uperturb})}. In the approximate range $60<x<120$ of figure \ref{fig_Q}(\textit{a}), turbulent spots are generated, which are surrounded by streamwise vortices and then convected away towards downstream. For a spatially fixed probe, intermittent signal of the physical quantity can be recorded, as shown in figure \ref{fig_st_history}.  Similar to the numerical finding of \cite*{marxen2019turbulence}, a core region of the turbulent spot can be developed, where the skin friction or heat flux can be escalated. The increase in heat transfer is also recorded in the turbulent window in figure \ref{fig_st_history} and later in figure \ref{fig_st_contour}. 

Figure \ref{fig_st_contour} shows the contours of the instantaneous Stanton number. For all blunt-plate cases, a notable observation is that streamwise streaks appear immediately downstream of the nose region. Following that, the turbulent spot surrounded by the streaky laminar flow emerges, which has a finite and gradually increasing spanwise width. In the core region of turbulent spots, high-frequency components become more evident (see figure \ref{fig_st_history}), which are associated with secondary instabilities under the streaky laminar flow. Eventually, the turbulent spot spans in the whole spanwise domain and develops to fully established turbulence.\added{ There are other similarities in the turbulent spot for different blunt-plate cases. The core high-$St$ part of the wave packet is always generated and travelling with a `slender-wedge' wave front. The initial semi-angle of the wedge front is around 10$^{\circ}$ on the $x$--$z$ plane, and slightly increased by less than 5$^{\circ}$ when the packet grows larger. Adjacent to both sides of the wave front, corrugated structures always appear with low or even negative instantaneous $St$ values. In figures \ref{fig_st_contour}(\textit{b},\textit{c}), the first and second high-$St$ regions represent the propagating turbulent spot and fully developed turbulent regions, respectively. Between them, the intermittent laminar-flow region is observed with low $St$, which corresponds to the `quiet' low-$St$ window in the time history at $x=200$ in figure \ref{fig_st_history}. Although there exist such instantaneous low-$St$ regions, the time-averaged $St$ rises monotonically in the streamwise direction due to gradually increasing intermittency in the transitional region. The downstream propagation of the turbulent spot ensures that the probes in the transitional region will experience high heat transfer sometime during long-period statistics.} 

\begin{figure*}
	\centering{ \includegraphics[width=11cm]{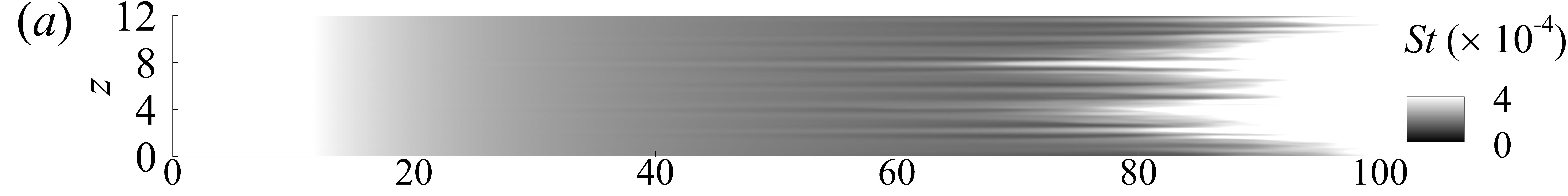}}
	\centering{ \includegraphics[width=11cm]{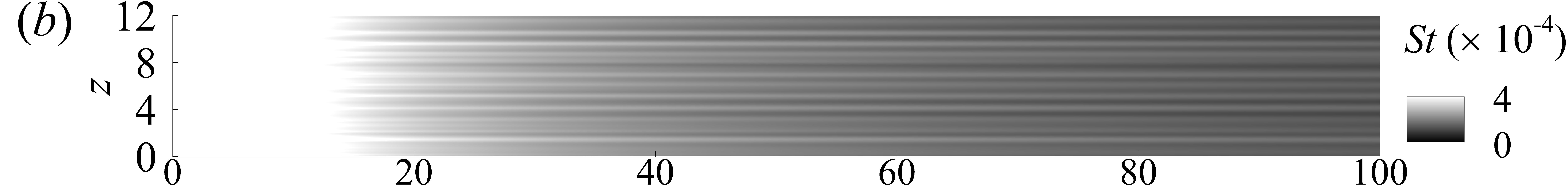}}
	\centering{ \includegraphics[width=11cm]{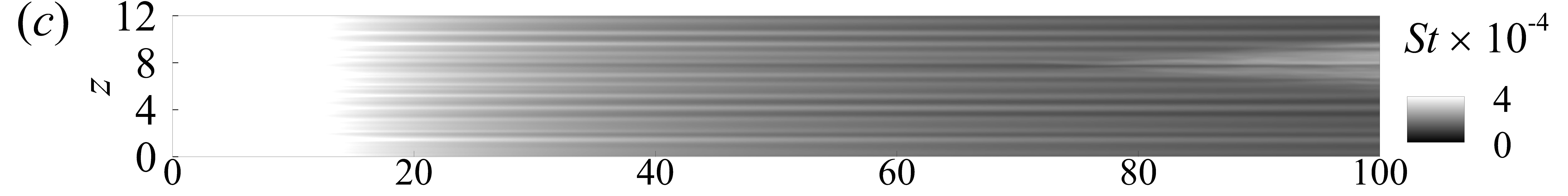}}
	\centering{ \includegraphics[width=11cm]{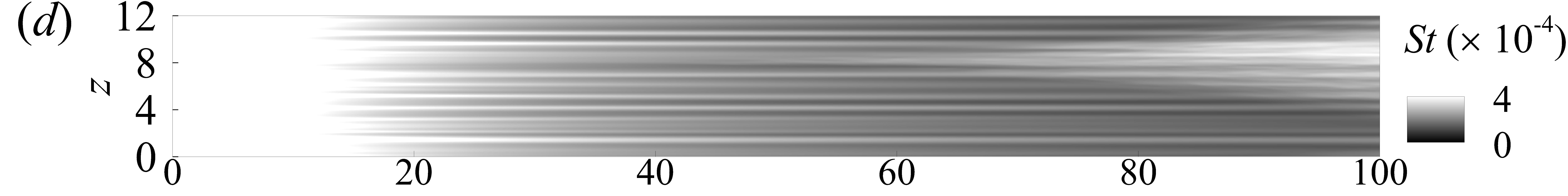}}
	\centering{ \includegraphics[width=11cm]{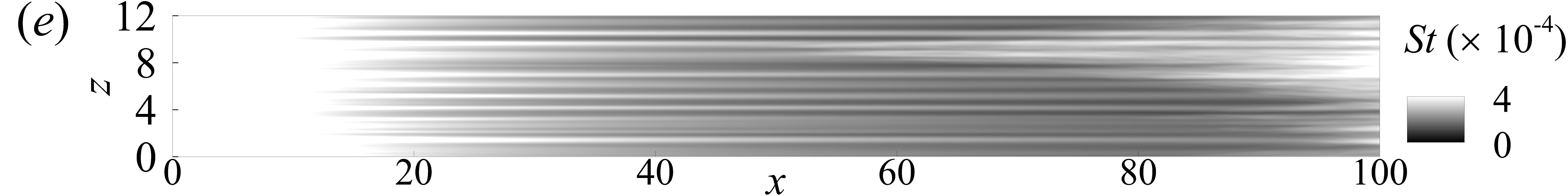}}
	\caption{Leading-edge streaks characterised by time-averaged Stanton number for cases (\textit{a}) R0, \replaced{(\textit{b}) R1.8, (\textit{c}) R2, (\textit{d}) R2.7 and (\textit{e}) R3}{(\textit{b}) R2, (\textit{c}) R2.7 and (\textit{d}) R3}. }
	\label{fig_averaged_st}
\end{figure*}

By contrast, for case R0 under the same environment of freestream disturbances, no strong signature of the streamwise streak is observed in the vicinity of the leading edge. Nonetheless, three-dimensionality is still seen in the pre-transitional and transitional regions of case R0, which might be related to first-mode instabilities. To clarify the physical cause, spanwise-wavenumber spectral analysis will be performed later in conjunction with LST.\added{ Since the surface heat transfer is escalated on the nose, the streak is not visible there under the current contour level. However, an unshown replotted contour near the nose illustrates that the streak originates from the curved nose, yet not at the stagnation point. The high-fluctuation region is initially detached from the wall near the stagnation point, and then appears in the nose boundary layer with the signature of streaks (see figure \ref{urms_contour}).} The streamwise streaks are further visualised by the mean Stanton number in figure \ref{fig_averaged_st}. The streaky spanwise spacings of cases\added{ R1.8,} R2, R2.7 and R3 seem to be close to each other.

\begin{figure*}
	\centering{ \includegraphics[width=4.4cm]{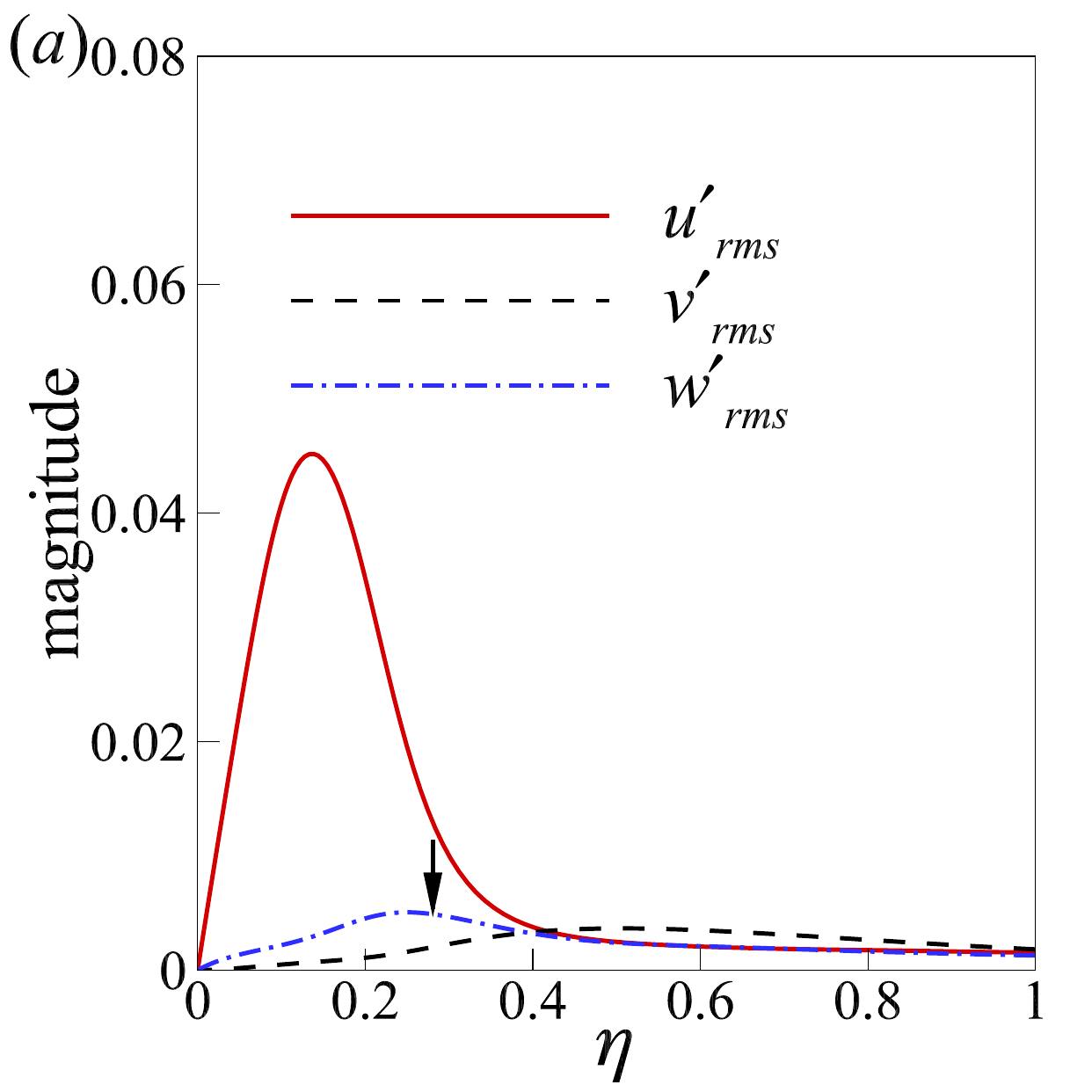}}
	\centering{ \includegraphics[width=4.4cm]{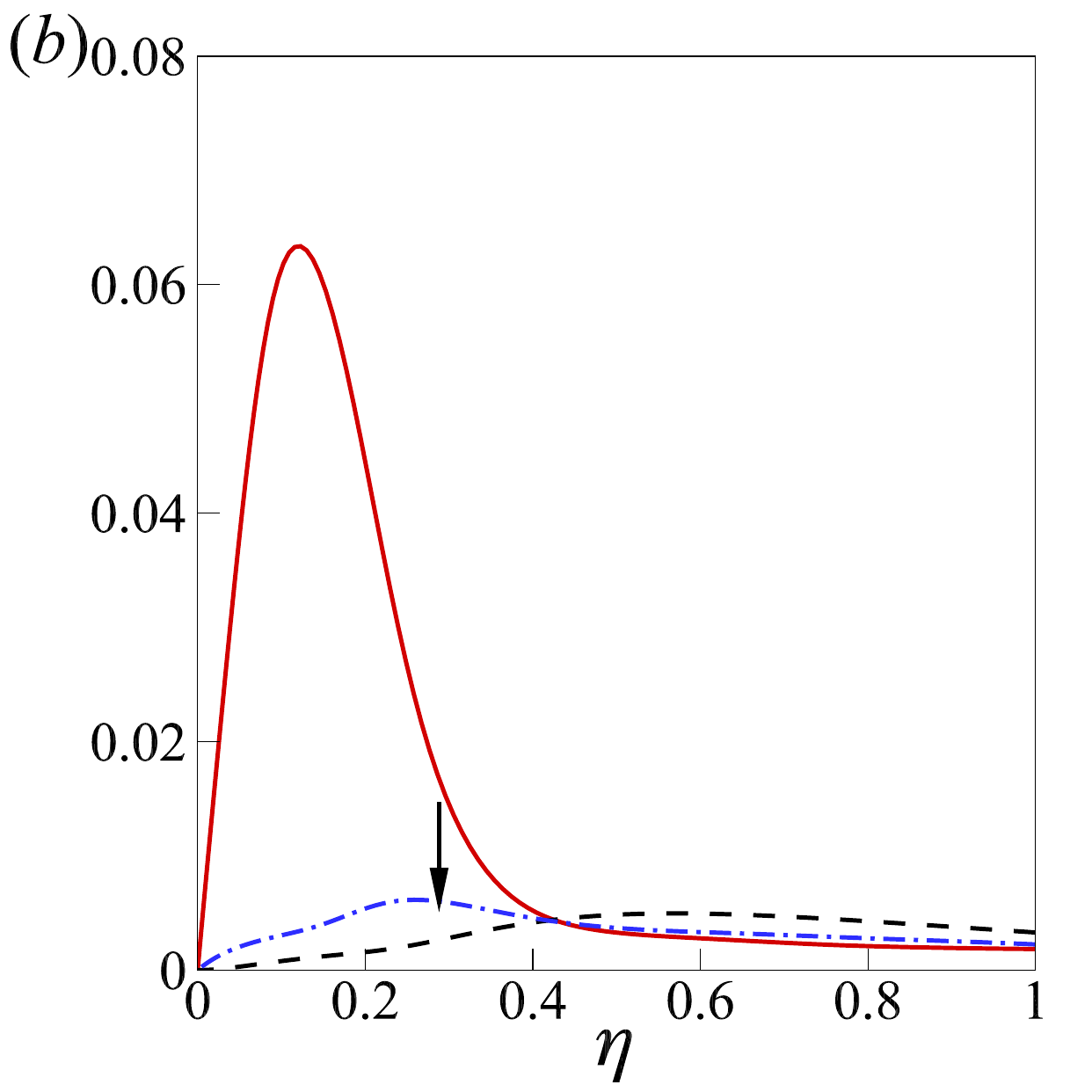}}
	\centering{ \includegraphics[width=4.4cm]{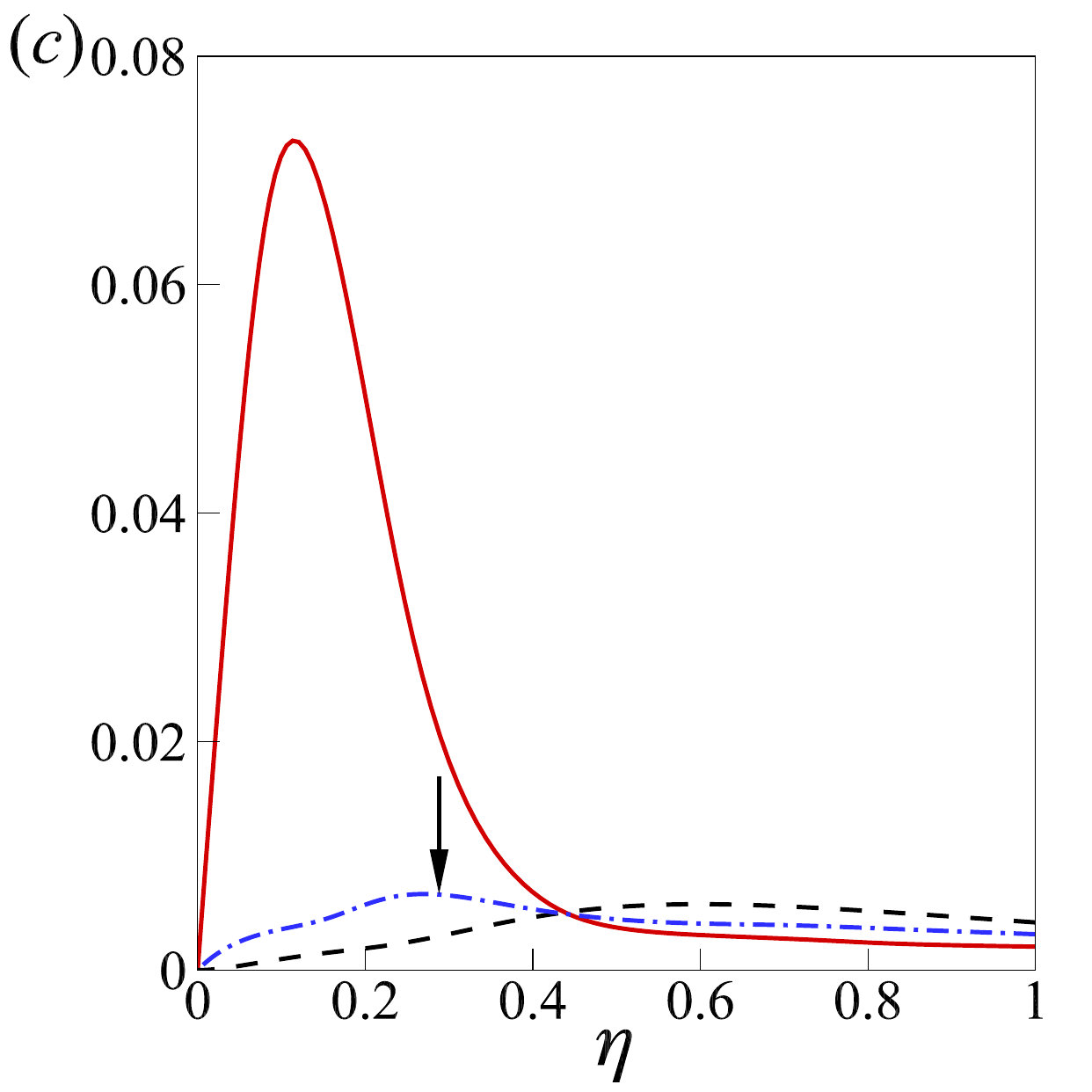}}
	\caption{Wall-normal profiles of spanwise-averaged r.m.s. of velocity fluctuations ($u'$, $v'$, $w'$) at the laminar-flow position $x=20$ for cases (\textit{a}) R2, (\textit{b}) R2.7 and (\textit{c}) R3. Arrows represent the locations of local laminar-boundary-layer thickness.}
	\label{fig_streak_rms}
\end{figure*}

Figure \ref{fig_streak_rms} shows the distribution of spanwise-averaged r.m.s. of the fluctuating velocities at the laminar-flow location $x=20$. The statistical feature $|u'|\gg|v'|$, $|u'|\gg|w'|$ is clear signature of streamwise streaks for blunt-plate flows.\added{ In comparison, streamwise vortices are more pronounced in $|v'|$ and $|w'|$, and the momentum transport of them can generate streaks \citep{orlandi1994generation}. Such energy transfer during the streamwise vortices to streaks is attributable to the lift-up mechanism \citep{landahl1980note}.} More importantly, the maximum of $u'_{\textit{rms}}$ gradually grows with increasing bluntness. The respective peak values are 0.045, 0.063 and 0.073 for cases R2, R2.7 and R3. The corresponding wall-normal heights $\eta$ are \replaced{0.135, 0.119 and 0.116}{0.067, 0.045 and 0.039}, respectively, which are inside the boundary layer. Therefore, as nose bluntness is increased, the streaks are enhanced, and simultaneously the peak fluctuation positions move \added{slightly} towards the wall. \added{For the unshown case R1.8, its peak magnitude of $u'_{\textit{rms}}$, $v'_{\textit{rms}}$ and $w'_{\textit{rms}}$ is close to that of case R2 at $x=20$. Nonetheless, the peak $v'_{\textit{rms}}$ and $w'_{\textit{rms}}$ of case R2 are about 3.1 times the respective values of case R1.8 at the downstream location $x=90$, where transition is soon triggered for case R2.} The finding of the strengthened streaks with nose bluntness is consistent with the resolvent analysis of blunt-cone flows by \cite{Melander2022Nose}. The above observation also accounts for the preceding enlarged deviation from the 2-D laminar flow in wall heat transfer with increasing nose bluntness (see figure \ref{fig_St_curves}(\textit{b--d})). In comparison, the difference between 3-D and 2-D laminar flows for case R0 is ignorable in figure \ref{fig_St_curves}(\textit{a}). This performance indicates that the streamwise streak does not play an important role in wall heat transfer of the sharp-leading-edge case.

\begin{figure*}
	\centering{ \includegraphics[width=6.6cm]{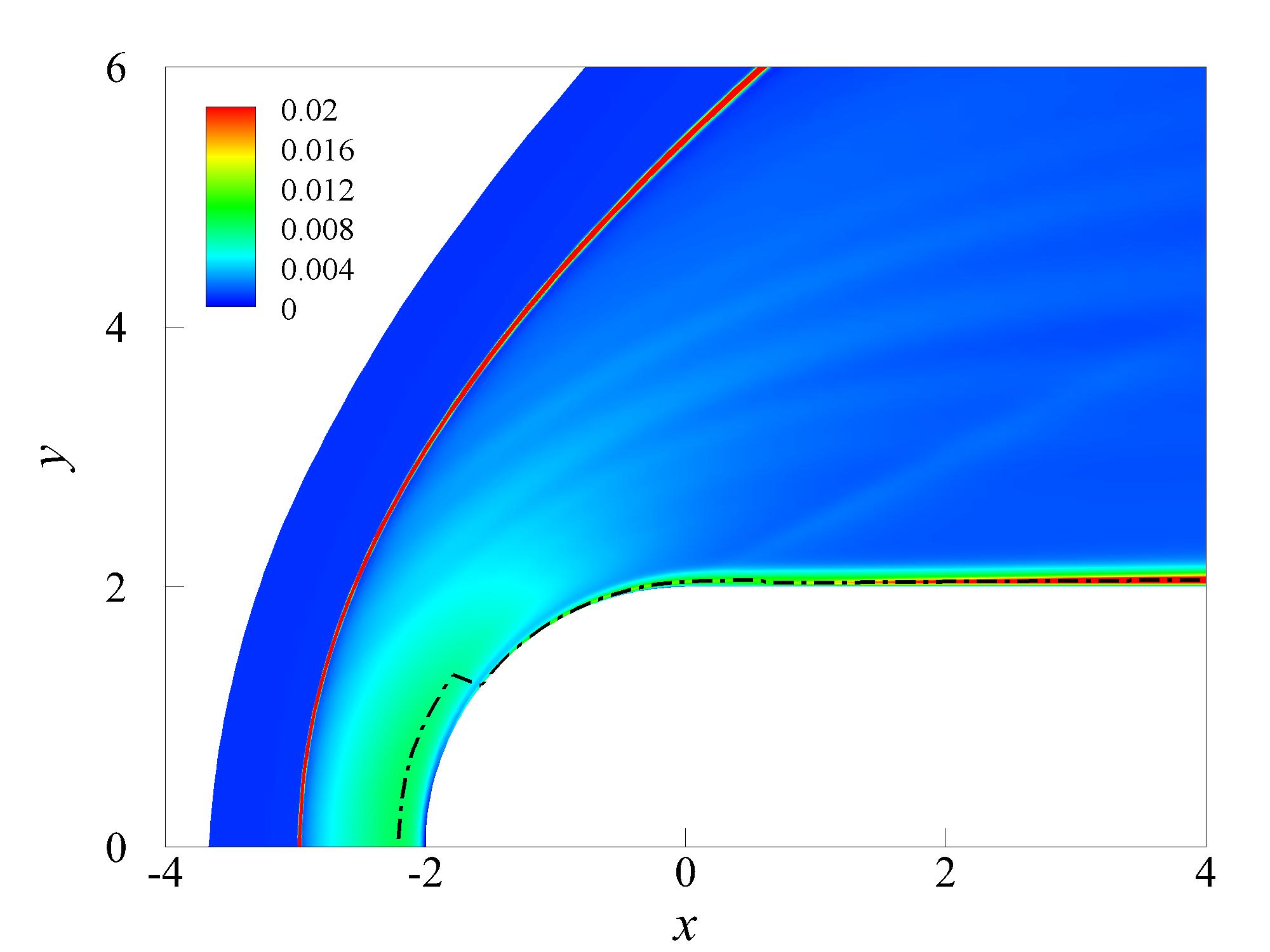}}
	\caption{\added{Contour of the spanwise-averaged $u'_\textit{rms}$ and the location of its local maximum (dashed-dotted line) for case R2.}}
	\label{urms_contour}
\end{figure*}

\begin{figure*}
	\centering{ \includegraphics[width=6.6cm]{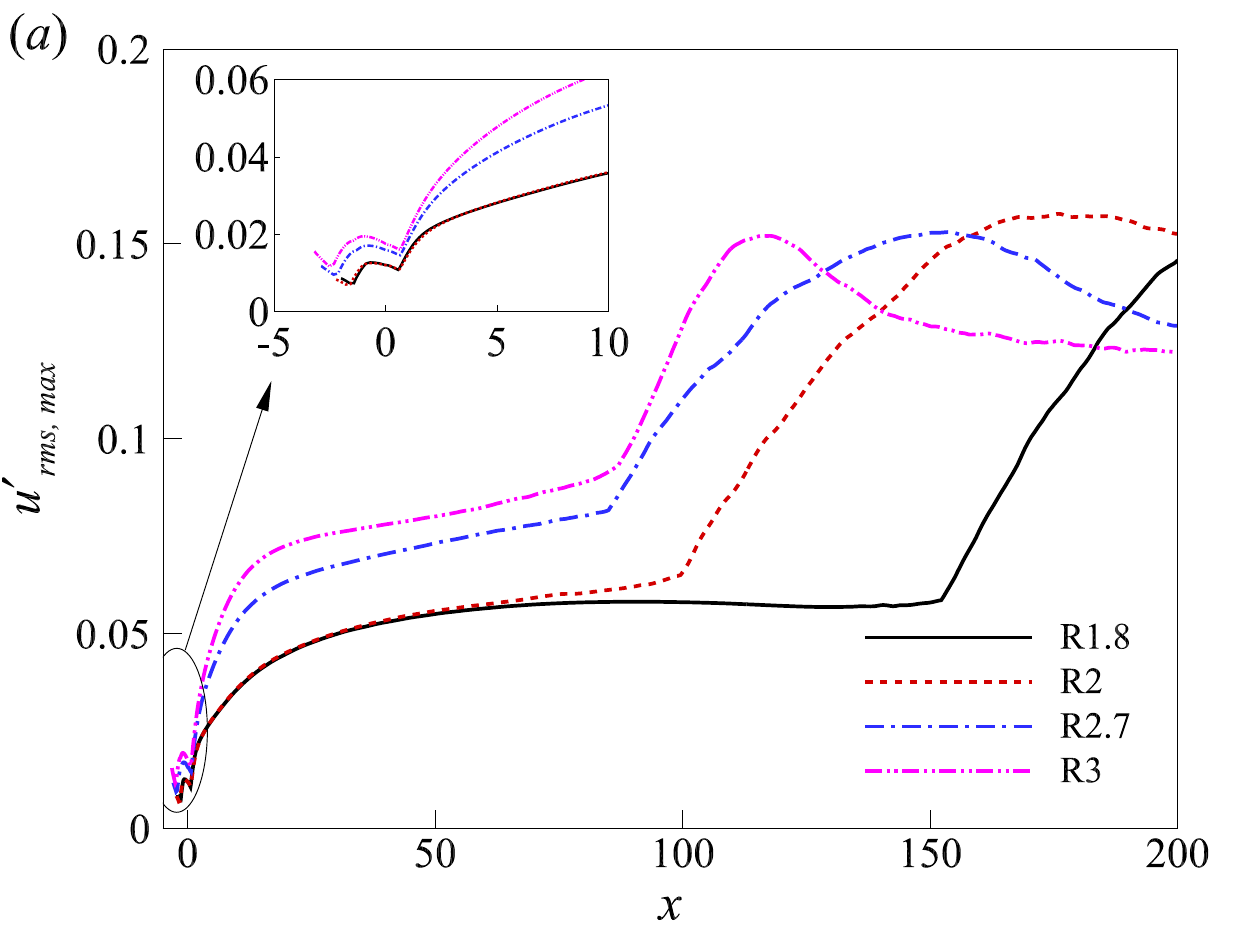}}
	\centering{ \includegraphics[width=6.6cm]{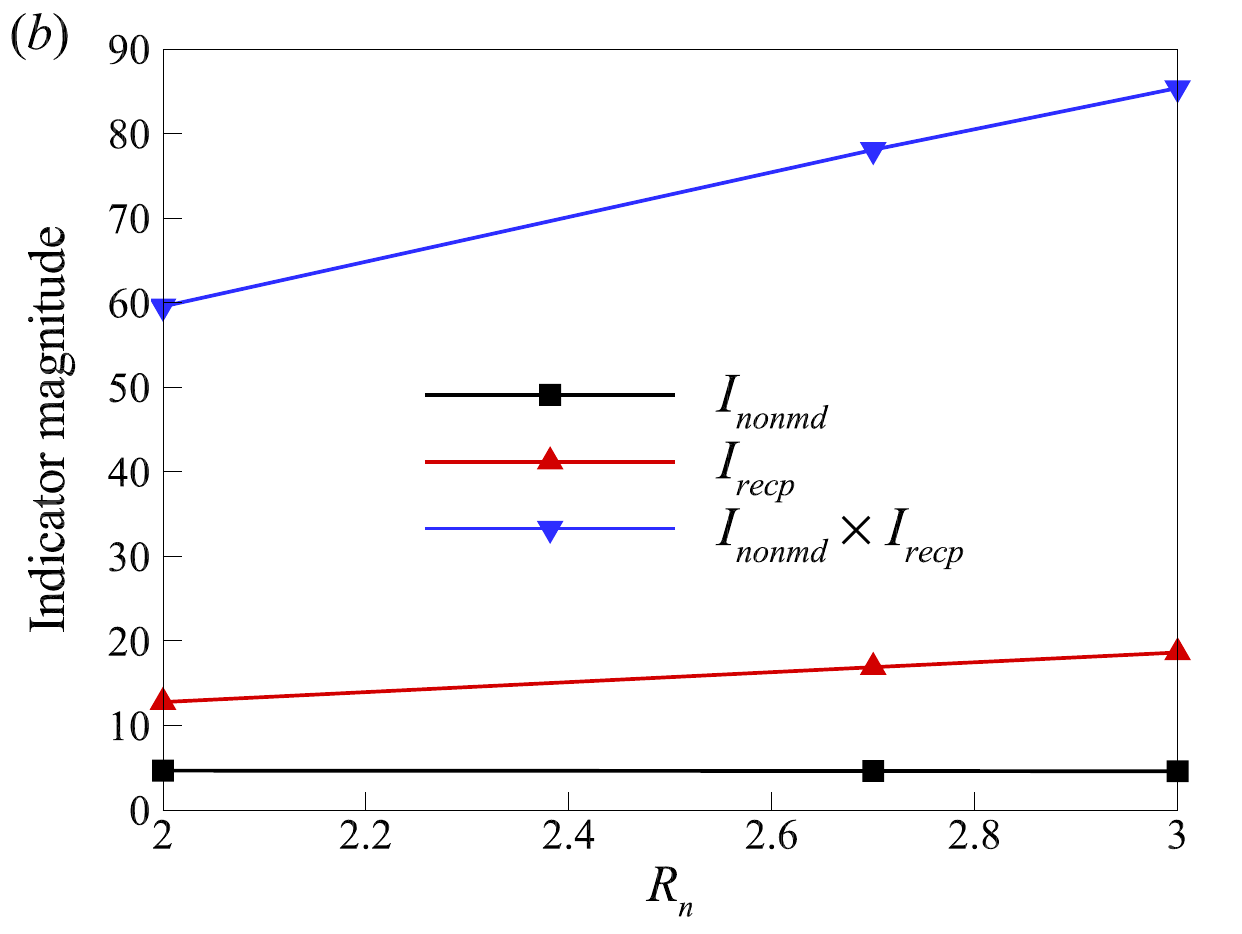}}
	\caption{\added{(\textit{a}) Local maximum of spanwise-averaged $u'_\textit{rms}$ versus $x$, and (\textit{b}) indicators of the significance of receptivity or nonmodal growth subject to the nose bluntness effect. The leading-edge reference location, $x_\textit{LE}=0$.}}
	\label{rms_max_develop}
\end{figure*}

\added{The augmented streaky response with increasing nose bluntness is attributable to either the strengthened leading-edge receptivity to freestream disturbances or the nonmodal amplification downstream. By mentioning `receptivity', we consider the process during which the freestream disturbances enter the boundary layer in the vicinity of the leading edge. By referring to `nonmodal growth', we examine the early evolution of the streaky strength. The streaky strength is quantified by the maximum of $u'_{\textit{rms}}$ of the local wall-normal profile based on statistical results. The maximum is taken in the smooth region downstream of the shock. Artificially, the significance of receptivity is defined and measured by the ratio of the leading-edge maximum $u'_{\textit{rms}}$ to the freestream value, $I_\textit{recp} = u'_{\textit{rms}, \textit{max}} (x = x_\textit{LE}) / u'_{\textit{rms},\infty}$. Herein, $x_\textit{LE}$ refers to a leading-edge reference location. Meanwhile, the importance of nonmodal growth is measured by $I_\textit{nonmd} = u'_{\textit{rms}, \textit{max}} (x = 50) / u'_{\textit{rms}, \textit{max}} (x = x_\textit{LE})$, where $x = 50$ is in the laminar-flow region. As a consequence, $I_\textit{recp} \times I_\textit{nonmd} = u'_{\textit{rms}, \textit{max}} (x = 50) / u'_{\textit{rms},\infty}$ characterises the downstream streaky strength under the same freestream disturbance intensity. If the increase of $I_\textit{recp}$ is more sensitive to  growing bluntness than that of $I_\textit{nonmd}$, it is suggested that the receptivity contributes more to the enhanced streaky response.}

\added{The results of the $u'_{\textit{rms}}$ contour and the marked maximum location are displayed in figure \ref{urms_contour}. The boundary layer response becomes visibly important starting from the 45-degree position of the cylindrically blunted nose. Subsequently, the rapid nonmodal growth of the streaky response is recorded by the maximum $u'_{\textit{rms}}$. This nonmodal growth stage is significant, and its sensitivity to the nose-tip radius in the large-bluntness regime should be further examined. Figure \ref{rms_max_develop}(\textit{a}) depicts the streamwise evolution of $u'_{\textit{rms}, \textit{max}}$ for different blunt-plate cases. The enhancement of the streaky response is clear from cases R2, R2.7 to R3. The response strength between cases R1.8 and R2 starts to deviate at a more downstream location (around $x > 70$), and thus the different upstream evolution is not evident for the two cases. Figure \ref{rms_max_develop}(\textit{b}) gives the two defined indicators and their multiplier from cases R2 to R3. The leading-edge reference location is taken as $x_\textit{LE} = 0$. As the nose-tip radius is increased, the receptivity indicator $I_\textit{recp}$ rises from 12.8 to 18.6, whereas the nonmodal-growth indicator $I_\textit{nonmd}$ is slightly reduced from 4.66 to 4.58. In the large-bluntness effect, the final response $I_\textit{recp} \times I_\textit{nonmd}$ at $x = 50$ due to the same freestream disturbance intensity is visibly increased from 59.6 to 85.4, which is contributed by $I_\textit{recp}$. Tests are also performed with different reference locations, e.g., $x_\textit{LE}=-0.1$ and $x_\textit{LE}=5$, and the qualitative trend remains aligned with figure \ref{rms_max_develop}(\textit{b}). The downstream probe is also adjusted from $x=50$ to $x=20$, where the curve slope in figure \ref{rms_max_develop}(\textit{a}) starts to decrease. The leading-edge reference location remains at $x_\textit{LE}=0$. In this case, $I_\textit{nonmd}$ is slightly increased by the nose-tip radius, whereas the degree of increase is still weaker than that of $I_\textit{recp}$. In fact, the different $u'_{\textit{rms}, \textit{max}}$ for varied nose radii has been largely initiated proceeding from the nose. Based on the above criterion, it is deduced that the enhanced streamwise streak with increasing nose bluntness is more attributed to the leading-edge receptivity than the nonmodal growth starting from the leading edge.}

\begin{figure*}
	\centering{ \includegraphics[width=13.5cm]{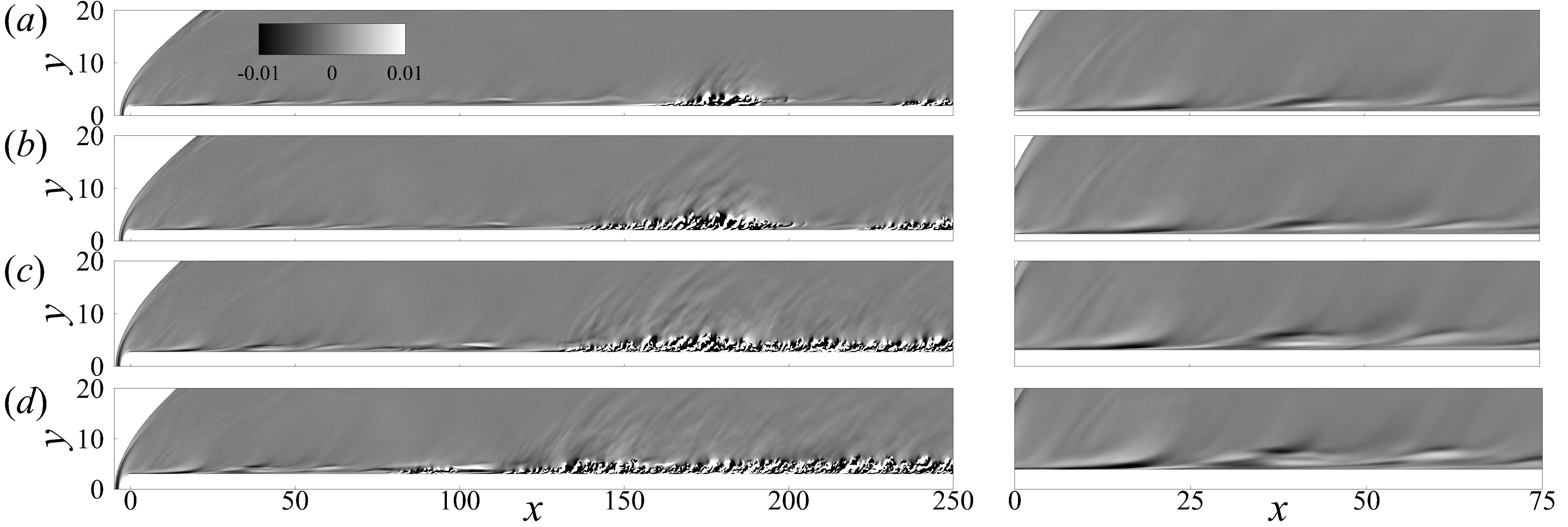}}
	\caption{Instantaneous spanwise velocity $w$ on the symmetry plane\added{ $z=L_z/2$} at $t=806$ for cases \replaced{(\textit{a}) R1.8, (\textit{b}) R2, (\textit{c}) R2.7 and (\textit{d}) R3}{(\textit{a}) R2, (\textit{b}) R2.7 and (\textit{c}) R3}. The characterization of the leading-edge streak is highlighted on the right column. The contour level ranges from $-0.01$ to $0.01$.}
	\label{fig_instant_w}
\end{figure*}

Figure \ref{fig_instant_w} presents the contour of the instantaneous spanwise velocity $w$ on the $x$--$y$ plane, which is able to characterise the three-dimensionality. Different from the cone geometry that was frequently reported, e.g. by \cite*{Hartman2021Nonlinear}, the blunt flat plate displays relatively weak signature of flow structures in the outer entropy layer. The discrepancy in the observed structure may be attributed to the different feature of the entropy layer. The mean streamline outside the boundary layer is oriented outward\added{ from} the wall in the blunt-flat-plate flow, whereas it is toward the wall before entropy swallowing in the blunt-cone flow. Thus, the perturbations outside the boundary layer appear to have a small impact on the boundary layer transition over the blunt flat plate. In the near-wall region of figure \ref{fig_instant_w}, the evolving streamwise streaks downstream of the nose tip ($0<x<75$) display staggered positive and negative regions of $w'$ on the plate.  As aforementioned, the plane stagnation flow is linearly stable to normal-mode perturbations. Therefore, the formation of the strong streamwise streaks in the vicinity of the nose tip is likely to be due to the nonmodal \replaced{growth}{instability}. Farther downstream, the presence of packets containing small-scale structures possibly implies secondary instabilities and intermittent breakdown to turbulence.

\subsection{Spatial and temporal spectra}\label{sec:spectra}

\begin{figure*}
	\centering{ \includegraphics[width=6.6cm]{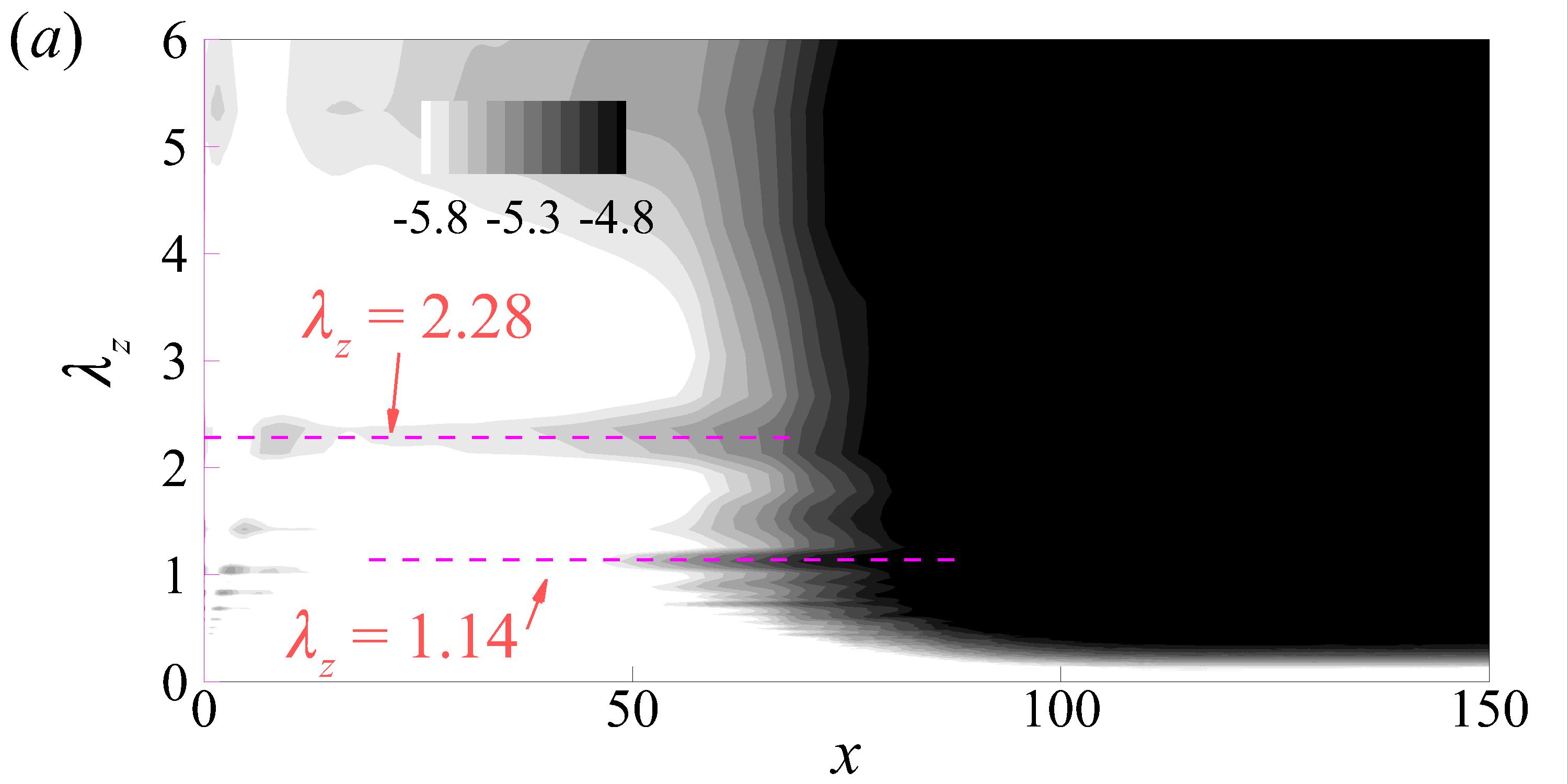}}
	\centering{ \includegraphics[width=6.6cm]{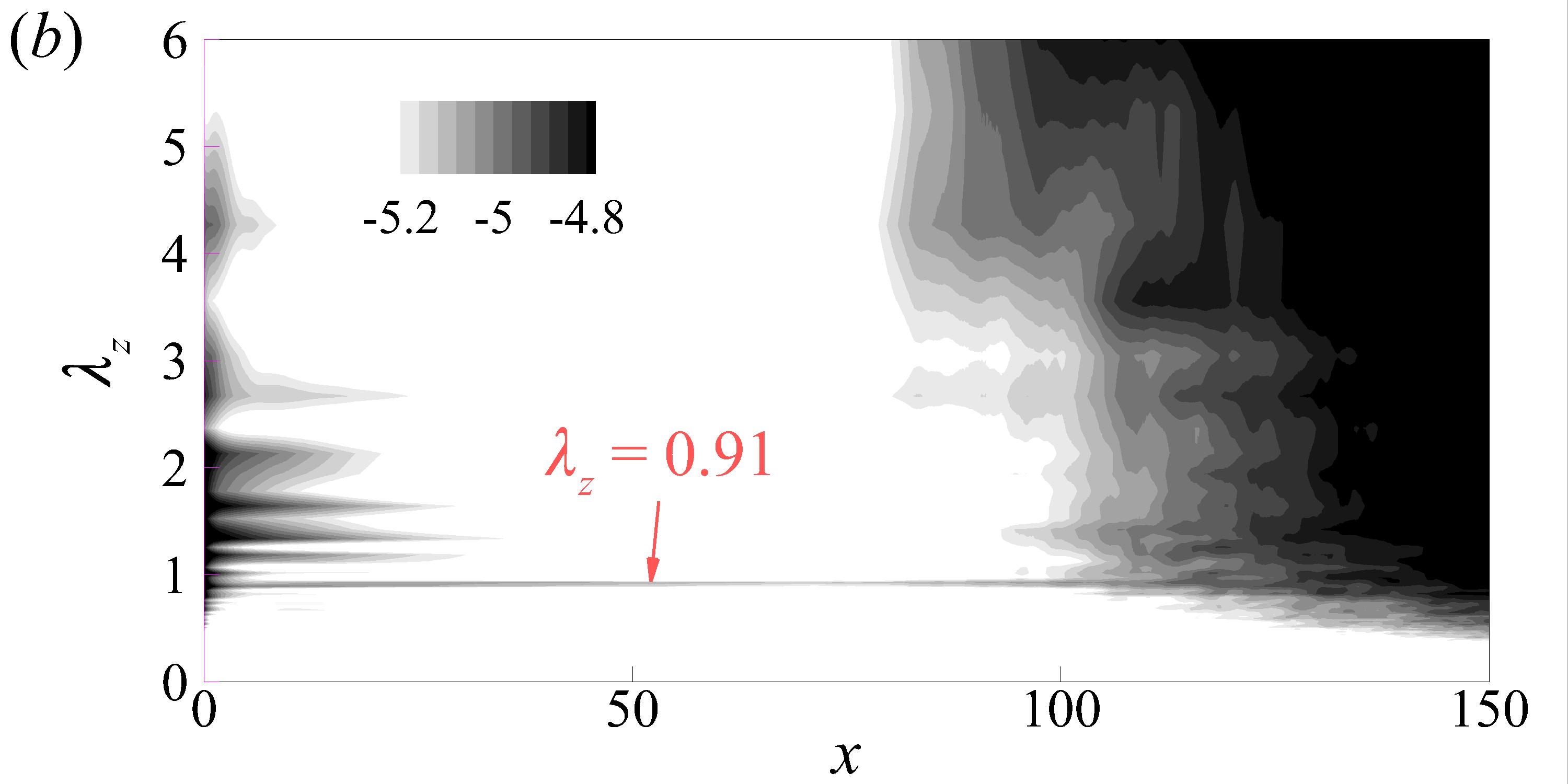}}
	\centering{ \includegraphics[width=6.6cm]{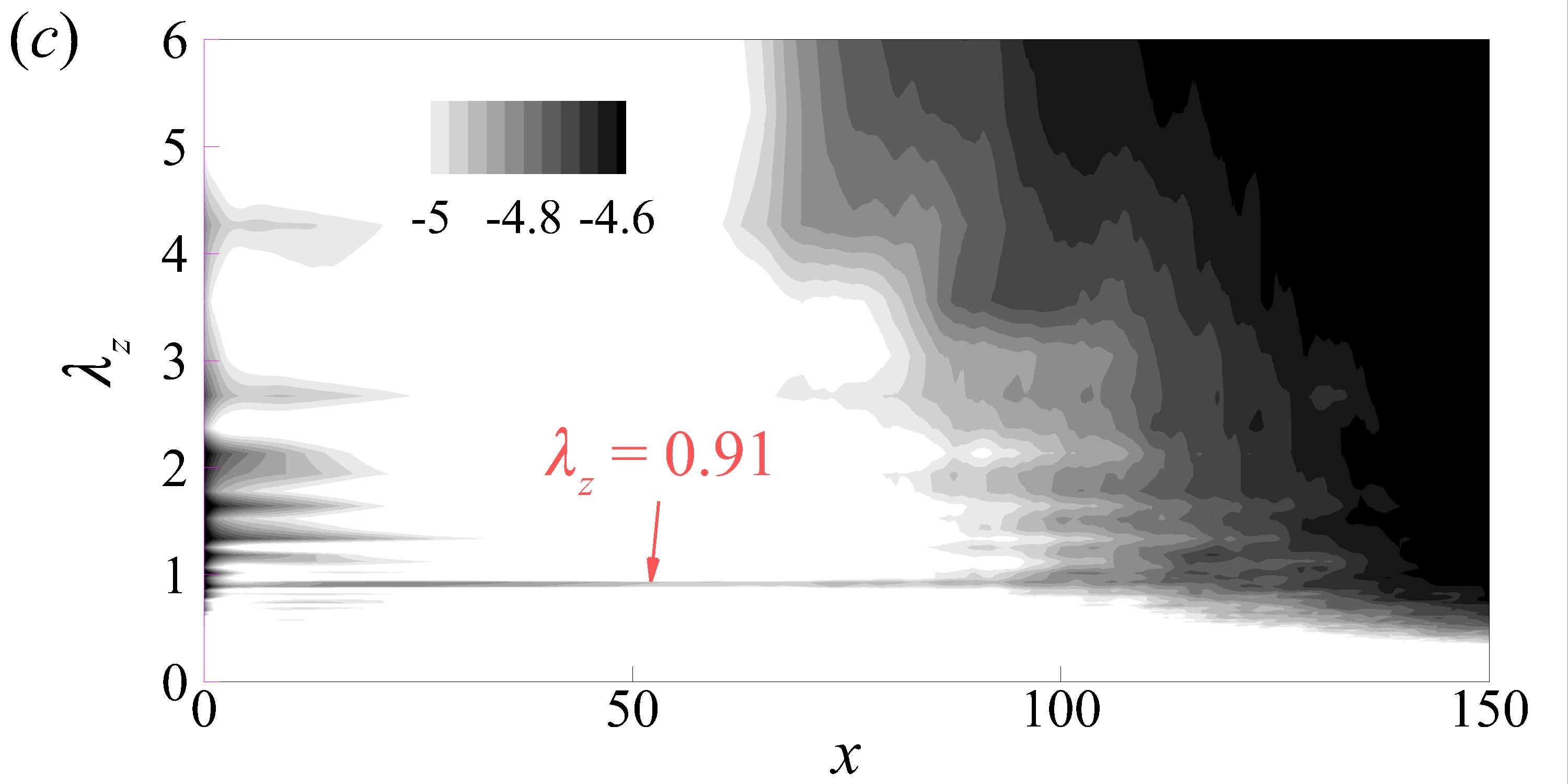}}
	\centering{ \includegraphics[width=6.6cm]{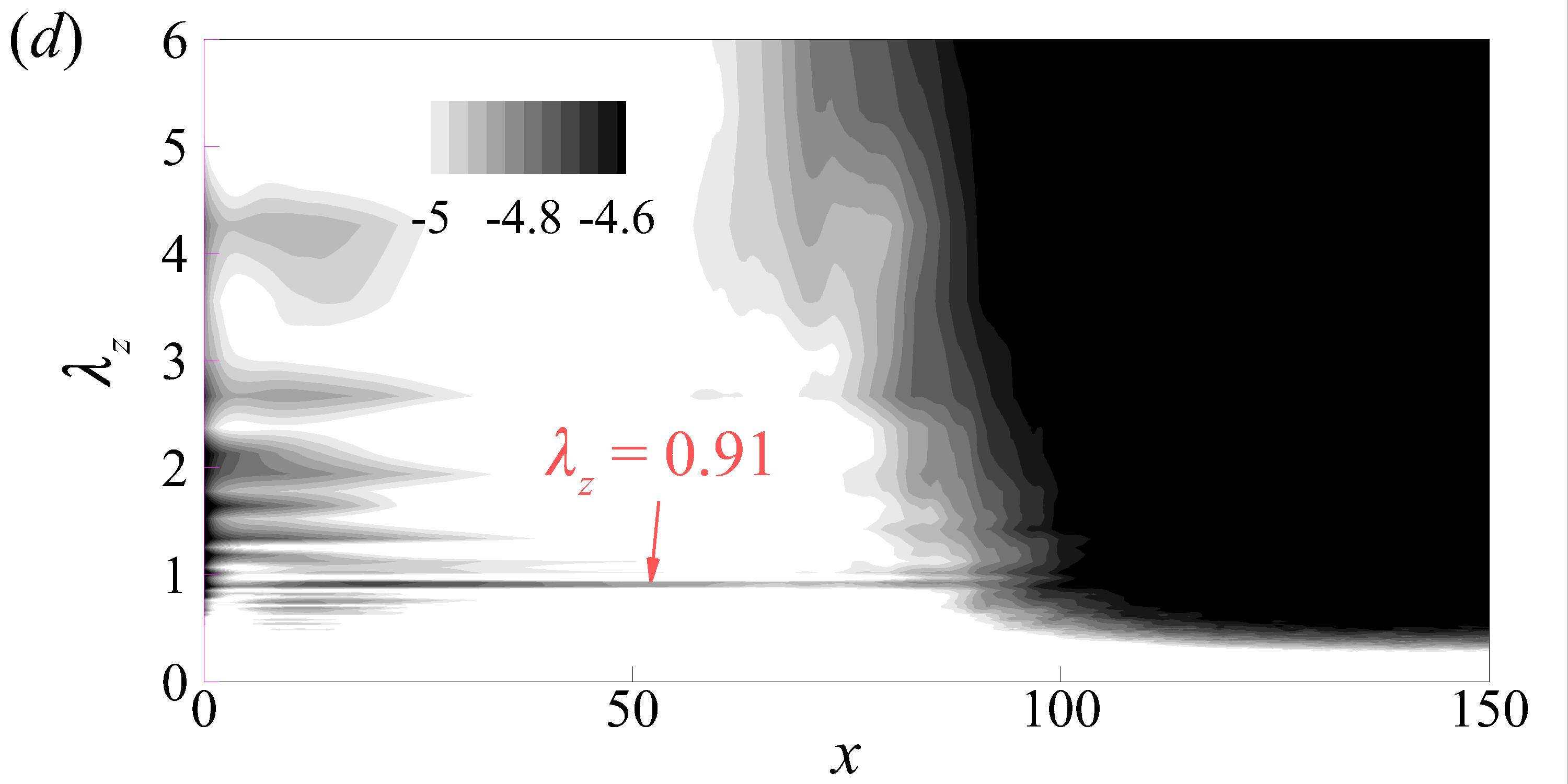}}
	\caption{Common logarithm of modulus of spanwise Fourier transform\added{s} of Stanton number $\log_{10}(|\widetilde{St}|)$ versus the steramwise coordinate $x$ and the spanwise wavelength $\lambda_z$ for cases (\textit{a}) R0, (\textit{b}) R2, (\textit{c}) R2.7 and (\textit{d}) R3. Ensemble average of the result is taken during $t_2-t_1=800$.}
	\label{fig_wavenumber_contour}
\end{figure*}

This subsection aims to quantify the spatial and temporal spectra, which point out the dominant scales in space and time. Figure \ref{fig_wavenumber_contour} gives the spanwise Fourier transform of the Stanton number, which is plotted against the streamwise coordinate and the spanwise wavelength $\lambda_z$. To reduce the randomness effect, ensemble average of the Fourier-transformed $St$ is taken with thousands of snapshots during a long time period. With regard to the sharp-leading-edge case R0, two pronounced spanwise wavelengths $\lambda_z=2.28$ and $\lambda_z=1.14$ appear successively in the pre-transitional region $x<x_t$, where the transition onset is $x_t=72$. The first leading wavelength $\lambda_z=2.28$ agrees very well with the most amplified first-mode wavelength by LST in \S\ \ref{sec:LST}. The second leading wavelength $\lambda_z=1.14$ emerges later, which has a weaker priority over other wavelengths compared to $\lambda_z=2.28$. It is conceivable that $\lambda_z=1.14$ arises from the nonlinear interaction of the dominant oblique first mode. Specifically, the mode--mode interaction in the Fourier space can be expressed by
\begin{equation}\label{eq_OB}
(f_1,\beta_1)-(f_1,-\beta_1)\rightarrow (0,2\beta_1),
\end{equation}
where $f_1$ and $\beta_1$ denote the first-mode frequency and spanwise wavenumber, respectively. Mode $(0,2\beta_1)$ is the generated streamwise \replaced{vortex}{vorticity} mode.\added{ One may substitute an arbitrary frequency from the broadband spectrum into (\ref{eq_OB}) to obtain the vortex mode. The most unstable first-mode frequency is likely to excite the strongest response  of the vortex mode.} This stationary mode possesses a spanwise wavenumber twice the first-mode one, and thus a spanwise spacing half of the first-mode spanwise wavelength. Therefore, the three-dimensionality in the pre-transitional region of case R0 is partly induced by the oblique first mode and the resulting streamwise \replaced{vortex}{vorticity} (or stationary streak) mode. For blunt-plate cases\added{ including the unshown case R1.8}, it is interesting to observe a preferential wavelength $\lambda_z=0.91$ that does not vary with the nose-tip radius. The preferential wavelength is formed in and  through out the pre-transitional regions. The value of this wavelength is visibly consistent with the streak visualisation in figure \ref{fig_averaged_st}. In this early stage, \replaced{incipient spots may be present in the streaky laminar flow}{turbulent spots have not yet appeared, or they are insignificant}. Note that the intensity of the added freestream disturbance is uniform with respect to the spanwise wavenumber/wavelength. It is thus inferred that the preferential wavelength is selected by the leading-edge receptivity process. Furthermore, different blunt-plate cases share similar spanwise spatial spectra. The main difference appears to be the transition location that is manifested by the appearance of the filled wavenumber spectra in figure \ref{fig_wavenumber_contour}. In that region with full spectra, the energy is concentrated on the large scales of $\lambda_z$ rather than the dissipative small scales, which is consistent with the energy cascade property of turbulent flows.

\begin{figure*}
	\centering{ \includegraphics[width=6cm]{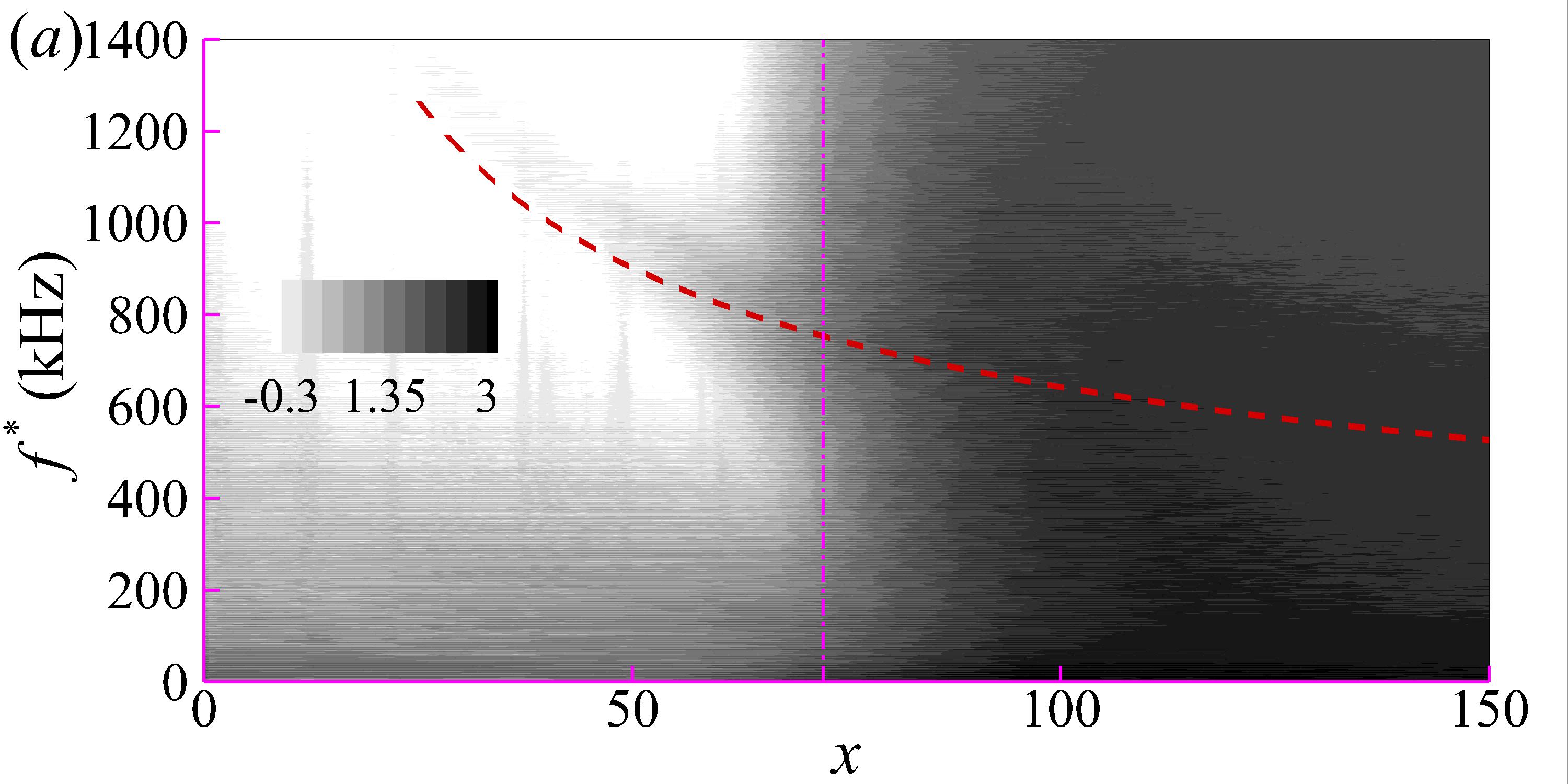}}
	\centering{ \includegraphics[width=6cm]{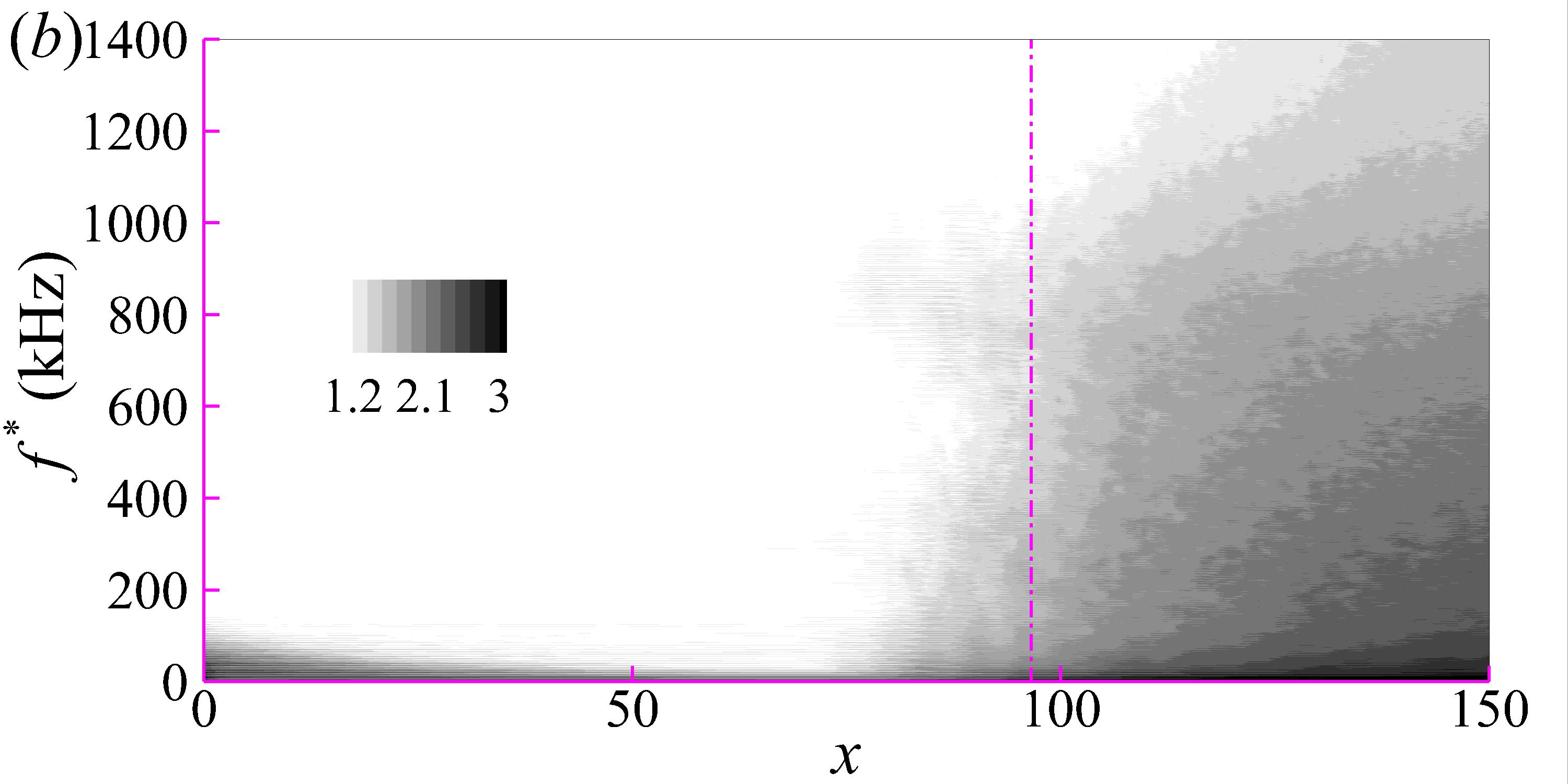}}
	\centering{ \includegraphics[width=6cm]{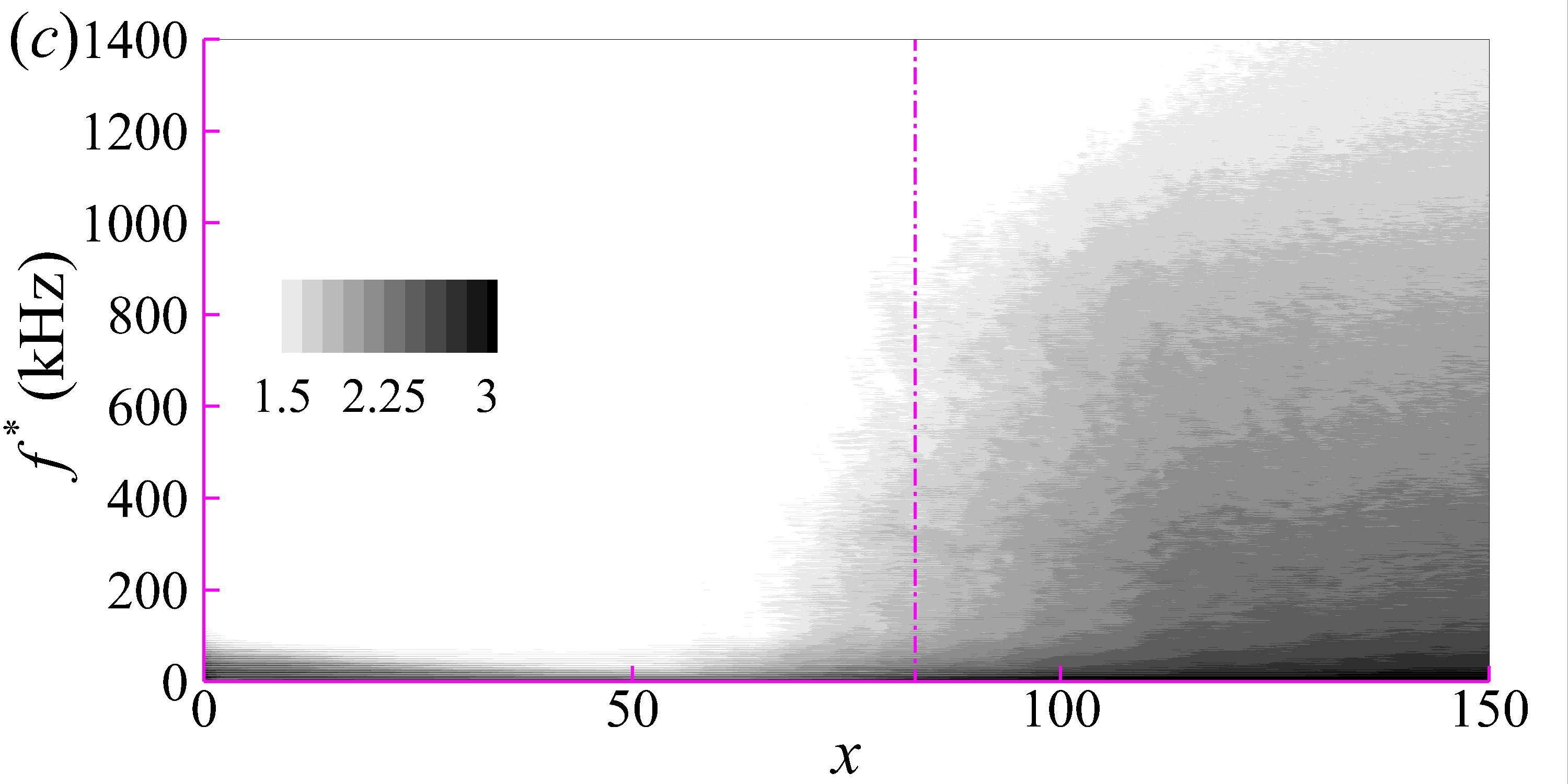}}
	\centering{ \includegraphics[width=6cm]{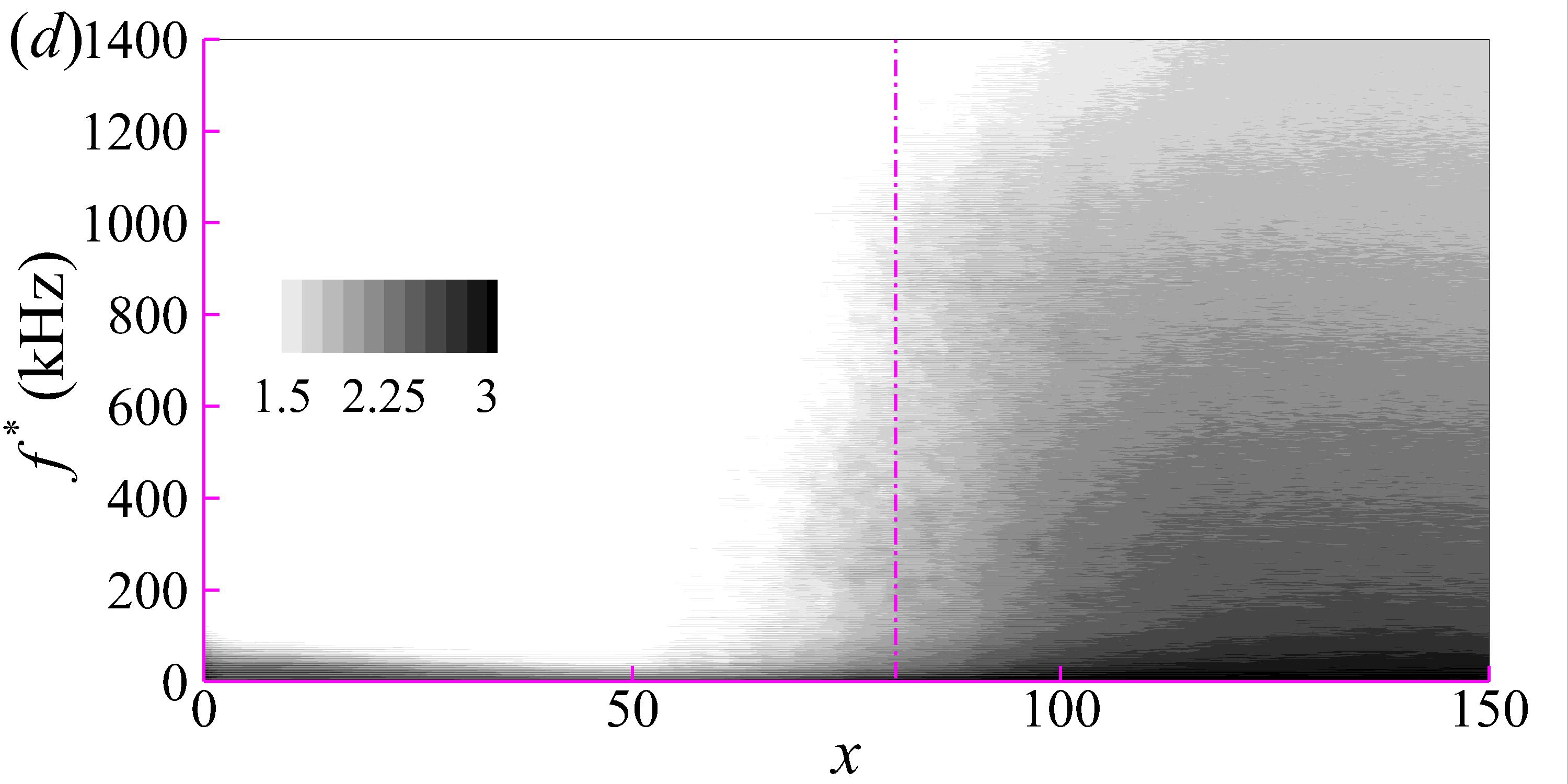}}
	\caption{Common logarithm of spanwise-averaged modulus of temporal Fourier transform\added{s} of Stanton number for cases (\textit{a}) R0, (\textit{b}) R2, (\textit{c}) R2.7 and (\textit{d}) R3. \added{Vertical dashed-dotted lines represent the transition onset locations in table \ref{table2}. }Dashed line in (\textit{a}) represents the frequency of the most unstable second mode calculated by LST. The Nyquist frequency is about 1540 kHz.}
	\label{fig_pFFT_2D}
\end{figure*}

In addition to the spatial spectra, temporal Fourier transform is also conducted and shown in figure \ref{fig_pFFT_2D}. For case R0, the most amplified second-mode frequency is calculated by LST via the Newton-Raphson method, which maximises the growth rate $\sigma$. The resulting frequency is represented by the dashed line in figure \ref{fig_pFFT_2D}(\textit{a}). A good agreement in the most pronounced second-mode frequency is reached between CFD and LST, and the difference is less than 100 kHz.\added{ Here, the most pronounced second-mode frequency corresponds to the locally most unstable one rather than the largest $x$-integrated growth factor.} For the remaining blunt-plate cases, similar spectral characteristic is observed again for the frequency spectrum. In the pre-transitional region, low-frequency components below 20 kHz are dominant for cases R2, R2.7 and R3. Farther downstream, higher-frequency components grow rapidly and fill the frequency spectrum. The growth of the high-frequency band is likely to be associated with secondary instabilities. No preferential frequency of hundreds of kilohertz is found\added{ using different contour levels}. The spectral characteristics in figures \ref{fig_pFFT_2D}(\textit{a}--\textit{d}) also resemble the experimental results of \cite{Marineau2014Mach}, where the signal was measured by PCB sensors over sharp and blunt cones.\added{ Note that the transition onset locations, marked by the vertical lines, appear slightly later than the spectral filling behaviour in figure \ref{fig_pFFT_2D}. This issue probably arises from the determination approach of the transition onset. The calculated onset location is more downstream than the location of the minimum Stanton number in figure \ref{appC_onset}. The latter location indicates the growing significant role of the high-frequency signal due to secondary instability, which contributes to spectral filling.}

\begin{figure*}
	\centering{ \includegraphics[width=6.6cm]{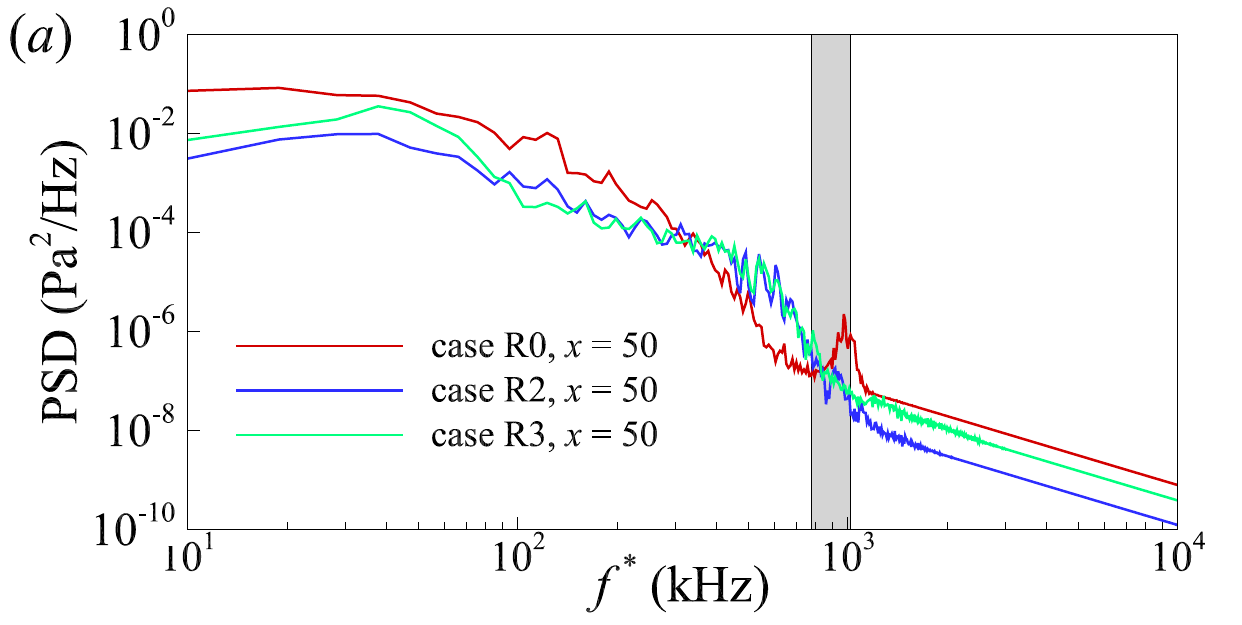}}
	\centering{ \includegraphics[width=6.6cm]{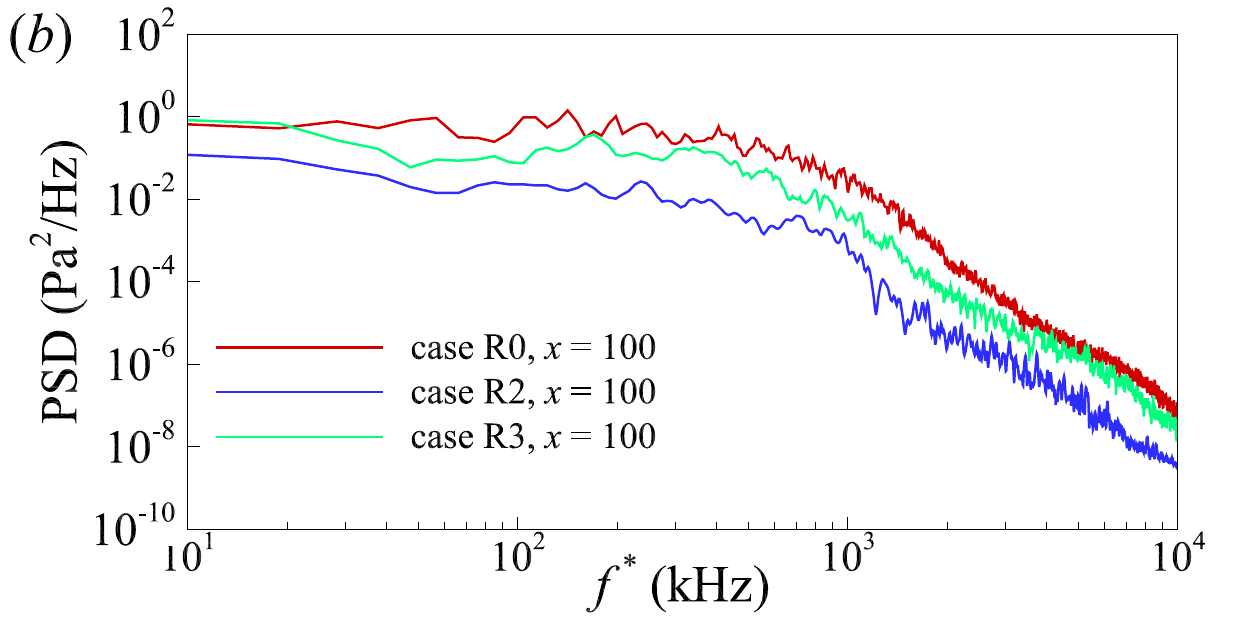}}
	\centering{ \includegraphics[width=6.6cm]{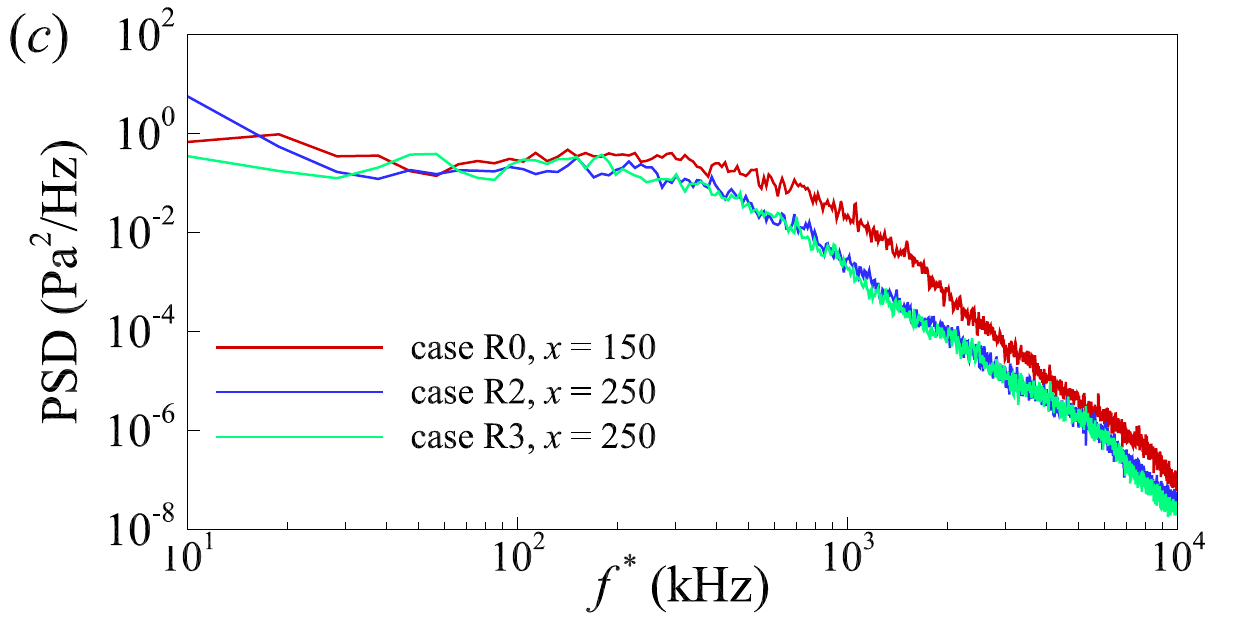}}
	\centering{ \includegraphics[width=6.6cm]{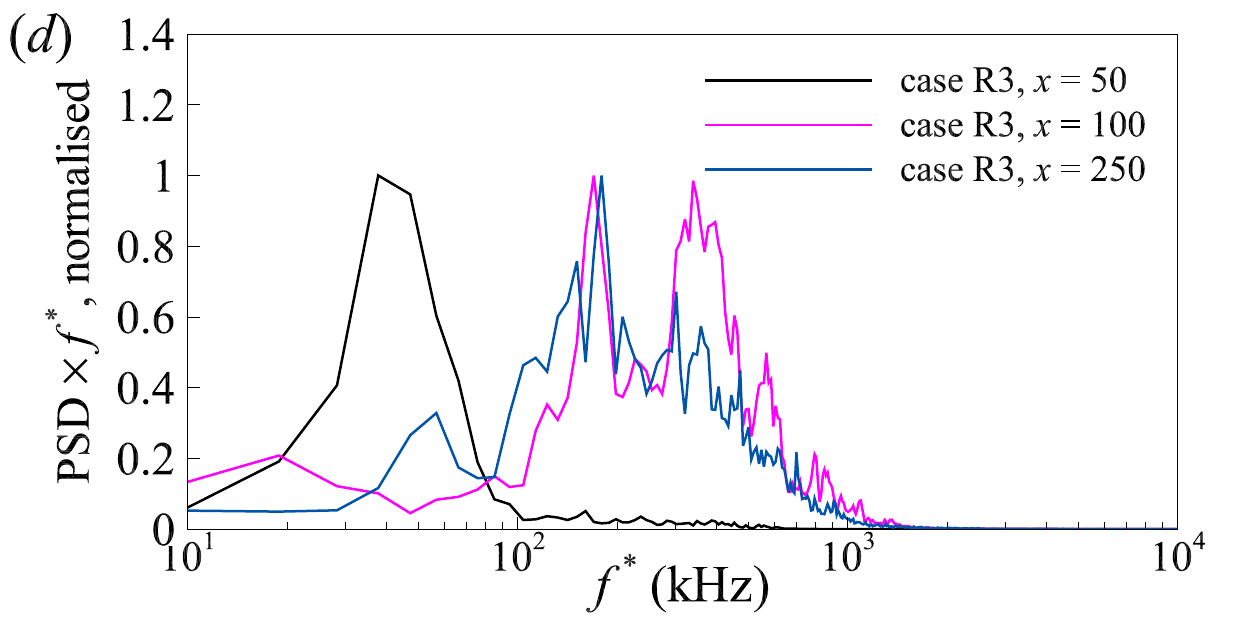}}
	
	\caption{Temporal PSD of the wall pressure fluctuation probed along $z=L_{z}/2$ in (\textit{a}) pre-transitional, (\textit{b}) transitional and (\textit{c}) nearly turbulent regions. \added{(\textit{d}) Premultiplied PSD normalised by each maximum for case R3.} The grey area in (\textit{a}) represents the frequency range of the unstable second mode given by LST.}
	\label{fig_PSD_1D}
\end{figure*}

The time history of wall pressure at several probes is further used to calculate power spectral density (PSD) for cases R0, R2 and R3. To this end, Welch's method is applied, and the time sequence is segmented with 50 \% overlapping. A Hamming windowing function is used to reduce spectral leakage. The PSD results in the pre-transitional, transitional and nearly turbulent regions are shown in figure\added{s} \ref{fig_PSD_1D}\added{(\textit{a}--\textit{c})}. The PSD of the Stanton number is also calculated, and the performance is similar to that of the wall pressure fluctuation. For case R0 in the laminar region, figure \ref{fig_PSD_1D}(\textit{a}) illustrates that the PSD peak at around 1000 kHz falls within the unstable second-mode frequency range. For other blunt-plate cases, no Mack-mode-like high-frequency component emerges at $x=50$. In the transitional region at $x=100$, no prominent frequency is identified for all the cases. The PSD of the large-bluntness case R3 exceeds that of case R2 throughout the frequency range at $x=100$. This noteworthy observation indicates that the energy advantage of case R3 over R2 is overall and broadband, rather than characterised by a particular frequency. The disappearance of a single Mack second mode for case R0 is possibly related to the fact that $x=100$ is already in the late nonlinear stage, as shown by figure \ref{fig_St_curves}(\textit{a}). At the nearly turbulent position $x=250$, the PSD curves of cases R2 and R3 become close to each other, which indicates proximity to a fully established turbulent state.  \added{Premultiplied spectra PSD $\times\ f^{\ast}$ normalised by each maximum are also shown for case R3 in figure \ref{fig_PSD_1D}(\textit{d}), which highlight the energy-containing scale. In the laminar-flow region ($x=50$), one peak is identified at 37 kHz for case R3 and also for cases R2 and R2.7, which implies the significance of low-frequency components. During the transition ($x=100$), double peaks emerge at hundreds of kilohertz, which suggests the growth of high-frequency components. In the fully developed turbulent region ($x=250$), one pronounced  peak is observed at about 200 kHz.}

To conclude, transition to turbulence in the sharp-leading-edge flat-plate flow is probably induced by Mack second mode and oblique first mode simultaneously. By contrast, the blunt-plate cases are dominated by low-frequency streamwise streaks with the same spanwise spacing and the subsequent high-frequency secondary instabilities. The characteristics of spatial and temporal spectra are found to be similar among cases R2, R2.7 and R3. The main quantitative difference is that the streaky response is increased by growing nose bluntness, as illustrated by \replaced{figures \ref{fig_streak_rms} and \ref{rms_max_develop}}{figure \ref{fig_streak_rms}}. It is thus deduced that the transition reversal is caused by enhanced receptivity to freestream disturbances without changing the transition mechanism essentially.

\subsection{Spectral proper orthogonal decomposition}
SPOD analysis is further conducted, which is a data-driven method to extract coherent structures or modes from flow fields. The data can be collected from experimental or high-fidelity computational studies. The conventional proper orthogonal decomposition (POD) seeks a set of deterministic modes, which are only spatially coherent. The SPOD modes are orthogonal in the context of a space--time inner product operation. As a result, the improved SPOD method captures the dominant modes that are coherent in space and time  \citep*{towne2018spectral}. Mathematically,
SPOD modes are the eigenvectors of a cross-spectral density (CSD)
tensor at each frequency. In practice, ${N_\textit{t}}$ snapshots of the fluctuating flow field are constructed into a snapshot matrix
\begin{equation}
	{\mathsfbi Q}=\left[{\bm q}_1, {\bm q}_2, ..., {\bm q}_{N_\textit{t}}\right].
\end{equation}
To estimate the CSD, the data are segmented into ${N_{{\textit{blk}}}}$
overlapping blocks with ${N_\textit{FFT}}$ snapshots in each, as
\begin{equation}
{\mathsfbi Q}^{(k)}=\left[{\bm q}_1^{(k)}, {\bm q}_2^{(k)}, ..., {\bm q}_{N_\textit{FFT}}^{(k)}\right],
\end{equation}
where the \textit{j}-th column in the \textit{k}-th block is
\begin{equation}
	{\bm q}_j^{(k)}={\bm q}_{j+(k-1)({N_\textit{FFT}}-{N_\textit{ovlp}})+1}.
\end{equation}
Here, $N_\textit{ovlp}$ is the number of overlapping snapshots for each block. The data segmentation serves to increase the number
of ensemble members. A Hamming windowing function is applied to each block. Subsequently, discrete Fourier transform is applied to each windowed block, and the CSD matrix is constructed for each discrete frequency. SPOD modes are then computed as the eigenvectors of the CSD matrix for each frequency, which are organised in a descending order of its energy. In this paper, Chu's energy norm is chosen to evaluate the SPOD mode energy. The SPOD analysis is performed based on the open-source code provided by \cite{schmidt2020guide}.

To compromise the need to converge the SPOD modes and the limitation of computational resources, we follow the parameteric setup by \cite{lin2024modal}, who carried out SPOD analysis of a transitional boundary layer dataset. To reach statistical convergence of the spectral density, the number of the flow realisation ${N_{{\textit{blk}}}}\ge20$ was considered to be enough. In this paper, we take ${N_{{\textit{blk}}}}=20$, and choose a reasonable resolution of ${N_\textit{FFT}}=512$  with 75\% overlapping. Based on the relationship
\begin{equation}
{N_{{\textit{blk}}}} = \left\lfloor {\frac{{{N_\textit{t}} - {N_\textit{ovlp}}}}{{{N_\textit{FFT}} - {N_\textit{ovlp}}}}} \right\rfloor, 
\end{equation}
where $\lfloor \cdot \rfloor$ represents the floor operator, the total snapshot number should be at least $N_\textit{t}=2944$. The snapshots are collected after the flow field reaches statistical stationary. For a low frequency of interest, say 20 kHz, the corresponding period is $t_1^{\ast}=5\times10^{-5}\ \rm{s}$. To obtain a good temporal resolution, 8 periods are fitted into one block, which yields  ${N_\textit{FFT}}/8=64$ snapshots for each period of interest. As a result, the physical snapshot sampling time $\Delta t^{\ast}_\textit{snap}=t_1^{\ast}/64$ and sampling frequency $f^{\ast}_s=1/\Delta t^{\ast}_\textit{snap}=1280\ \rm{kHz}$. The Nyquist frequency is thus 640 kHz, and the minimum resolved frequency is $f_\textit{min}^{\ast}=1/(N_\textit{FFT}\Delta t^{\ast}_\textit{snap})=2.5\ \rm{kHz}$.  

\begin{figure*}
	\centering{\includegraphics[width=13.5cm]{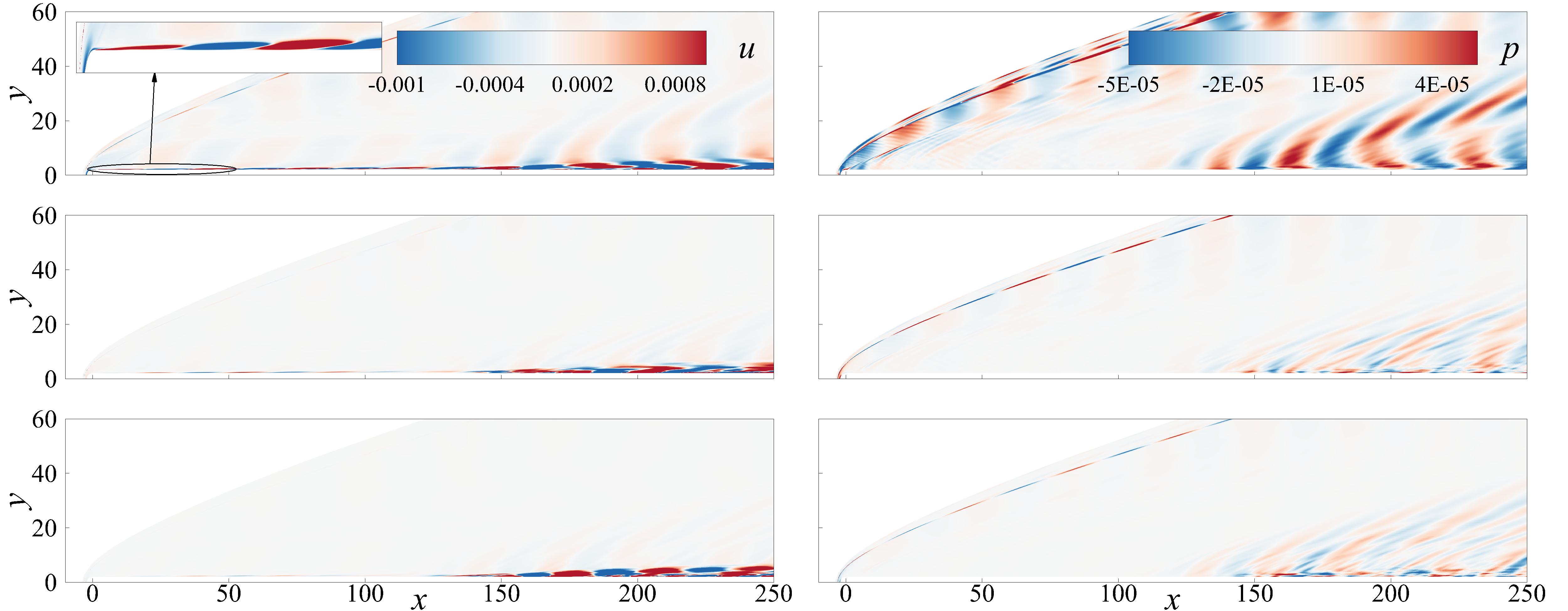}}
	\caption{Streamwise velocity (left column) and \replaced{pressure}{density} (right column) for the first, second and third leading SPOD $x$--$y$ modes from top to bottom at $f^\ast=20\ \rm{kHz}$\added{ for case R2C}.}
	\label{XYSPOD_freq20}
\end{figure*}

\begin{figure*}
	\centering{\includegraphics[width=13.5cm]{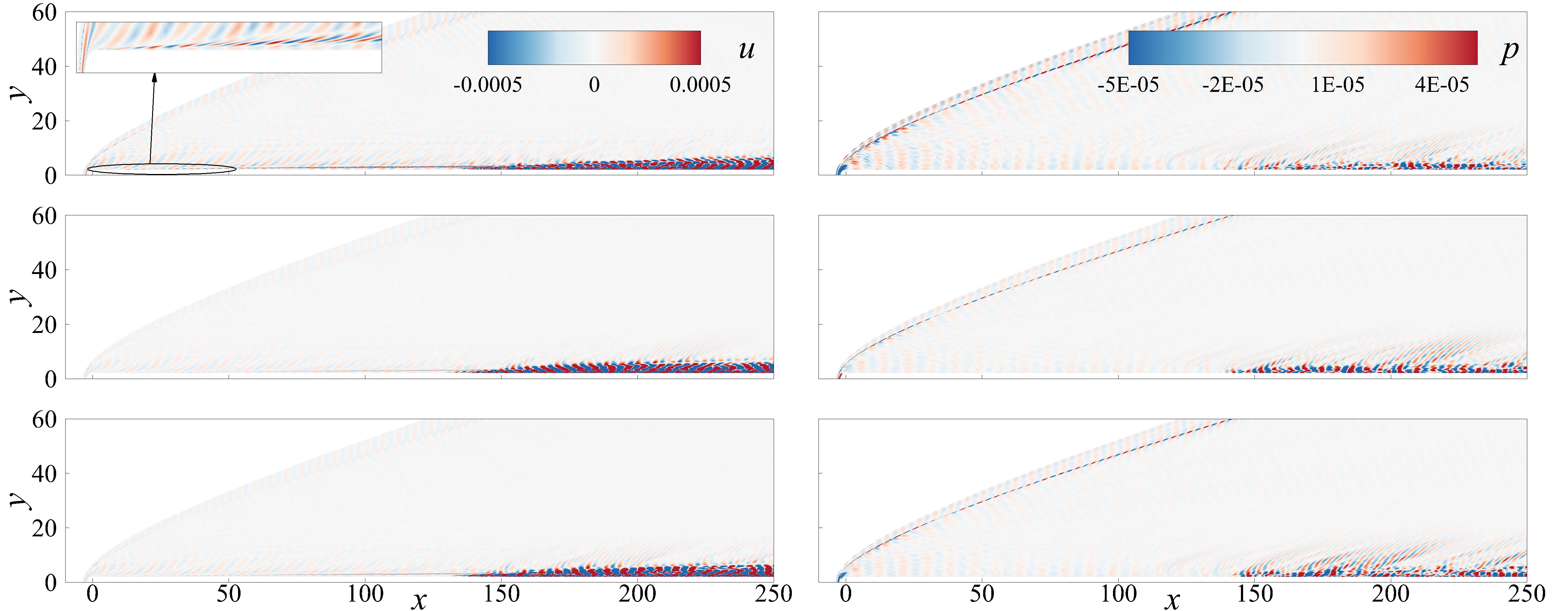}}
	\caption{Streamwise velocity (left column) and \replaced{pressure}{density} (right column) for the first, second and third leading SPOD $x$--$y$ modes from top to bottom at $f^\ast=120\ \rm{kHz}$\added{ for case R2C}.}
	\label{XYSPOD_freq120}
\end{figure*}

\begin{figure*}
	\centering{\includegraphics[width=13.5cm]{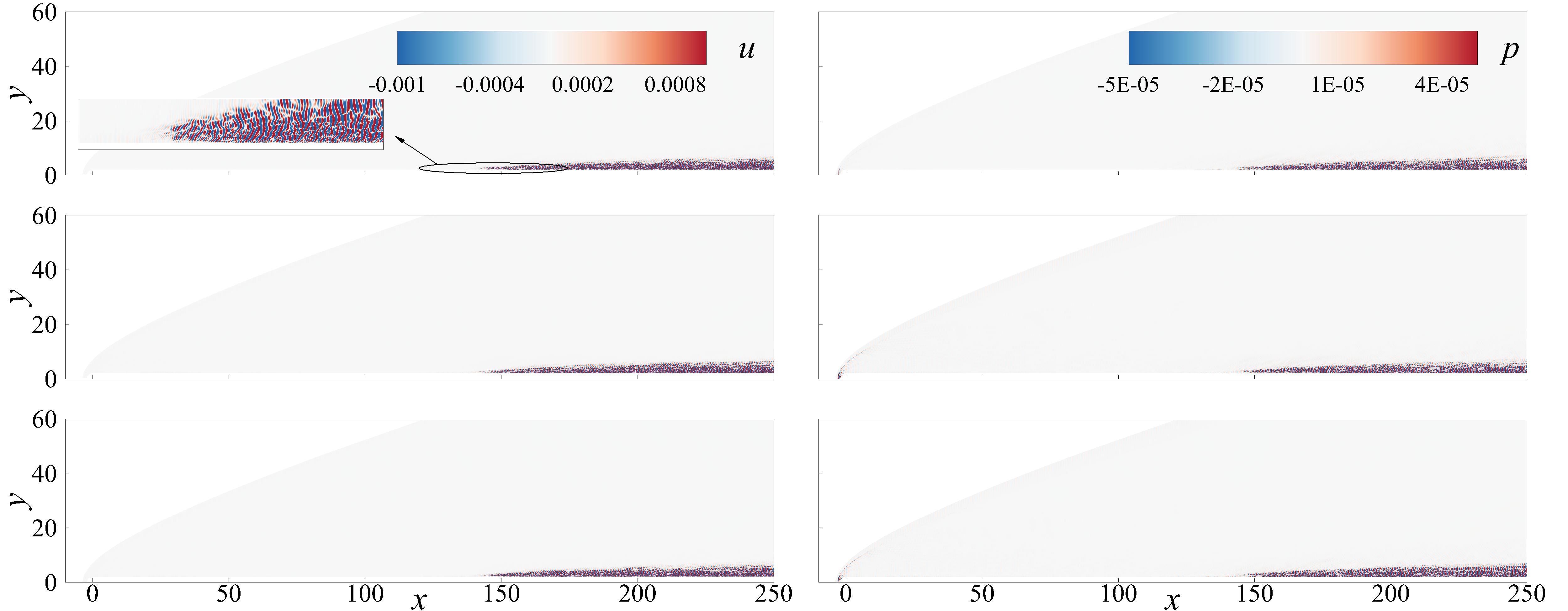}}
	\caption{Streamwise velocity (left column) and \replaced{pressure}{density} (right column) for the first, second and third leading SPOD $x$--$y$ modes from top to bottom at $f^\ast=550\ \rm{kHz}$\added{ for case R2C}.}
	\label{XYSPOD_freq550}
\end{figure*}

\begin{figure*}
	\centering{ \includegraphics[width=7cm]{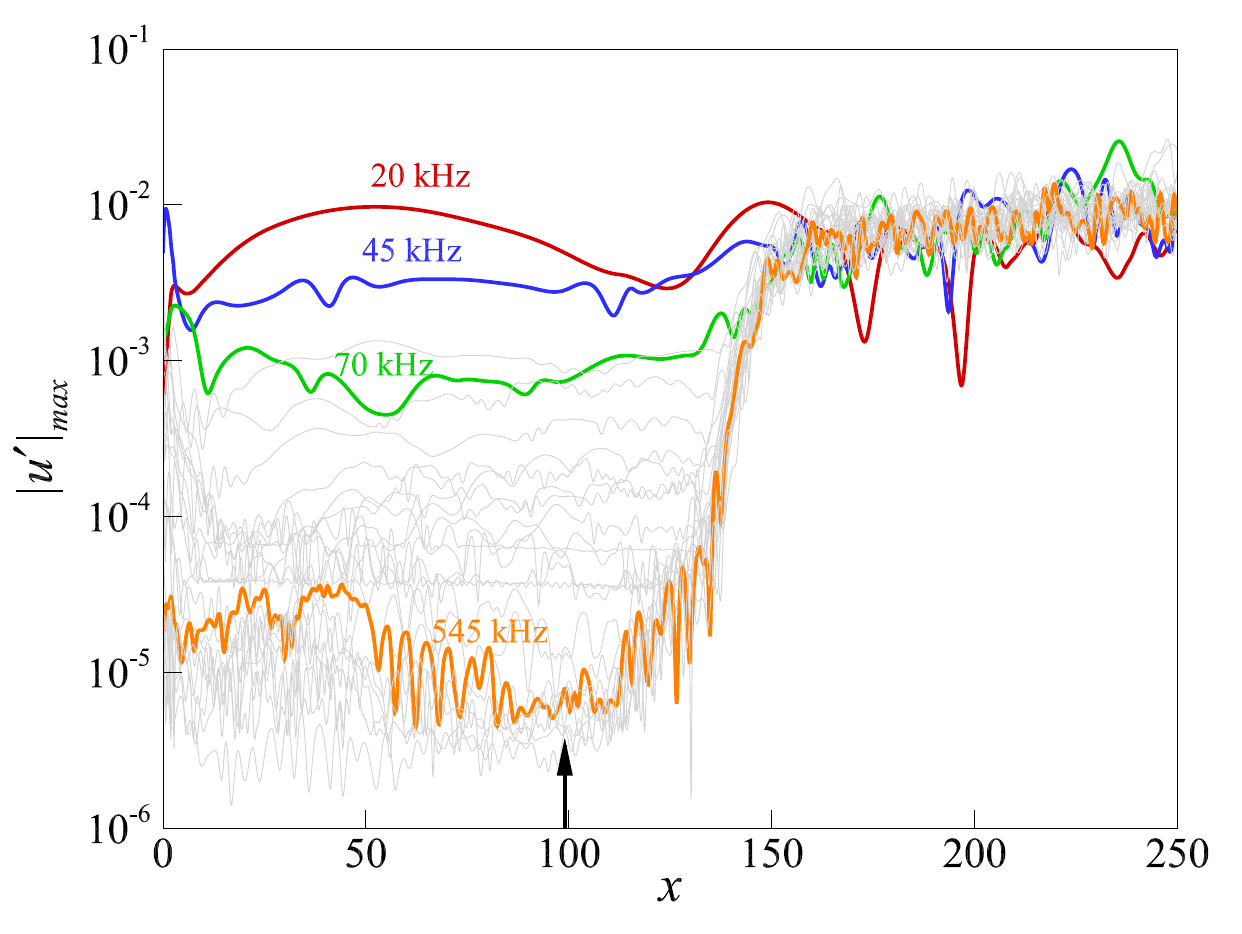}}
	\caption{Development of the local maximum of the streamwise velocity fluctuation $|u'_\textit{max}|$ for the first leading SPOD $x$--$y$ mode\added{ for case R2C}. Different frequencies are displayed with an interval of 25 kHz starting from 20 kHz, among which four frequencies are highlighted (20 kHz, 45 kHz, 70 kHz, 545 kHz). Envelope of the signal is taken from the original SPOD $|u'_\textit{max}|$ to highlight the overall evolution. The arrow represents the transition onset location.}
	\label{umax_envelope}
\end{figure*}

Since the general flow characteristics are similar for all the blunt-plate cases, it is able to provide sufficient physical insights by examining SPOD modes of case R2C only. Figures \ref{XYSPOD_freq20}--\ref{XYSPOD_freq550} show the three leading SPOD modes on the $x$--$y$ plane for three chosen frequencies, namely low frequency 20 kHz, medium frequency 120 kHz and high frequency 550 kHz, respectively. In general, the three frequencies display distinct disturbance patterns. The low-frequency SPOD mode displays signature of streamwise streaks immediately downstream of the nose region. The streamwise length of the streak is around $\lambda_x=30$ near the leading edge, which is evidently larger than the boundary layer thickness. For the medium frequency, the leading-edge SPOD mode structure looks more complicated. The disturbances are developed in both the entropy layer and the boundary layer. The pattern in the entropy layer somewhat resembles the wisp-like structure in the experimental measurement of \cite{Kennedy2019Visualizations}. In general, the SPOD mode energy is still more concentrated in the boundary layer for the medium frequency. In terms of the low and medium frequencies, there is visible signature in the relatively upstream region $x<100$ for the first leading SPOD mode. This observation possibly suggests that the receptivity to the upstream disturbance is not ignorable for the low- and medium-frequency components. In comparison, the high-frequency SPOD mode disappears in the upstream region and emerges suddenly in the range $x>140$. Therefore, the high-frequency component is more likely to originate from secondary instabilities in the developed distorted base flow rather than from the freestream receptivity.

To quantify the instability evolution, figure \ref{umax_envelope} displays the streamwise dependence of the maximal streamwise velocity fluctuation $|u'_\textit{max}|$ for the first leading SPOD mode. For relatively low-frequency modes, the order of magnitude of the velocity fluctuation is already in the range $10^{-3}\sim10^{-2}$, which is maintained in the transitional region (downstream of the arrow). For the highlighted high-frequency (545 kHz) mode, the order of magnitude of $|u'_\textit{max}|$ undergoes an escalation from around $10^{-6}$ to $10^{-3}$ during the transition. In other words, high-frequency SPOD modes are rather weak in the pre-transitional region and grow rapidly in the transitional region. Physical factors other than the secondary instabilities do not seem convincing to account for such a sudden growth. The described instability evolution further supports that the high-frequency SPOD mode emerges from secondary-instability-like mechanisms.

\begin{figure*}
	\centering{ \includegraphics[width=7cm]{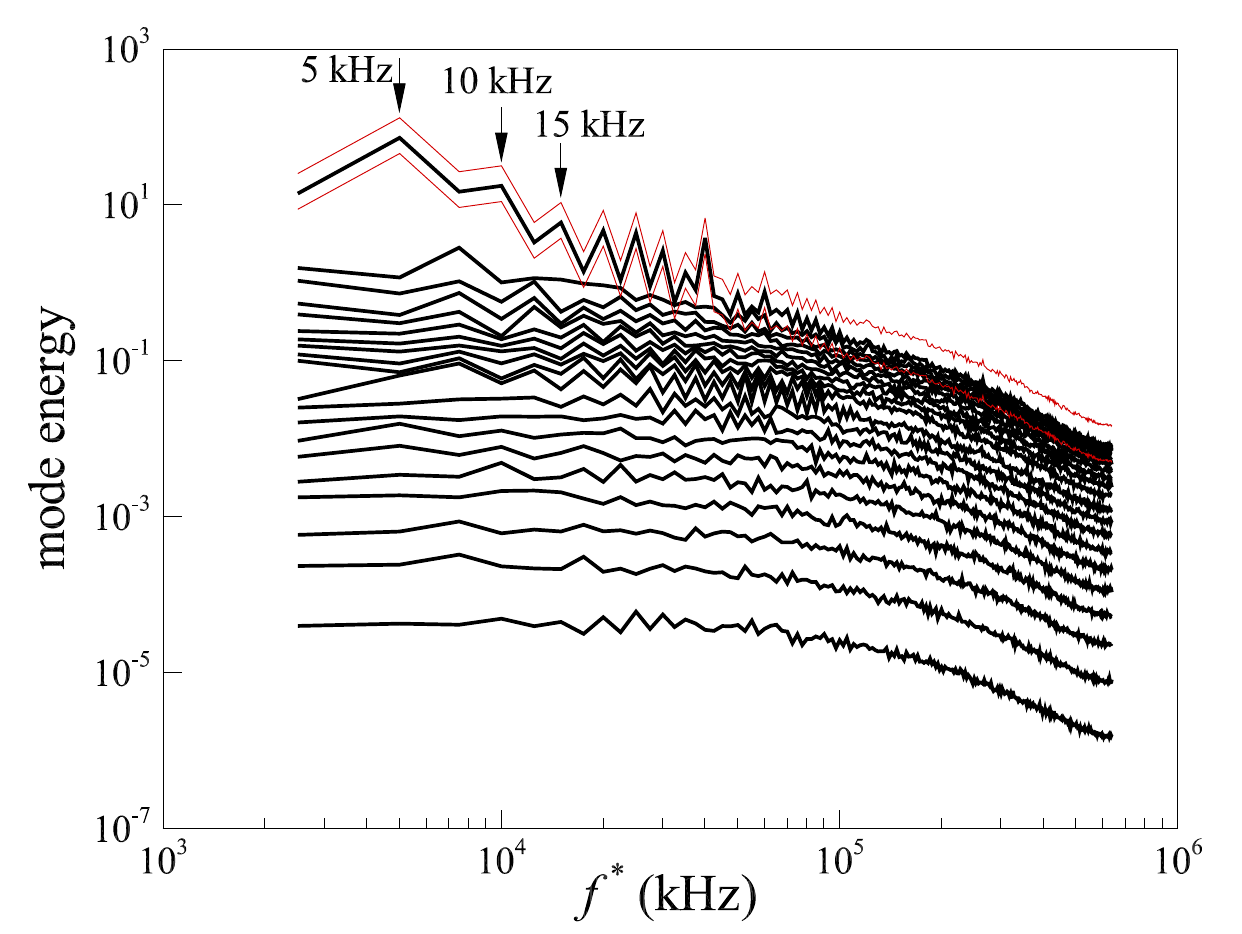}}
	\caption{Energy spectra for the 20 leading SPOD $x$--$y$ modes\added{ for case R2C}. Red lines represent the upper and lower bounds with a 99 \% confidence level for the first leading mode.}
	\label{fig_spod_energy}
\end{figure*}

\begin{figure*}
	\centering{ \includegraphics[width=12cm]{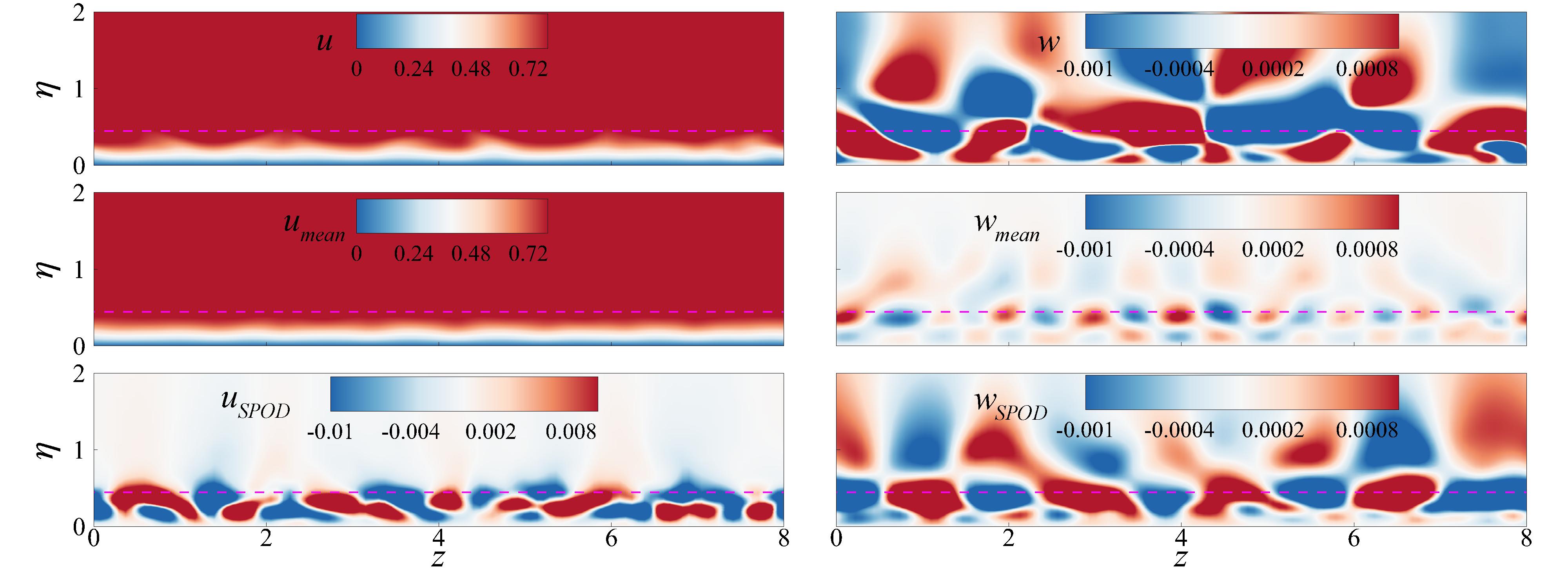}}
	\caption{Instantaneous (top), time-averaged (middle) and SPOD-modal (bottom) quantities of the first leading SPOD $z$--$y$ mode with $f^\ast=20\ \rm{kHz}$\added{ for case R2C}, including streamwise velocity (left column) and spanwise velocity (right column). The slice is extracted in the pre-transitional region $x=50$. Dashed line represents the location of the 2-D laminar boundary layer thickness.}
	\label{fig_SPOD_YZ}
\end{figure*}

Figure \ref{fig_spod_energy} shows the frequency spectra for the 20 leading SPOD modes.\added{ The 99 \% confidence level for the first leading mode is estimated based on the assumption that SPOD eigenvalues follow a chi-squared distribution (see details in \citealt{schmidt2020guide}).}  For the first leading SPOD mode, the mode energy peaks at discrete frequencies of a series of harmonics, which resembles the SPOD energy feature in the controlled transition by \cite{lin2024modal}. These peak frequencies are integer times the frequency interval $\Delta f^{\ast}=f^{\ast}_{1}/2=5\ \rm{kHz}$. Here, $f^{\ast}_{1}$ is the fundamental (minimal) frequency that is imposed in the 3-D broadband disturbance model. Since 10 kHz, 15 kHz, 20 kHz, etc. are added in the model, the lowest frequency peak 5 kHz in figure \ref{fig_spod_energy} should originate from the nonlinear difference interaction between 15 kHz and 10 kHz. In terms of the second and higher SPOD modes, the signature of the tuned frequencies becomes weaker, and the impact of the initial harmonic behaviour in the freestream disturbance model fades away.

In addition to the symmetry plane, the snapshots of the transverse plane were also stored for SPOD analysis. Figure \ref{fig_SPOD_YZ} provides the first leading SPOD mode on the $z$--$y$ plane in the streaky laminar flow region. Instantaneous, time-averaged and SPOD-modal flow fields are listed from top to bottom. The contours of the streamwise and spanwise velocities are shown on the left and right columns, respectively. Based on the time-averaged results, the 3-D laminar flow is visibly distorted and displays spanwise periodic distributions in $u_\textit{mean}$ and $w_\textit{mean}$. The spanwise spacing manifested in the contour of $w_\textit{mean}$ is close to the preferential spanwise wavelength $\lambda_z=0.91$ in \S\ \ref{sec:spectra}. The SPOD mode in figure \ref{fig_SPOD_YZ} exhibits staggered high- and low-speed streaks mainly inside the boundary layer. Unlike the aligned pattern of $w_\textit{mean}$, the SPOD modes look inclined and more irregular on the $z$--$y$ plane. The two-stage transition scenario over the present blunt flat plates becomes apparent based on the above SPOD analysis.

\section{Conclusions}\label{sec:Conclusions}
In the present paper, the stability and transition to turbulence of high-speed boundary layers over sharp and blunt flat plates are investigated. To initiate the boundary layer transition, 3-D broadband slow-acoustic-wave disturbances are added on the freestream boundary. The frequency spectrum obeys the distribution obtained from tunnel measurements, while no preferential spanwise wavenumber is imposed in the combination of harmonics. As the nose-tip radius exceeds 2 mm, the delay trend of the transition onset location reverses. A good agreement is reached between the numerical and experimental results with respect to the reversal tendency, which is reported for the first time.

A prior 2-D receptivity simulation displays a plateau of energy growth downstream of the blunt nose region. By contrast, Mack-mode like structures appear in the sharp-plate flow, which undergo remarkable energy amplification. Linear stability analysis identifies unstable Mack second mode and oblique first mode simultaneously for the sharp-plate flow, while no significant unstable normal modes are found for the blunt-plate flow. As a result, the transition to turbulence for the considered blunt-plate flow is due to nonmodal instabilities. Following that, 3-D high-resolution simulations report a two-stage transition mechanism, including the formation of a 3-D streaky laminar flow and a subsequent secondary instability stage. In stage I, low-frequency streamwise streaks are formed immediately downstream of the nose for blunt-plate cases, whereas their signature is not evident in the sharp-plate flow. A unified preferential spanwise wavelength 0.91 mm is observed for blunt-plate cases with different nose-tip radii. In comparison, the wavelengths for the most amplified oblique first mode and the resulting streamwise-\replaced{vortex}{vorticity} mode are pronounced in the sharp-plate flow. With regard to the frequency spectra, low-frequency components within 20 kHz are dominant for the blunt-plate cases, while Mack second mode with around 1 MHz emerges in the sharp-plate flow. In stage II, intermittent turbulent spots appear in the streaky flow, convect downstream and span the spanwise domain. High-frequency components grow rapidly during this process, which arise from secondary instabilities. In this stage, the spatial and temporal spectra as well as the intermittency feature of blunt-plate flows with various nose-tip radii are similar. The crucial difference is the growing streaky response in stage I as the nose-tip radius is increased. This initial factor gives rise to earlier birth of turbulent spots and thus an advance of the transition onset with large nose bluntness. However, the transition reversal does not suggest an essentially different transition mechanism beyond the critical nose-tip radius.

The present computational data do not either contradict or confirm the roughness explanation by experimental investigators. The possible nose-tip roughness is expected to increase the streaky response in stage I, which further moves the transition onset forward with large nose bluntness. The present work also demonstrates that the employed 3-D broadband disturbance model is a useful choice for similar transition studies in the future. 


\appendix

\section{Mesh convergence study}\label{appA}
\begin{figure*}
	\centering{ \includegraphics[width=6.6cm]{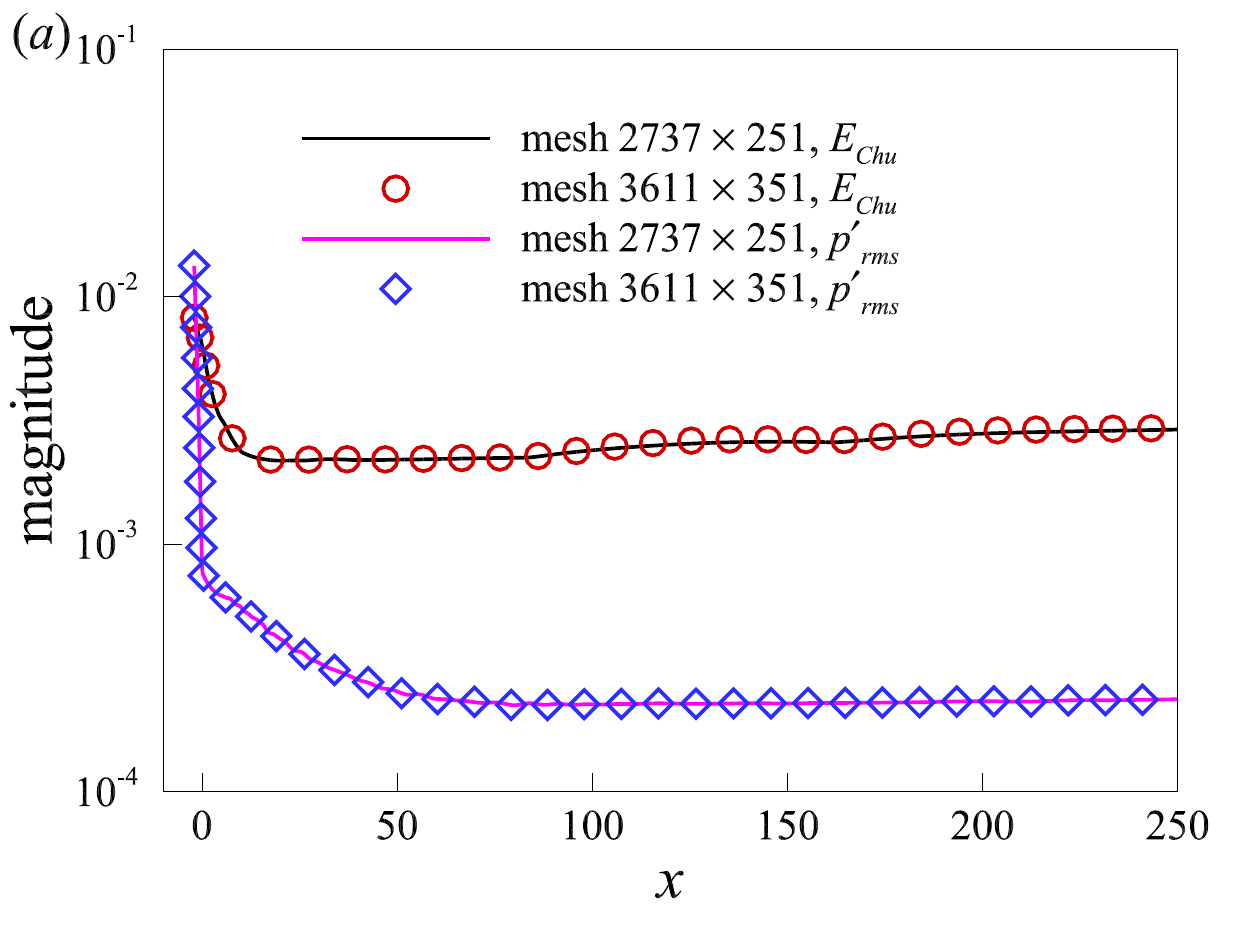}}
	\centering{ \includegraphics[width=6.6cm]{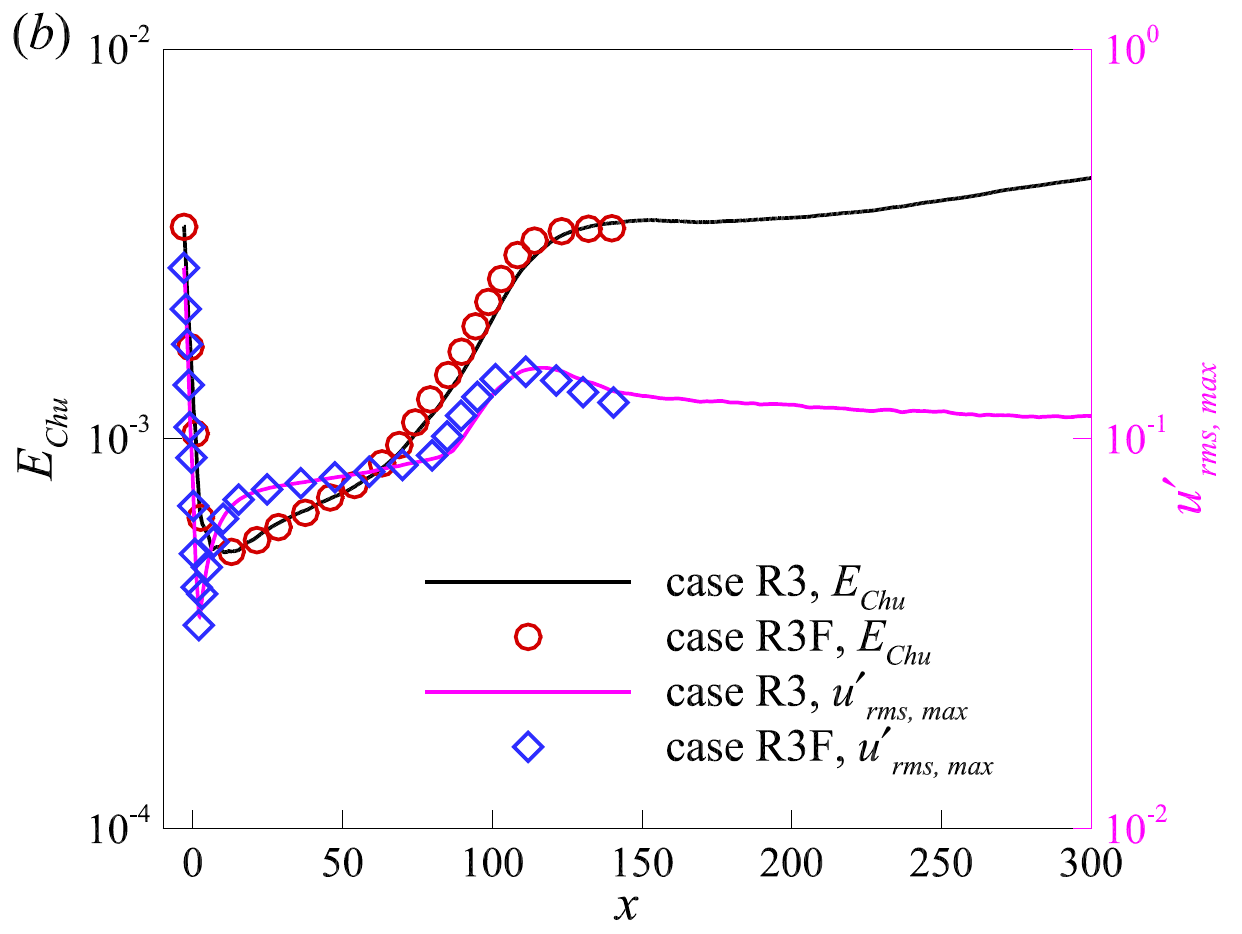}}
	\caption{Mesh resolution effect on (\textit{a}) Chu's energy and r.m.s. of the wall pressure fluctuation for 2-D case R2, and (\textit{b}) spanwise-averaged Chu's energy and maximum of the streamwise velocity fluctuation for 3-D transitional case R3.}
	\label{fig_app1}
\end{figure*}

Figure \ref{fig_app1}(\textit{a}) shows the mesh resolution effect on the 2-D receptivity study of case R2. The convergence in the evolutions of the Chu's energy and the wall pressure fluctuation r.m.s. is confirmed. This observation also implies the convergence of the base flow. For 3-D studies, the spanwise wavelength of the dominant streak, the breakdown scenario, the transition location, and other statistical quantities are examined successively. The main results are not obviously affected by increasing the mesh resolution to that of case R3F, which is close to the previous hypersonic DNS resolution of \cite{Huang2017} and \cite{duan2019characterization}. Figure \ref{fig_app1}(\textit{b}) shows the streamwise development of the Chu's energy and the maximal streamwise velocity fluctuation. The energy growth in the late transitional stage appears to be slightly affected. Nonetheless, the overall deviation is acceptable. Hence, the main conclusions keep unchanged with increasing mesh resolution.

\section{Law of the wall for mean velocity profiles}\label{appB}
\begin{figure*}
	\centering{ \includegraphics[width=8cm]{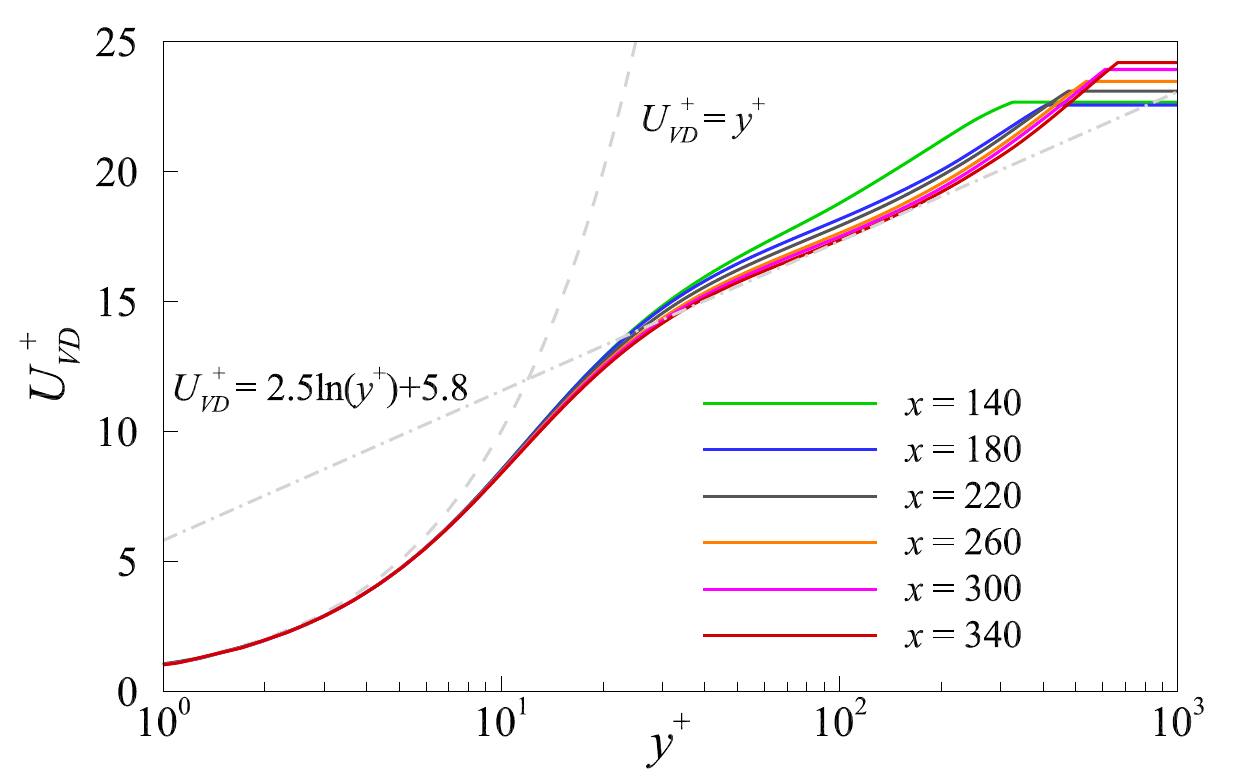}}
	\caption{The van Driest transformed mean velocity profile of case R3.}
	\label{UVD}
\end{figure*}

To further examine the fully developed turbulent region, the mean velocity profile is compared with the law of the wall. Case R3 is adopted for analysis, which possesses the longest fully developed turbulent region among the considered cases. In order to remove the influence of mean density variation due to the compressibility effect, the van Driest transformation for the mean velocity is utilised as
\begin{equation}\label{eq_UVD}
U_\textit{VD}^ +  = \int_0^{{{\bar u}^ + }} {\sqrt {{{\bar \rho } \mathord{\left/
				{\vphantom {{\bar \rho } {{{\bar \rho }_\textit{w}}}}} \right.
				\kern-\nulldelimiterspace} {{{\bar \rho }_\textit{w}}}}} } d{\bar u^ + }(y),
\end{equation}
where $\bar u^ +=\bar u/\bar u_{\tau}$. Figure \ref{UVD} plots the transformed mean velocity profiles at different streamwise locations. The viscous sublayer law $U_\textit{VD}^+=y^+$ and the log law $U_\textit{VD}^+=2.5 \text{ln}(y^+)+5.8$ are also shown. Through the van Driest transformation, the intercept of the log law is higher than the incompressible counterpart 5.0, which has been encountered in previous DNS of hypersonic wall-bounded flows \citep{Franko_Lele_2013,Guo_Heat_2022,zhu2023direct}. The slope appears to collapse onto that of the standard log law starting from around $x=180$. Combining with the Stanton number result in figure \ref{fig_St_curves}(\textit{d}), we deduce that a fully developed turbulent state has been established since then for case R3.

\section{Determination of transition onset locations}\label{appC}
\begin{figure*}
	\centering{ \includegraphics[width=8cm]{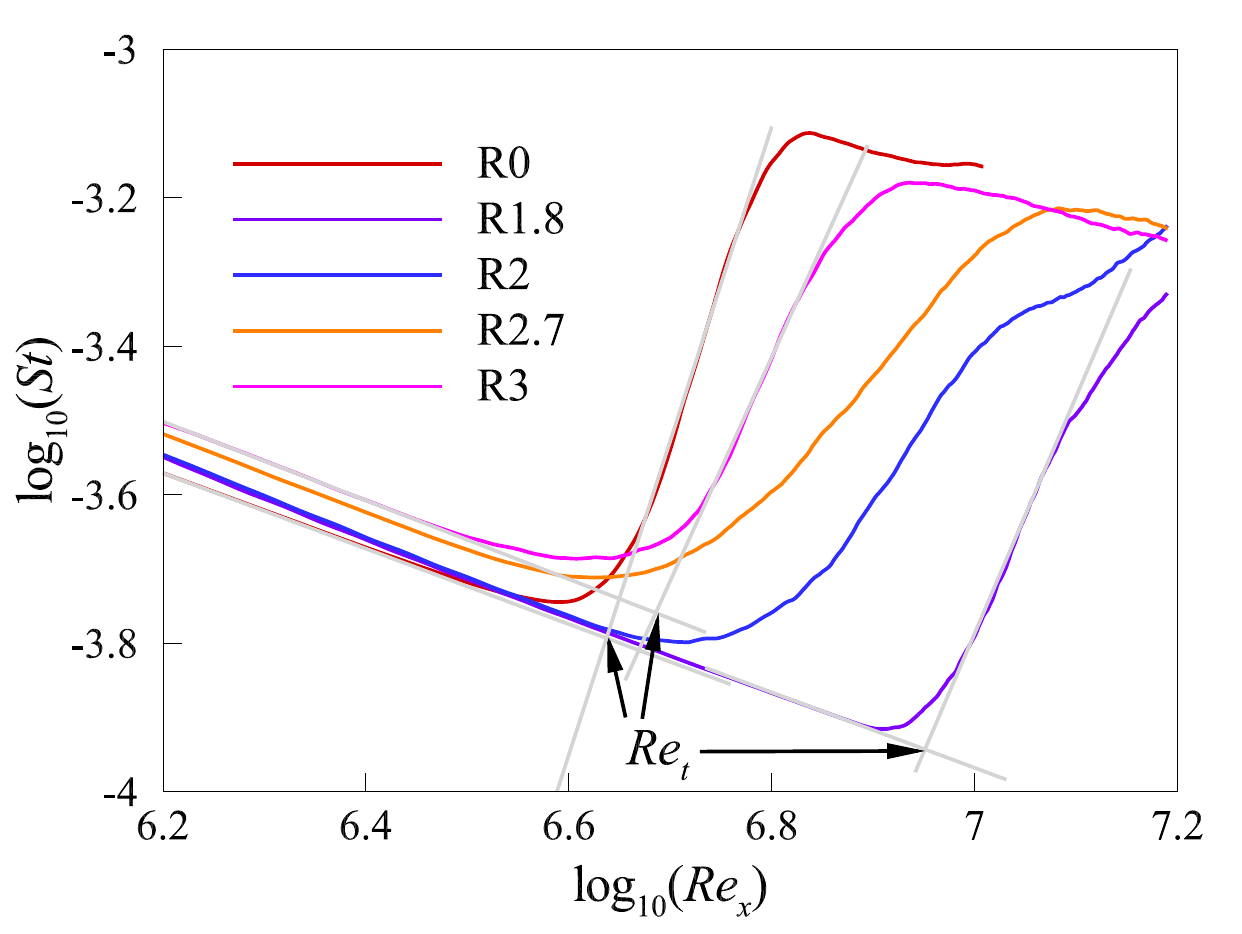}}
	\caption{Determination of the transition onset location from the $\log_{10}(\Rey_x)$--$\log_{10}(St)$ plot \citep{borovoy2022laminar}. Cases R0\added{, R1.8} and R3 are shown for an example.}
	\label{appC_onset}
\end{figure*}

To determine the transition onset location as the experimental investigator did, figure \ref{appC_onset} plots $\log_{10}(St)$ versus $\log_{10}(\Rey_x)$ for different cases. The transition onset Reynolds numbers $\Rey_t$ correspond to the displayed intersection points. Transition onset locations of cases R0\added{, R1.8} and R3 are given as an example.

\begin{figure*}
	\centering{ \includegraphics[width=8cm]{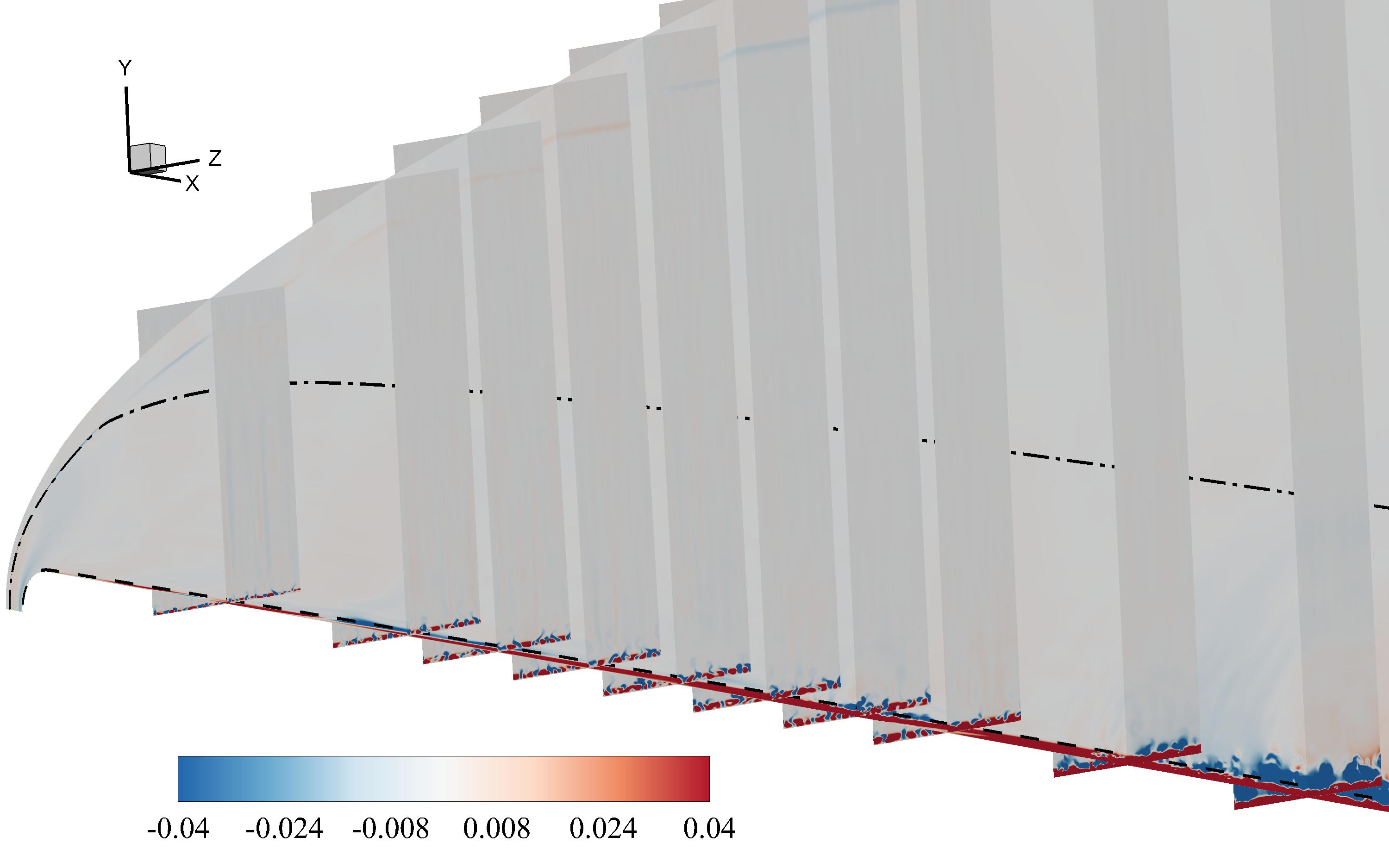}}
	\caption{Contour of the streamwise velocity perturbation $u_{\textit{perturb}}$ compared to the laminar flow for case R3 at $t=806$. Dashed and dashed-dotted lines mark the boundary layer edge and the entropy layer edge, respectively. Transverse slices are extracted and shown with $x$-coordinate of 20, 40, 50, 60, 70, 80, 90, 120 and 140 successively.}
	\label{fig_uperturb}
\end{figure*}

\begin{figure*}
	\centering{ \includegraphics[width=7cm]{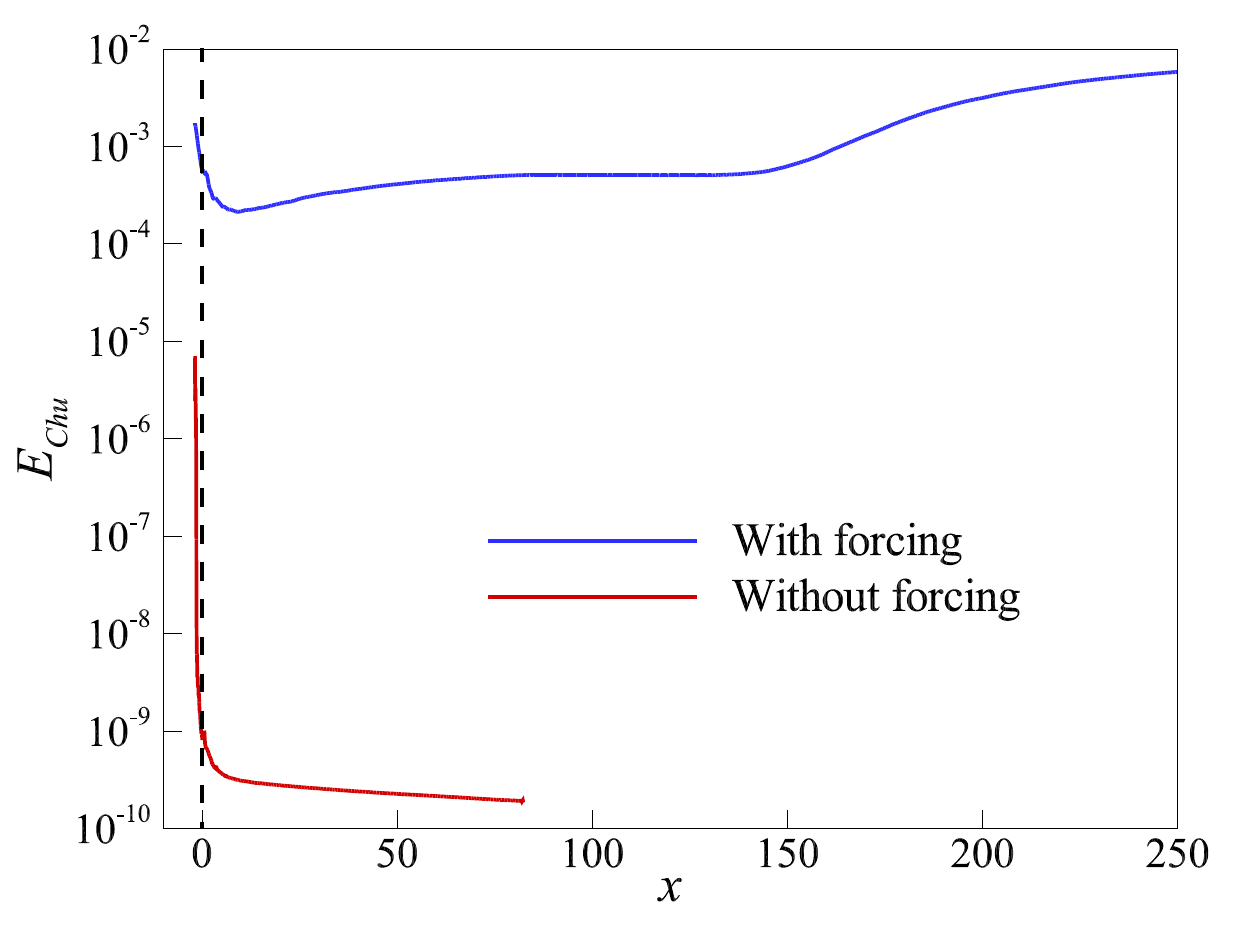}}
	\caption{\added{Chu's energy for 3-D case R1.8 with and without the freestream forcing. The vertical dashed line indicates the junction location $x =0$.}.}
	\label{appE_Chu_energy}
\end{figure*}

\section{A global view of the streamwise velocity perturbation}\label{appD}
\added{A global view of streamwise velocity perturbation is provided in figure \ref{fig_uperturb}. The result illustrates that the outer entropy layer is only marginally disturbed, and the excited disturbance is mainly concentrated inside the boundary layer.}

\section{Case test without freestream disturbances}\label{appE}
\added{Tests are conducted for case R1.8 if there is not any disturbance added on the farfield boundary. The streamwise length of the 3-D computational domain is reduced to about $x=80$. Laminar state is then confirmed in the 3-D case without external forcings. Small-amplitude high-frequency numerical noise is formed near the nose stagnation point and decays rapidly on the nearby nose. As shown by figure \ref{appE_Chu_energy}, Chu's energy without the freestream forcing is 3$\sim$6 orders of magnitude lower than the baseline case with forcing. This level of background noise should have no evident influence on the transition.}

\backsection[Funding]{This research is supported by the Hong Kong Research Grants Council (no. 15216621, no. 15217622, no. 15204322 and no. 25203721)\replaced{,}{ and} the National Natural Science Foundation of China (no. 12102377)\added{ and the Start-up Fund for RAPs by the Hong Kong Polytechnic University}.}

\backsection[Declaration of interests]{The authors report no conflict of interest.}

\bibliographystyle{jfm}

\bibliography{jfm}


\end{document}